\RequirePackage{fix-cm}

\documentclass[twocolumn]{svjour3}          
\smartqed  
\usepackage{amsmath}

\usepackage[conf={restate,no link to proof}]{proof-at-the-end}
\usepackage{comment}
\usepackage[normalem]{ulem}
\usepackage[linesnumbered,ruled,vlined]{algorithm2e}
\usepackage{etoolbox}
\usepackage{makecell}
\usepackage{graphicx}
\usepackage{float}
\usepackage{bbm}

\newbool{ARXIV}
\booltrue{ARXIV}
\usepackage{enumitem}

\usepackage{color} 
\usepackage{xargs} 
\usepackage{xspace} 

\def\notationcolor{black} 

\newcommand{\notation}[2]{\newcommand{#1}{{\textcolor{\notationcolor}{\ensuremath{#2}}}}}

\newcommand{\term}[2]{\newcommand{#1}{\textcolor{\notationcolor}{#2}\xspace}}

\notation{\permspace}{\Pi} 
\notation{\perm}{\pi} 
\notation{\e}{e} 
\notation{\one}{\vec{1}}
\notation{\zero}{\mat{0}}
\notation{\rec}{r} 
\notation{\attr}{Att} 
\notation{\numattr}{n_a} 
\notation{\attrvalue}{a} 
\notation{\datasize}{n} 
\notation{\data}{\mathcal{D}}
\notation{\mech}{\mathcal{M}} 
\notation{\datavec}{\vec{x}} 
\notation{\covar}{\mathbf{\Sigma}} 
\notation{\bmat}{\mathbf{B}} 
\notation{\comvar}{\covar_{*}} 
\notation{\comb}{\bmat_{*}} 
\notation{\commech}{\mech_{*}} 
\notation{\amat}{\mathbf{A}} 
\notation{\pcostmat}{\mathbf{\Gamma}}
\notation{\shared}{\zeta}
\notation{\nullspace}{\mathbf{H}}
\notation{\dimsize}{d}
\notation{\query}{\vec{q}}
\notation{\numquery}{m}
\notation{\senstwo}{\Delta_2}
\notation{\brank}{k}
\notation{\randalg}{\mathcal{A}}
\notation{\loewner}{\preceq}
\notation{\rloewner}{\succeq}
\notation{\rowspace}{\text{rowspace}}
\notation{\trace}{\text{trace}}
\notation{\kron}{\otimes}
\notation{\Kron}{\bigotimes}
\notation{\genericworkload}{\mat{W}}
\notation{\margworkload}{Wkload}
\notation{\varfun}{Var}
\notation{\varvec}{v}

\notation{\submat}{\mat{Sub}}
\notation{\margset}{\mat{A}} 

\notation{\marginal}{\mat{Q}} 
\notation{\identity}{\mat{\mathcal{I}}}
\notation{\onevec}{\vec{1}}
\notation{\range}{\text{range}}
\notation{\outp}{\omega}
\notation{\resid}{\mat{R}}
\notation{\closure}{\text{closure}}
\notation{\ncells}{n_{cells}} 
\notation{\cellcount}{\#cells} 
\notation{\loss}{\mathcal{L}}
\notation{\pcost}{pcost}
\notation{\sov}{SoV}
\notation{\avgv}{\text{avgvar}}

\term{\obj}{\textbf{OBJ}}

\term{\myalg}{CM}
\term{\newrp}{ResidualPlanner+}

\notation{\attwork}{\mat{W}}
\notation{\strat}{\mat{S}}
\notation{\stratcov}{\mat{C}}
\notation{\strathalf}{\mat{\Gamma}}

\let\vec\mathbf  
\let\mat\mathbf  

\AtBeginDocument{%
  \providecommand\BibTeX{{%
    \normalfont B\kern-0.5em{\scshape i\kern-0.25em b}\kern-0.8em\TeX}}}

\def\algoname{ResPlan\xspace}

\newcommand{\addcolor}[1]{\textcolor{black}{#1}}

\usepackage[utf8]{inputenc} 
\usepackage[T1]{fontenc}    
\usepackage{hyperref}  
\hypersetup{colorlinks=true,
urlcolor     = blue, 
  linkcolor    = blue, 
  citecolor   = red 
  }
\usepackage{url}            
\usepackage{booktabs}       
\usepackage{amsfonts}       
\usepackage{nicefrac}       
\usepackage{microtype}      
\usepackage{xcolor}         

\allowdisplaybreaks

\newcommand{\mystrut}{\rule[1em]{0pt}{\baselineskip}}

\DeclareMathOperator{\argmin}{\arg\min}

\begin{document}

\title{ResidualPlanner+: a scalable matrix mechanism for marginals and beyond. \thanks{This work was supported by National Science Foundation awards CNS-1931686, CNS-2317232, CNS-2317233, and a gift from Facebook.}
}


\author{Guanlin He   \thanks{First and second authors have equal contribution to this work.}        \and Yingtai Xiao  \and Levent Toksoz \and Zeyu Ding \and Danfeng Zhang \and Daniel Kifer
}


\institute{
           Guanlin He \at
              The Pennsylvania State University
              \\
              \email{gbh5146@psu.edu}
              \and
Yingtai Xiao \at 
              The Pennsylvania State University
 \\
              \email{yingtai0323@gmail.com}            \\
             \emph{Present address:} TikTok Inc.
              \and
           Levent Toksoz \at
              The Pennsylvania State University
              \\
              \email{lkt5297@psu.edu} 
              \and
           Zeyu Ding \at
              Binghamton University
              \\
              \email{dding1@binghamton.edu} 
              \and
           Danfeng Zhang \at
              Duke University
              \\
              \email{dz132@duke.edu} 
              \and
              Daniel Kifer \at
              The Pennsylvania State University
              \\
              \email{duk17@psu.edu}
}

\date{}

\maketitle

\begin{abstract}

Noisy marginals are a common form of confidentiality protecting data release and are useful for many downstream tasks such as contingency table analysis, construction of Bayesian networks, and even synthetic data generation. Matrix mechanisms, an important class of privacy mechanisms that provide unbiased noisy answers to linear queries, are often used to answer marginal queries.

We propose ResidualPlanner and \newrp, two highly scalable matrix mechanisms. ResidualPlanner is both optimal (among matrix mechanisms that use Gaussian noise) and scalable for answering marginal queries, while \newrp provides support for more general workloads, such as combinations of marginals and range queries or prefix-sum queries. ResidualPlanner can optimize for many loss functions  that can be written as a convex function of marginal variances (prior work was restricted to just one predefined objective function). ResidualPlanner can optimize the accuracy of marginals in large scale settings in seconds, even when the previous state of the art (HDMM) runs out of memory. It even runs on datasets with 100 attributes in a couple of minutes. Furthermore, ResidualPlanner can efficiently compute variance/covariance values for each marginal (prior methods quickly run out of memory, even for relatively small datasets). 

\newrp provides support for more complex workloads that combine marginal and range/prefix-sum queries (e.g., a marginal on race, a range query on age, and a combined race/age tabulation that answers age range queries for each race). It  even supports custom user-defined workloads on different attributes. With this added flexibility, \newrp is not necessarily optimal, however it is still extremely scalable and outperforms the prior state-of-the-art (HDMM) on prefix-sum queries both in terms of accuracy and speed.

\end{abstract}

\section{Introduction}\label{sec:intro}
Marginals are tables of counts on a set of attributes (e.g., how many people  there are for each combination of race and gender). They are one of the most common formats for the dissemination of statistical data \cite{census2010,tdahdsr}, studying correlations between attributes, and are  sufficient statistics for loglinear models, including Bayesian networks and Markov random fields. For this reason, a lot of work in the differential privacy literature has considered how to produce a set of noisy marginals that is both privacy-preserving and accurate.

One line of work, called the \emph{matrix mechanism} \cite{LMHMR15,YZWXYH12,li2012adaptive,yuan2015optimizing,mckenna2018optimizing,YYZH16,xiao2020optimizing,edmonds2020power,nikolov2014new} designs differentially private algorithms for answering linear queries (such as marginals). They add noise to a set of linear \emph{strategy} queries and then answer desired workload queries via linear postprocessing of the strategy noisy answers. As a result, the privacy-preserving noisy answers to the workload queries are accurate, unbiased, and have a simple distribution (e.g., multivariate normal). These crucial properties allow statisticians to work with the data, model error due to data collection (sampling error) and error due to privacy protections. It enables valid confidence intervals and hypothesis tests  and other methods for quantifying the uncertainty of a statistical analysis (e.g., \cite{gaboardi2016differentially,lei2011differentially,yu2014scalable,kakizaki2017differentially,kifer2016new}). Incidentally, sets of noisy marginals are also used to generate differentially private synthetic data (e.g., \cite{privbayes,aydore2021differentially,mckenna2019graphical,cai2021data}).

For the case of marginals, significant effort has been spent in designing optimal or nearly optimal matrix mechanisms for just a single objective function (total variance of all the desired marginals) \cite{LMHMR15,yaroslavtsev2013accurate,dingcube,YZWXYH12,yuan2015optimizing,li2013optimal,YYZH16} and each new objective function requires significant additional effort  \cite{barak2007privacy,edmonds2020power,nikolov2014new,xiao2020optimizing}.
However, existing optimal solutions do not scale and additional effort is needed to design scalable, but suboptimal, matrix mechanisms for marginals \cite{mckenna2018optimizing,mckenna2021hdmm}. Furthermore,  computing the individual variances of the desired noisy marginals is a slow process and more difficult is computing the covariance between cells in the same marginal.

Additionally, there is a need to provide support for more generalized marginals, where each attribute has its own workload type. For example, for numeric attributes like age, end-users are often interested in accurate answers to range queries on those attributes rather than only individual counts for each age value. A pure marginal on age, in contrast, would guarantee an accurate count for each age, but would not guarantee accurate counts for all age ranges (larger ranges would have proportionally more error). An alternative to range queries is the set of prefix-sum queries, which ask questions such as ``how many records have age less than 30?'' Any range query can be answered by subtracting one prefix-sum query from another.
Thus, when a numeric attribute is used, an end-user may want it to always appear in the form of a range query or a prefix-sum. For example, such a ``generalized'' marginal on (age, race) would support \emph{accurate} age ranges for each race, a generalized marginal on (age, income) would support 2-d range queries on those attributes, while (race, location) would represent a standard 2-d marginal (e.g., a count of the number of people of each race in each location).

For workloads consisting of pure marginals, we present ResidualPlanner, a novel differentially private matrix mechanism that achieves both optimality (among matrix mechanisms that use Gaussian noise) and scalability for marginal queries. Unlike prior matrix mechanisms, which only supported  specific objectives, ResidualPlanner can optimize for a wide variety of loss functions  (subject to constraints on the privacy budget) while running in seconds on datasets where existing methods exhaust memory. For more general workloads, we present \newrp. It allows users to customize the workload type for different attributes. It maintains scalability and can optimize for a wide variety of objective functions. Although not necessarily optimal, it is still very accurate and outperforms the previous state of the art (HDMM \cite{mckenna2021hdmm}) on workloads involving prefix-sum queries -- its query answers are more accurate and can quickly optimize workloads for which prior work runs out of memory.

This paper is an extension of our previous conference paper \cite{rplanner} which introduced ResidualPlanner. The current work includes the following substantial new content: (1) \newrp, a scalable generalization to more complex workloads, (2) we show how to integrate ResidualPlanner with secure noise generation (the discrete Gaussian \cite{discgauss}), which is challenging because ResidualPlanner uses \emph{correlated} multivariate Gaussian noise, but the discrete Gaussian generation algorithm does not support correlation, and (3) we show how the scalability of ResidualPlanner can allow data curators to quickly explore the unintended accuracy-loss consequences of commonly used objective functions such as the total variance of reconstructed query answers. 

To summarize, our contributions are the following.
\begin{enumerate}
    \item \textbf{ResidualPlanner}: A scalable and optimal matrix mechanism for marginals that leverages the following insights.  Since a dataset can be represented as a vector $\datavec$ of counts, and since a marginal query on a set $\margset$ of attributes can be represented as a matrix $\marginal_{\margset}$ (with $\marginal_{\margset}\datavec$ being the true answer to the marginal query), we find a new linearly independent basis that can parsimoniously represent both a marginal $\marginal_{\margset}$ and the ``difference'' between two marginals $\marginal_{\margset}$ and $\marginal_{\margset^\prime}$ (subspace spanned by the rows of $\marginal_{\margset}$ that is orthogonal to the rows of $\marginal_{\margset^\prime}$). Using parsimonious linear bases, instead of overparametrized mechanisms,  accounts for the scalability. Optimality results from a deep analysis of the symmetry that marginals impose on the optimal solution --  the same linear basis  is optimal for a wide variety of loss functions.
    \item \textbf{Discrete Gaussian Implementation}: Well-designed deployments of differentially private algorithms require the use of hardened noise-generation mechanisms such as the Discrete Gaussian \cite{discgauss}. Simply replacing the continuous Gaussian in ResidualPlanner with the discrete Gaussian distribution does not result in the same provable privacy guarantee (in fact, we show that it can be up to $2^k$ times worse for $k$-way marginals). However, we show how carefully designed transformations of the linear queries used by ResidualPlanner can support the use of Discrete Gaussian noise to obtain the same privacy and utility properties. 
    \item \textbf{Cell Fairness Analysis}: A common, and mathematically convenient, practice in the matrix mechanism literature is to design mechanisms that try to minimize the (weighted) sum of variances of queries in a workload subject to a privacy budget. We show that ResidualPlanner can optimize this objective function in closed form, and provide direct formulas for the sum of the variances of workload queries. This fast calculation allows data curators, for the first time, to rapidly analyze how this objective function impacts individual queries in the workload. As an example, we provide such an analysis to identify types of workload marginals whose accuracy gets neglected by such an objective function.
    \item \textbf{\newrp}: We propose a scalable and customizable extension of ResidualPlanner that can optimize for workloads that generalize marginals, such as mixtures of marginal and multivariate range/prefix-sum queries (e.g., instead of a 3-way marginal on age, race, and income, we can support 3-way ``generalized'' marginals that provide 2-d age/income range queries for each race). This generalization achieves state-of-the-art accuracy and scalability for workloads containing prefix-sum queries.  
\end{enumerate}

This paper is organized as follows.
In Section \ref{sec:prelim}, We present notation and background information.
We discuss related work in Section \ref{sec:related}.
We present ResidualPlanner in Section \ref{sec:rplanner}.
We show how to use discrete Gaussian noise with ResidualPlanner in Section \ref{sec:dgm}.
We show that ResidualPlanner can optimize the sum of variance objective function in closed
form and provide a formula for the reconstructed variance of different marginals in Section \ref{sec:fair}. 
In that section, we also apply these results to study how the accuracy of individual marginals of different sizes are affected by this objective function.
We present \newrp, an extension to more complex workloads, in Section \ref{sec:extension}.
We present experiments for ResidualPlanner in Section \ref{sec:experiments} and for \newrp in
Section \ref{sec:exprange}. We present conclusions and discuss future work in Section \ref{sec:conc}.
The proofs not included in the main text can be found in the appendix.
Source code is available at \url{https://github.com/cmla-psu/ResidualPlannerPlus}.

\section{Preliminaries}\label{sec:prelim}
A dataset  $\data=\{\rec_1,\dots,\rec_\datasize\}$ is a collection of records. Each record $\rec_i$ contains $\numattr$ attributes $\attr_1,\dots, \attr_{\numattr}$ and each attribute $\attr_j$ can take values $\attrvalue^{(j)}_1,\dots,\attrvalue^{(j)}_{|\attr_j|}$. An attribute value $\attrvalue^{(j)}_i$ for attribute $\attr_j$ can be represented as a vector using one-hot encoding. Specifically, let $\e^{(j)}_i$ be a row vector of size $|\attr_j|$ with a 1 in component $i$ and $0$ everywhere else. In this way $\e^{(j)}_i$ represents the attribute value $\attrvalue^{(j)}_i$.
A record $\rec$ with attributes $\attr_1=\attrvalue^{(1)}_{i_1}$, $\attr_2=\attrvalue^{(2)}_{i_2}$, $\dots, \attr_{\numattr}=\attrvalue^{(\numattr)}_{i_{\numattr}}$ can  be represented as the Kronecker product $\e^{(1)}_{i_1}\kron \e^{(2)}_{i_2}\kron \cdots \kron \e^{(\numattr)}_{i_{\numattr}}$. This vector has a 1 in exactly one position and 0s everywhere else. The position of the 1 is the \emph{index} of record $\rec$. With this notation, a dataset $\data$ can be represented as a vector $\datavec$ of integers. The value at index $i$ is the number of times the record associated with index $i$ appears in $\data$. The number of components in this vector is denoted as $\dimsize=\prod_{i=1}^{\numattr} |\attr_i|$.
Given a subset $\margset$ of attributes, a \emph{marginal query} on $\margset$ is a table of counts: for each combination of values for the attributes in $\margset$, it provides the number of records in $\data$ having those attribute value combinations. The marginal query can be represented as a Kronecker product $\marginal_{\margset} =\mat{V}_1 \kron\cdots\kron\mat{V}_{\numattr}$ where $\mat{V}_i$ is the row vector of all ones (i.e, $\one^T_{|\attr_i|}$) if $\attr_i\notin \margset$ and $\mat{V}_i$ is the identity matrix $\identity_{|\attr_i|}$ if $\attr_i\in\margset$. The answer to the marginal query is obtained by evaluating the matrix-vector product $\marginal_{\margset}\datavec$.
For convenience, the notation we will be using in this paper is summarized in Table \ref{tab:notation}.

\begin{table}[ht!]
\begin{center}
\caption{Table of Notation}\label{tab:notation}
\begin{tabular}{|cp{0.7\linewidth}|}\hline
$\data$:& Dataset \\
$\rec_i$: & $i^\text{th}$ record in $\data$\\
$\numattr:$ &number of attributes each record has\\
$\attr_j$: &$j^\text{th}$ attribute, whose possible values are $\attrvalue_1^{(j)},\dots, \attrvalue^{(j)}_{|\attr_j|}$\\
$\dimsize$:     & Number of possible records: $\dimsize = \prod\limits_{j=1}^{\numattr} |\attr_j|$\\
$\datavec$: & Representation of $\data$ as a $\dimsize$-dimensional vector of counts (e.g., histogram) \\
$\margset$: &(Sub)set of attributes\\
$\one_{k}$: &the $k$-dimensional vector whose entries are all $1$.\\
$\identity_{k}$: &the $k\times k$ identity matrix\\
$\pcost(\mech)$: &Privacy cost of a Gaussian linear mechanism $\mech(\datavec)\equiv\bmat\datavec + N(\zero, \covar)$. It is defined as the largest diagonal of $\bmat^T\covar^{-1}\bmat$. Differential privacy parameters can be computed from $\pcost(\mech)$.\\
$\margworkload$: & A query workload represented as a collection of subsets of attributes. \\
$\closure$: & $\closure(\margworkload)$ is the collection of all subsets of elements in $\margworkload$.\\
$\dagger$: & The operator that gives the pseudo-inverse of a matrix\\
$\submat_m$: & Subtraction matrix used to define residual bases\\
$\resid_{\margset}$: &  Residual matrix used to define base mechanisms\\
$\sigma_{\margset}$:& Data independent noise scale parameter\\
$\mech_{\margset}$:& The base mechanism defined as $\mech_{\margset}(\datavec)\equiv\resid_{\margset}\datavec + N(\zero, \sigma^2_{\margset}\covar_{\margset})$. It uses a data independent noise parameter $\sigma^2_{\margset}$\\
$\outp_{\margset}$: & noisy output of mechanism $\mech_{\margset}$\\
$\attwork_i$: & Basic matrix for $\attr_i$ (Section \ref{sec:extension})\\
$\strat_i$: & Strategy replacement matrix (Section \ref{sec:extension})\\
\hline
\end{tabular}
\end{center}
\end{table}

\begin{example}\label{ex:running}
As a running example, consider a dataset in which there are two attributes: $\attr_1$ with values ``\emph{yes}'' and ``\emph{no}'', and $\attr_2$ with values ``\emph{low}'', ``\emph{med}'', ``\emph{high}''. The record (\emph{no}, \emph{med}) is represented by the kron product 
$\left[\begin{smallmatrix}0 & 1
\end{smallmatrix}\right] \kron \left[\begin{smallmatrix}0 & 1 & 0
\end{smallmatrix}\right]$ and the marginal query on the set $\margset=\{\attr_1\}$ is represented as 
$\marginal_{\{\attr_1\}} = \left[\begin{smallmatrix}1 & 0\\0 & 1
\end{smallmatrix}\right]\kron \left[\begin{smallmatrix}1 & 1 & 1
\end{smallmatrix}\right]$. Similarly, the marginal on attribute $\attr_2$ is represented as $\marginal_{\{\attr_2\}}=\left[\begin{smallmatrix}1 1
\end{smallmatrix}\right]\kron \left[\begin{smallmatrix}1 & 0 & 0\\0 & 1 & 0\\0 & 0 & 1
\end{smallmatrix}\right]$. The query representing all one-way marginals is obtained by stacking them: $\marginal^{\text{1-way}}=\left[\begin{smallmatrix}\marginal_{\{\attr_1\}}\\\marginal_{\{\attr_2\}}\end{smallmatrix}\right]$ and $\marginal^{\text{1-way}}\datavec$ consists of the five query answers (number of records with $\attr_1=yes$, number with $\attr_1=no$, number with $\attr_2=$low, etc.).
\end{example}

\subsection{Differential Privacy}\label{subsec:prelim:dp}
A mechanism $\mech$ is an algorithm whose input is a dataset and whose output provides privacy protections. Differential privacy is a family of privacy definitions that guide the behavior of mechanisms so that they can inject enough noise to mask the effects of any individual. There are many versions of differential privacy that support  Gaussian noise, including approximate DP, zCDP, and Gaussian DP.

\begin{definition}[Differential Privacy]
Let $\mech$ be a mechanism.  For every pair of datasets $\data_1,\data_2$ that differ on the presence/absence of a single record and for all (measurable) sets $S\subseteq \range(\mech)$,
\begin{itemize}[leftmargin=*,parsep=0pt]
\item If $P(\mech(\data_1)\in S)\leq e^\epsilon P(\mech(\data_2)\in S) + \delta$ then $\mech$ satisfies $(\epsilon,\delta)$-approximate differential privacy \cite{dworkKMM06:ourdata};
\item If $\Phi^{-1}(P(\mech(\data_1)\in S))\leq \Phi^{-1}(P(\mech(\data_2)\in S))+\mu$, where  $\Phi$ is the cdf of the standard Gaussian distribution, then $\mech$ satisfies $\mu$-Gaussian DP \cite{fdp}.
\item If the R\'{e}nyi divergence $D_{\alpha}(\mech(\data_1)||\mech(\data_2))$ between the output distributions of $\mech(\data_1)$ and $\mech(\data_2)$ satisfies $D_{\alpha}(\mech(\data_1)||\mech(\data_2))\leq \rho\alpha$ for all $\alpha>1$, then $\mech$ satisfies $\rho$-zCDP \cite{zcdp}.
\end{itemize}
\end{definition}

Queries that are linear functions of the data vector $\datavec$ can be answered privately  using the \emph{linear Gaussian mechanism}, which adds correlated Gaussian noise to a linear function of $\datavec$, as follows.

\begin{definition}[Linear Gaussian Mechanism \cite{xiao2020optimizing}]\label{def:lgm}
Given a $\numquery\times \dimsize$ matrix $\bmat$ and $\numquery\times\numquery$ covariance matrix $\covar$, the (correlated) linear Gaussian mechanism $\mech$ is defined as $\mech(\datavec)=\bmat \vec{x} + N(\vec{0},\covar)$. The privacy cost matrix of $\mech$ is defined as $\bmat^T\covar^{-1}\bmat$. The privacy cost of $\mech$, denoted by $\pcost(\mech)$, is the largest diagonal of the privacy cost matrix and is used to compute the privacy parameters: 
\begin{itemize}
\item $\mech$ satisfies $\rho$-zCDP with $\rho=\pcost(\mech)/2$ \cite{xiao2020optimizing}, 
\item $\mech$ satisfies $(\epsilon,\delta)$-approximate DP with 
$\delta=$\\$\Phi\left(\frac{\sqrt{\pcost(\mech)}}{2}-\frac{\epsilon}{\sqrt{\pcost(\mech)}}\right)$\\$- e^\epsilon \Phi\left(-\frac{\sqrt{\pcost(\mech)}}{2}-\frac{\epsilon}{\sqrt{\pcost(\mech)}}\right).$ \\This is an increasing function of $\pcost(\mech)$ \cite{BW18},
\item $\mech$  satisfies $\mu$-Gaussian DP with $\mu=\sqrt{\pcost(\mech)}$ \cite{fdp,xiao2020optimizing}.  
\end{itemize}
\end{definition}

The use of a non-identity covariance matrix will help simplify the description of the optimal choices of $\bmat$ and $\covar$.
%
We note that an algorithm $\mech^*$ that releases the outputs of multiple linear Gaussian mechanisms $\mech_1,\dots, \mech_k$ (with $\mech_i(\datavec)=\bmat_i\datavec + N(\zero,\covar_i)$ )
is again a linear Gaussian mechanism. It is represented as $\mech^*(\datavec)=\bmat^*\datavec+N(\zero,\covar^*)$ with the matrix $\bmat^*$ obtained by vertically stacking the $\bmat_i$ and with covariance $\covar^*$ being a block-diagonal matrix where the blocks are the $\covar_i$. Its privacy cost $\pcost(\mech^*)=\pcost(\mech_1,\dots,\mech_k)$ is the largest diagonal entry of $\sum_{i=1}^k \bmat_i^T\covar_i^{-1}\bmat_i$.

\subsection{Matrix Mechanism}
The Matrix Mechanism \cite{LMHMR15,YZWXYH12,li2012adaptive,yuan2015optimizing,mckenna2018optimizing,mckenna2021hdmm,YYZH16,xiao2020optimizing,edmonds2020power,nikolov2014new} is a framework for providing unbiased privacy-preserving answers to a workload of linear queries, represented by a matrix $\genericworkload$ (so that the true non-private answer to the workload queries is $\genericworkload\datavec$). The matrix mechanism framework consists of 3 steps: \emph{select}, \emph{measure}, and \emph{reconstruct}. The purpose of the \emph{select} phase is to determine \emph{what} we add noise to and \emph{how much} noise to use. More formally, when a user's preferred noise distribution is Gaussian, the select phase  chooses a  Gaussian linear mechanism $\mech(\datavec)\equiv \bmat\datavec+N(\zero, \covar)$ whose noisy output can be used to estimate the true query answer $\genericworkload\datavec$. Ideally, $\mech$ uses the least amount of noise subject to privacy constraints (specified by a privacy definition and settings of its privacy parameters). 
The \emph{measure} phase runs the mechanism on the data to produce (noisy) privacy-preserving outputs $\outp=\mech(\datavec)$. The \emph{reconstruct} step uses $\outp$ to compute an unbiased estimate of $\genericworkload\datavec$. The unbiased estimate is typically $\genericworkload(\bmat^T\covar^{-1}\bmat)^{\dagger}\bmat^T\covar^{-1}\outp$, where $\dagger$ represents the Moore-Penrose pseudo-inverse. This is the best linear unbiased estimate of $\genericworkload\datavec$ that can be obtained from $\outp$ \cite{LMHMR15}. This means that the goal of the select step is to optimize the choice of $\bmat$ and $\covar$ so that the reconstructed answer is as accurate as possible, subject to privacy constraints. Ideally, a user would specify their accuracy requirements using a loss function, but existing matrix mechanisms do not allow this flexibility -- they hard-code the loss function. The reason is each loss function requires significant research and new optimization algorithms \cite{yuan2015optimizing,xiao2020optimizing,edmonds2020power}. On top of this, existing optimal matrix mechanism algorithms do not scale, while scalable matrix mechanisms are not guaranteed to produce optimal solutions \cite{mckenna2018optimizing}. Additionally, the reconstruction phase should also compute the variance of \underline{each} workload answer. The variances are the diagonals of $\genericworkload(\bmat^T\covar^{-1}\bmat)^{\dagger}\genericworkload^T$ and making this computation scale is also challenging.

\section{Additional Related Work}\label{sec:related}
The marginal release mechanism by Barak et al. \cite{barak2007privacy} predates the matrix mechanism \cite{LMHMR15,YZWXYH12,li2012adaptive,yuan2015optimizing,dingcube,qardaji2014priview,mckenna2018optimizing,YYZH16,xiao2020optimizing,edmonds2020power,nikolov2014new,mckenna2021hdmm} and adds noise to the Fourier decomposition of marginals. We add noise to a \emph{different} basis, resulting in the scalability and optimality properties. The SVD bound \cite{li2013optimal} is a lower bound on total matrix mechanism error when the loss function is the sum of variances. This lower bound is tight for marginals and we use it as a sanity check for our results and implementation (note ResidualPlanner  provides optimal solutions even when the SVD bound is infeasible to compute).

Alternative approaches to the matrix mechanism can produce privacy preserving marginal query answers that reduce variance by adding bias. This is often done by generating differentially private synthetic data or other such data synopses from which marginals can be computed. State-of-the-art approaches iteratively ask queries and fit synthetic data to the resulting answers \cite{hardt2012simple,liu2021leveraging,aydore2021differentially,gaboardi2014dual,mckenna2022aim,liu2021iterative,vietri2020new,zhang2021privsyn}. For such mechanisms, it is difficult to estimate error of a query answer but recently AIM \cite{mckenna2022aim} has made progress in upper bounding the error. PGM \cite{mckenna2019graphical} provides a connection between the matrix mechanism and this line of work, as it can postprocess noisy marginals into synthetic data. It is a better alternative to sampling a synthetic dataset from models fit to carefully chosen marginals \cite{privbayes,chen2015differentially,zhang2019differentially,cai2021data}. Synthetic data for answering marginal queries can also be created from random projections \cite{xu2017dppro},  copulas \cite{li2014differentially,asghar2019differentially}, and deep generative models \cite{jordon2019pate,abay2019privacy,liu2021iterative}.

Although data-dependent noise approaches can sometimes be more accurate than matrix mechanisms, we focus on matrix mechanisms for the following reasons:
\begin{itemize}[leftmargin=*]
    \item Many important applications, like the 2020 Decennial Census \cite{tdahdsr} used a matrix mechanism to create noisy query answers and then converted them into microdata. This post processing improved accuracy in ways that are similar to data-dependent algorithms. 
    \item The noisy query answers provided by matrix mechanisms are unbiased and have simple distributions that are easier for statisticians to analyze. In particular, matrix mechanisms that use Gaussian noise provide query answers whose distributions are multivariate Gaussians.
    \item Matrix mechanisms can provide tight accuracy guarantees before accessing any data. This can be useful for planning purposes when designing disclosure avoidance systems.
\end{itemize}

With respect to the matrix mechanism, the reconstruction step is often one of the bottlenecks to scalability. While PGM \cite{mckenna2019graphical} provides one solution, another proposal by McKenna et al. \cite{mckenna2021relaxed} is to further improve scalability by sacrificing some consistency (the answers to two different marginals may provide conflicting answers to submarginals they have in common).
Work on differential privacy marginals has also seen extensions to hierarchical datasets, in which records form meaningful groups that need to be queried. That is, in addition to marginals on characteristics of people,  marginals can be computed in different hierarchies such as geographic level (state, county, etc) and marginals on household composition (or other groupings of people) \cite{tdahdsr,kuo2018differentially,liu2022private}.

ResidualPlanner was introduced in our prior conference paper \cite{rplanner}. A recent extension by Mullins et al. \cite{mullins} shows how to add nonnegativity constraints when reconstructing the noisy measurements produced by ResidualPlanner. Our extensions over the ResidualPlanner conference paper \cite{rplanner} are listed in Section \ref{sec:intro}.

\section{ResidualPlanner}\label{sec:rplanner}
 ResidualPlanner is our proposed matrix mechanism  for optimizing the accuracy of marginal queries with Gaussian noise. It is optimal and more scalable than existing approaches. It supports optimizing the accuracy of marginals under a wide variety of loss functions and provides exact variances/covariances of the noisy marginals in closed form. In this section, we first explain the loss functions it supports. We then describe the base mechanisms it uses to answer marginal queries. We next show how to reconstruct the marginal queries from the outputs of the base mechanisms and how to compute their variances in closed form. We then explain how to optimize these base mechanisms for different loss functions. The reason this selection step is presented last is because it depends on the closed form variance calculations. Then we analyze computational complexity.

\subsection{Loss Functions Supported by ResidualPlanner}\label{sec:residloss}
The loss functions we consider are based on the following principle: different marginals can have different relative importance but within a marginal, its cells are equally important. That is, a loss function can express that the two-way marginal on the attribute set $\{$Race, MaritalStatus$\}$ is more important (i.e., requires more accuracy) than the 1-way marginal on $\{$EducationLevel$\}$, but all cells within the $\{$Race, MaritalStatus$\}$ marginal are equally important. This is a commonly accepted principle for answering differentially private marginal queries (e.g., \cite{LMHMR15,YZWXYH12,li2012adaptive,yuan2015optimizing,mckenna2018optimizing,YYZH16,xiao2020optimizing,edmonds2020power,nikolov2014new,mckenna2022aim,aydore2021differentially,liu2021leveraging}) and is also true for the 2020 Census redistricting data \cite{tdahdsr}.

Let $\margworkload=\{\margset_1,\dots,\margset_k\}$ be a workload consisting of marginals, where each $\margset_i$ is a subset of attributes and represents a marginal. E.g., $\margworkload=\{\{$Race, MaritalStatus$\}$, $\{$EducationLevel$\}\}$ consists of 2 marginals, a two-way marginal on Race/MaritalStatus, and a one-way marginal on Education.  Let $\mech$ be a Gaussian linear mechanism whose output can be used to reconstruct unbiased answers to the marginals in $\margworkload$.
For each $\margset_i\in\margworkload$, let $\varfun(\margset_i;\mech)$ be the function that returns the variances of the reconstructed answers to the marginal on $\margset_i$; the output of $\varfun(\margset_i;\mech)$ is a vector $\varvec_i$ with one component for each cell of the marginal on $\margset_i$. A loss function $\loss$ aggregates all of these vectors together: $\loss(\varvec_1,\dots,\varvec_k)$. We have the following regularity conditions on the loss function.

\begin{definition}[Regular Loss Function]\label{def:regular}
We say the loss function $\loss$ is \emph{regular} if: (1) $\loss$ is convex and continuous; (2)
$\loss(\varvec_1,\dots,\varvec_k)$ is minimized when all the $\varvec_i$ are the 0 vectors; and (3)
for any $i$, permuting just the components of  $\varvec_i$ does not affect the value of $\loss(\varvec_1,\dots,\varvec_k)$. This latter condition just says that cells within the \emph{same} marginal are equally important.
\end{definition}
Loss functions used on prior work are all regular. For example, weighted sum of variances \cite{LMHMR15,YZWXYH12,li2012adaptive,yuan2015optimizing,mckenna2018optimizing,YYZH16} can be expressed as $\loss(\varvec_1,\dots, \varvec_k)=\sum_i c_i\one^T\varvec_i$, where the $c_i$ are the nonnegative weights that indicate the relative importance of the different marginals. Another popular loss function is maximum (weighted) variance \cite{xiao2020optimizing,edmonds2020power,nikolov2014new}, which is expressed as $\loss(\varvec_1,\dots, \varvec_k)=\max\left\{\frac{\max(\varvec_1)}{c_1},\dots, \frac{\max(\varvec_k)}{c_k} \right\}$.
Thus, the optimization problem that the selection step needs to solve is either privacy constrained: minimize  loss while keeping  privacy cost (defined at the end of Section \ref{subsec:prelim:dp}) below a threshold  $\gamma$; or utility constrained: minimize privacy cost such that the loss is at most $\gamma$.
\begin{align}
\lefteqn{\textbf{Privacy constrained:}}\nonumber\\
&\quad\arg\min_{\mech} \loss(\varfun(A_1;\mech),\dots, \varfun(A_k;\mech))\label{eqn:privacyconstrained}\\
&\qquad \textbf{ s.t. }\pcost(\mech)\leq \gamma \nonumber\\
\lefteqn{\textbf{Utility constrained:}} \nonumber\\
&\quad\arg\min_{\mech} \pcost(\mech)\label{eqn:accuracyconstrained}\\
&\qquad\textbf{ s.t. }\loss(\varfun(A_1;\mech),\dots, \varfun(A_k;\mech))\leq \gamma\nonumber
\end{align}

\subsection{Base Mechanisms used by ResidualPlanner}\label{subsec:margbase}
As long as the loss function $\loss$ is regular, we will show that an optimal mechanism can be constructed from a set of base mechanisms that we describe here. We define a \emph{subtraction matrix} $\submat_m$ to be an $(m-1)\times m$ matrix where the first column is filled with  1, entries of the form $(i,i+1)$ are -1, and all other entries are 0. For example, $\submat_3= \left[\begin{smallmatrix} 
 1 & -1 & 0 \\
 1  & 0 & -1 
 \end{smallmatrix}\right]$ and $\submat_2=\left[\begin{smallmatrix} 
 1 & -1 
 \end{smallmatrix}\right]$. 
We use these subtraction matrices to define special matrices called \emph{residual matrices} that are important for our algorithm.

 For any subset $\margset\subseteq\{\attr_1,\dots,\attr_{\numattr}\}$ of attributes,
we define the \emph{residual matrix} $\resid_{\margset}$ as the Kronecker product $\resid_{\margset}=\mat{V}_1\kron\cdots\kron\mat{V}_{\numattr}$, where $\mat{V}_i=\one^T_{|\attr_i|}$ if $\attr_i\notin \margset$ and $\mat{V}_i=\submat_{|\attr_i|}$ if $\attr_i\in\margset$. Continuing Example \ref{ex:running}, we have 
$\resid_{\emptyset}=\left[\begin{smallmatrix}1 & 1\end{smallmatrix}\right]\kron\left[\begin{smallmatrix}1 & 1 & 1\end{smallmatrix}\right]$, and
$\resid_{\{\attr_1\}}=\left[\begin{smallmatrix}1 & -1\end{smallmatrix}\right]\kron\left[\begin{smallmatrix}1 & 1 & 1\end{smallmatrix}\right]$, and
$\resid_{\{\attr_2\}}=\left[\begin{smallmatrix}1 & 1\end{smallmatrix}\right]\kron\left[\begin{smallmatrix}1 & -1 & 0\\1 &0 &-1\end{smallmatrix}\right]$, and
$\resid_{\{\attr_1,\attr_2\}}=\left[\begin{smallmatrix}1 & -1\end{smallmatrix}\right]\kron\left[\begin{smallmatrix}1 & -1 & 0\\1 &0 &-1\end{smallmatrix}\right]$.

Using subtraction matrices, we also define the matrix $\covar_{\margset}$  as the Kronecker product $\bigotimes\limits_{\attr_i\in\margset} (\submat_{|\attr_i|}\submat_{|\attr_i|}^T)$ and we note that it is proportional to $\resid_{\margset}\resid_{\margset}^T$. $\covar_{\emptyset}$ is defined as 1. 
Each subset $\margset$ of attributes can be associated with a ``base'' mechanism $\mech_{\margset}$ that
takes as input the data vector $\datavec$ and  a scalar parameter $\sigma_{\margset}^2$ for controlling how noisy the answer is. $\mech_{\margset}$ is defined as:
 \begin{align}
 \mech_{\margset}(\datavec; \sigma_{\margset}^2)&\equiv \resid_{\margset}\datavec + N(\zero,\sigma_{\margset}^2\covar_{\margset})\label{eqn:base}
 \end{align}
The residual matrices $\resid_{\margset}$ used by base mechanisms form a linearly independent basis that compactly represent marginals, as the next result shows.

\begin{theoremEnd}[category=base]{theorem}\label{thm:linspace} Let $\margset$ be a set of attributes and let $\marginal_{\margset}$ be the matrix representation of the marginal on $\margset$. Then the rows of the matrices $\resid_{\margset^\prime}$, for all $\margset^\prime\subseteq \margset$, form a linearly independent basis of the row space of $\marginal_{\margset}$. Furthermore, if $\margset^\prime\neq \margset^{\prime\prime}$ then  $\resid_{\margset^\prime}\resid^T_{\margset^{\prime\prime}}=\zero$ (they are mutually orthogonal).
\end{theoremEnd}
\begin{proofEnd}
Consider two sets $\margset^\prime\neq \margset^{\prime\prime}$ and represent there respective residual matrices as:
\begin{align*}
\resid_{\margset^\prime}&=\mat{V}^\prime_1\kron\cdots\kron\mat{V}^\prime_{\numattr}\\
\resid_{\margset^{\prime\prime}}&=\mat{V}^{\prime\prime}_1\kron\cdots\kron\mat{V}^{\prime\prime}_{\numattr}\\
\resid_{\margset^\prime}\resid_{\margset^{\prime\prime}}^T &= (\mat{V}^\prime_1(\mat{V}^{\prime\prime}_1)^T)\kron\cdots\kron(\mat{V}^\prime_\numattr(\mat{V}^{\prime\prime}_\numattr)^T)
\end{align*}
Since $\margset^\prime\neq \margset^{\prime\prime}$ then one of them contains an attribute, say $\attr_i$, that the other doesn't have. Therefore  either $\mat{V}^\prime_i$ or $\mat{V}^{\prime\prime}_i$ is the vector $\one^T_{|\attr_i|}$ and the other is $\submat_{|\attr_i|}$. However, $\one^T_{|\attr_i|}\submat_{|\attr_i|}^T=\zero$ and $\submat_{|\attr_i|}\one_{|\attr_i|}=\zero$ and hence $\resid_{\margset^\prime}\resid_{\margset^{\prime\prime}}^T=\zero$. 

Next, for any set $\margset^\prime$, it is clear that the row space of $\resid_{\margset^\prime}$ is contained in the row space of the marginal matrix $\marginal_{\margset^\prime}$. It is also clear that if $\margset^\prime\subseteq \margset$ then the row space of the marginal matrix $\marginal_{\margset^\prime}$ is contained in the row space of $\marginal_{\margset}$ (because $\marginal_{\margset^\prime}$ represents a sub-marginal of $\marginal_{\margset}$). Thus the rows of the matrices $\resid_{\margset^\prime}$, for all $\margset^\prime\subseteq \margset$, are contained in the rowspace of $\marginal_{\margset}$. Thus we just need to show that 
the combined rows of $\resid_{\margset^\prime}$, for all $\margset^\prime\subseteq \margset$,
 are  linearly independent and that the number of rows is the same as the number of rows of $\marginal_{\margset}$. 

First, each $\resid_{\margset^\prime}$ is a kronecker product of matrices with full row rank, and so $\resid_{\margset^\prime}$  has full row rank (therefore its rows are linearly independent). Furthermore, since $\resid_{\margset^\prime}\resid_{\margset^{\prime\prime}}^T=\zero$ whenever $\margset^\prime\neq \margset^{\prime\prime}$ this means that the row space  of $\resid_{\margset^\prime}$ is orthogonal to the row space of $\resid_{\margset^{\prime\prime}}$. Hence the combined rows of the $\resid_{\margset^\prime}$, for all $\margset^\prime\subseteq \margset$, are linearly independent.

Next, the number of rows in $\resid_{\emptyset}$ is 1 and the number of rows in $\resid_{\margset^\prime}$ is equal to $\prod\limits_{\attr_i\in\margset^\prime} (|\attr_i|-1)$ for $\margset^\prime\neq\emptyset$ and so the total number of rows in the residual matrices is $1+\sum\limits_{\substack{\margset^\prime\subseteq \margset\\\margset^\prime\neq \emptyset}}\prod\limits_{\attr_i\in\margset^\prime} (|\attr_i|-1)$. By the distributive property of multiplication, this is exactly the same as the product:
\begin{align*}
\prod\limits_{\attr_i\in\margset} \left((|\attr_i|-1) + 1\right) = \prod\limits_{\attr_i\in\margset} |\attr_i|
\end{align*}
which is the number of rows in $\marginal_{\margset}$ and that proves that the combined rows of $\resid_{\margset^\prime}$, for all $\margset^\prime\subseteq \margset$, form a linearly independent basis for the row span of $\marginal_{\margset}$.
\end{proofEnd}
\begin{remark}To build an intuitive understanding of residual matrices, consider again Example \ref{ex:running}. Both $\resid_{\emptyset}$ and $\marginal_{\emptyset}$ are the sum query (marginal on no attributes). The rows of $\resid_{\{\attr_1\}}$ span the subspace of $\marginal_{\{\attr_1\}}$ that is orthogonal to $\marginal_{\emptyset}$ (and similarly for $\resid_{\{\attr_2\}}$). The rows of $\resid_{\{\attr_1,\attr_2\}}$ span the subspace of $\marginal_{\{\attr_1,\attr_2\}}$ that is orthogonal to both $\marginal_{\{\attr_1\}}$ and $\marginal_{\{\attr_2\}}$. Hence a residual matrix spans the subspace of a marginal that is orthogonal to its sub-marginals.
\end{remark}

Theorem \ref{thm:linspace} has several important implications. If we define the downward closure of a marginal workload $\margworkload=\{\margset_1,\dots,\margset_k\}$ as the collection of all subsets of the sets in $\margworkload$ (i.e., $\closure(\margworkload)=\{\margset^\prime~:~\margset^\prime\subseteq \margset \text{ for some $\margset\in\margworkload$}\}$) 
then the theorem implies that the combined rows from $\{\resid_{\margset^\prime}~:~\margset^\prime\in \closure(\margworkload)\}$ forms a linearly independent basis for the marginals in the workload. In other words, it is a linearly independent bases for the space spanned by the rows of the marginal query matrices $\marginal_{\margset}$ for $\margset\in\margworkload$. 
Thus, in order to provide privacy-preserving answers to all of the marginals represented in $\margworkload$, we need all the mechanisms $\mech_{\margset^\prime}$ for $\margset^\prime\in\closure(\margworkload)$ -- \underline{any} other matrix mechanism that provides fewer noisy outputs cannot reconstruct unbiased answers to the workload marginals. This is proved in Theorem \ref{thm:optimal}, which also states  that optimality is achieved by carefully setting the $\sigma_{\margset}$ noise parameter for each $\mech_{\margset}$.

\begin{theoremEnd}[category=optmain]{theorem}\label{thm:optimal}
Given a marginal workload $\margworkload$ and a regular loss function $\loss$, suppose the optimization problem (either Equation \ref{eqn:privacyconstrained} or \ref{eqn:accuracyconstrained}) is feasible.
Then there exist nonnegative constants $\sigma^2_{\margset}$ for each $\margset \in \closure(\margworkload)$ (the constants do not depend on the data),  such that the optimal linear Gaussian mechanism $\mech_{opt}$ releases $\mech_{\margset}(\datavec; \sigma^2_{\margset})$ for all $\margset\in\closure(\margworkload)$. Furthermore, any matrix mechanism for this workload must release at least this many noise query answers.
%
\end{theoremEnd}
\begin{proofEnd}
Let $ALL$ represent \\$\closure(\{\attr_1,\dots, \attr_{\numattr}\})$ -- all possible subsets of attributes.
Theorem \ref{thm:pcosteigen} guarantees that there is an optimal mechanism whose privacy cost matrix $\pcostmat$ has eigenvectors equal to the rows of the residual matrices. Rows within the same residual matrix have the same eigenvalues. Since privacy cost matrices are symmetric positive semidefinite, this means that for every $\margset\in ALL$, there exists a nonnegative number $\beta_{\margset}$ such that: 
\begin{align*}
\pcostmat \resid_{\margset}^T &= \beta_{\margset}\resid_{\margset}^T
\end{align*}

By Theorem 3.5 of \cite{arxivcommon}, if two Gaussian linear mechanisms have the same privacy cost matrix then each can be obtained by linearly processing the other. Thus they have the same privacy properties (under any postprocessing invariant privacy definition) and can be used to answer the same queries with the same exact accuracies (under any measure of accuracy). Thus we just need to construct the appropriate mechanism having privacy cost matrix $\pcostmat$. 

For each $\margset$, let $\mat{Z}_{\margset}$ be a matrix with orthonormal rows that span the row space of $\resid_{\margset}$. Thus the rows of $\mat{Z}_{\margset}$ are also eigenvectors of $\pcostmat$ (having common eigenvalue $\beta_{\margset}$) and the rows of $\mat{Z}_{\margset}$ are orthogonal to the rows of $\mat{Z}_{\margset^\prime}$ for $\margset\neq \margset^\prime$ (a consequence of Theorem \ref{thm:linspace}). Thus the set of rows of the $\mat{Z}_{\margset}$ for all $\margset\in ALL$ are a complete list of the eigenvecotrs of $\pcostmat$ (the are linearly independent and span $\mathbb{R}^\dimsize$). Thus the (symmetric positive semidefinite) privacy cost matrix $\pcostmat$ can be expressed as:
\begin{align*}
\pcostmat = \sum_{\margset\in ALL} \beta_{\margset} \mat{Z}_{\margset}^T\mat{Z}_{\margset}
\end{align*}
and one mechanism that achieves this privacy cost matrix is the one that releases $\mat{Z}_{\margset}\datavec +N(\zero, \frac{1}{\beta_{\margset}}\identity)$ for each $\margset\in ALL$ for which $\beta_{\margset}\neq 0$ (i.e., we can drop the eigenvectors with eigenvalue equal to 0 as they make no difference to the privacy cost matrix). 

Now, since the rows of $\resid_{\margset}$ and $\mat{Z}_{\margset}$ are independent linear bases of the same subspace, then  there exists an invertible matrix $\mat{Y}_\margset$ such that $\resid_{\margset}=\mat{Y}_\margset\mat{Z}_\margset$. Furthermore, $\resid_{\margset}\resid_{\margset}^T$ is invertible and $\mat{Z}_{\margset}\mat{Z}_\margset^T=\identity$ by orthonormality of its rows. Therefore
\begin{align*}
\resid_{\margset}^T(\resid_\margset\resid_\margset^T)^{-1}\resid_\margset &= \mat{Z}_{\margset}^T\mat{Y}_{\margset}^{T}(\mat{Y}_{\margset}\mat{Z}_{\margset}\mat{Z}_{\margset}^T\mat{Y}^{T})^{-1}\mat{Y}_{\margset}\mat{Z}_{\margset}\\
&=\mat{Z}_{\margset}^T\mat{Y}_{\margset}^{T}\mat{Y}_{\margset}^{-T}(\mat{Z}_{\margset}\mat{Z}_{\margset}^T)^{-1}\mat{Y}_{\margset}^{-1}\mat{Y}_{\margset}\mat{Z}_{\margset}\\
&=\mat{Z}_{\margset}^T(\mat{Z}_{\margset}\mat{Z}_{\margset}^T)^{-1}\mat{Z}_{\margset}\\
&=\mat{Z}_{\margset}^T\mat{Z}_{\margset}\\& \quad\text{ by orthonormality of the rows of $\mat{Z}_{\margset}$}
\end{align*}

Thus we have 
\begin{align*}
\pcostmat = \sum_{\margset\in ALL} \beta_{\margset} \resid_{\margset}^T(\resid_{\margset}\resid_{\margset}^T)^{-1}\resid_{\margset}
\end{align*}
and a mechanism that achieves this privacy cost matrix is the one that releases $\resid_{\margset}\datavec +N(\zero, \frac{1}{\beta_{\margset}}\resid_{\margset}\resid_{\margset}^T)$ for each $\margset$ for which $\beta_{\margset}\neq 0$. 

We next note that each covariance matrices we propose to use, $\covar_{\margset}$, is proportional to $\resid_{\margset}\resid_{\margset}^T$ (they are equal up to positive rescaling). If we define the positive constants $\kappa_{\margset}$ such that $\resid_{\margset}\resid_{\margset}^T= \kappa_{\margset}\covar_{\margset}$ then we note that the  $\sigma^2_{\margset}$ in the theorem statement are equal to  $\kappa_{\margset}/\beta_{\margset}$.

Next, we show that  the eigenvalues $\beta_{\margset}>0$ for $\margset\in\closure(\margworkload)$ and 0 otherwise, so that the optimal mechanism would not make use of any submechanism $\mech_{\margset}$ for $\margset\notin\closure(\margworkload)$.

First, by Theorem \ref{thm:linspace}, the rows of the matrices $\resid_{\margset}$, for all $\margset\in\closure(\margworkload)$ form an independent linear basis for the space spanned by the rows of the marginals $\marginal_{\margset}$ for $\margset\in\margworkload$. If a noisy $\resid_{\margset}\datavec$ is not released for some $\margset\in\closure(\margworkload)$, then an unbiased  noisy answer to at least one of the workload marginals could not be computed. Hence, they must all be part of the optimal mechanism (and thus, because of linear independence, \emph{any} mechanism needs to get at least as many scalar noisy answers as this). This shows that $\beta_{\margset}>0$ for all $\margset\in\closure(\margworkload)$. On the other hand since the rows of $\resid_{\margset}$ are orthogonal to the rows of $\resid_{\margset^\prime}$ for $\margset\neq \margset^\prime$, getting answers to $\resid_{\margset^\prime}\datavec$, for $\margset^\prime\notin\closure(\margworkload)$, cannot help estimate the answers to the marginals $\marginal_{\margset}$ for $\margset\in\margworkload$ (by Theorem \ref{thm:linspace}, $\resid_{\margset^\prime}$ are orthogonal to the matrices representing these marginals when $\margset^\prime\notin\closure(\margworkload)$). Hence an optimal privacy mechanism cannot waste privacy budget on these irrelevant queries. This shows that  $\beta_{\margset^\prime}=0$for $\margset^\prime\notin\closure(\margworkload)$ and concludes the proof.
\end{proofEnd}


%

\begin{algorithm}[h]
   \DontPrintSemicolon
    $\vec{v} \gets $ compute marginal  on $\margset$ from $\datavec$\tcp{equals $\marginal_{\margset}\datavec$} \label{line:eval}
    $m\gets \prod_{\attr_i\in \margset} |\attr_i|$\;
    $\mat{H}\gets \bigotimes\limits_{\attr_i\in\margset} \submat_{|\attr_i|}$\tcp{Use implicit representation, don't expand}\label{line:measureH}
    $\vec{z}\gets N(\zero, \identity_{m})$\tcp{ independent noise}\label{line:measureZ}
    \Return $\mat{H}\vec{v} + \sigma_{\margset}\mat{H}\vec{z}$\tcp{use  kron-product/vector multiplication from \cite{mckenna2018optimizing}}
\caption{Efficient implementation of $\mech_{\margset}(\datavec; \sigma^2_{\margset})\equiv\resid_{\margset}\datavec + N(\zero,\sigma^2_{\margset}\covar_{\margset})$}\label{alg:measure}
\end{algorithm}

$\mech_{\margset}$ can be evaluated efficiently, directly from the marginal of $\datavec$ on attribute set $\margset$, as shown in Algorithm \ref{alg:measure}. It uses the  technique from \cite{mckenna2018optimizing} to perform fast multiplication between a Kronecker product and a vector. The privacy cost $\pcost(\mech_{\margset})$  of each base mechanism $\mech_{\margset}$ is also easy to compute and is given by the following theorem.

\begin{theoremEnd}[category=base,all end]{lemma}\label{lem:pcostlemma}
$\submat_{|\attr_i|}^T(\submat_{|\attr_i|}\submat_{|\attr_i|}^T)^{-1}\submat_{|\attr_i|}$ \\$= \identity_{|\attr_i|}- \frac{1}{|\attr_i|}\one_{|\attr_i|}\one^T_{|\attr_i|}$
for any $i$.
\end{theoremEnd}
\begin{proofEnd}
We note that $\submat_{|\attr_i|}$ has size $|\attr_i|-1 \times |\attr_i|$, rank $|\attr_i|-1$ and its rows are orthogonal to $\one_{|\attr_i|}$. Using the SVD decomposition, express $\submat_{|\attr_i|}=\mat{U}\mat{D}\mat{V}^T$, where $\mat{U}$ is an $|\attr_i|-1\times |\attr_i|-1$ orthogonal matrix, $\mat{D}$ is a $|\attr_i|-1\times |\attr_i|-1$ diagonal matrix, and $\mat{V}$ is an $|\attr_i|\times |\attr_i|-1$ matrix with orthogonal columns.

We note that $\mat{D}$ is invertible because the rank of $\submat_|\attr_i|$ is $|\attr_i|-1$ and the columns of $\mat{V}$ must be orthogonal to $\one_{|\attr_i|}$ because $\submat_{|\attr_i|}\one_{|\attr_i|}=\vec{0}$.
Then 
\begin{align*}
\lefteqn{    \submat_{|\attr_i|}^T(\submat_{|\attr_i|}\submat_{|\attr_i|}^T)^{-1}\submat_{|\attr_i|}}\\
&= \mat{V}\mat{D}\mat{U}^T (\mat{U}\mat{D}\mat{V}^T \mat{V}\mat{D}\mat{U}^T)^{-1} \mat{U}\mat{D}\mat{V}^T\\
&= \mat{V}\mat{D}\mat{U}^T (\mat{U}\mat{D}\mat{D}\mat{U}^T)^{-1} \mat{U}\mat{D}\mat{V}^T\\
&= \mat{V}\mat{D}\mat{U}^T \mat{U}\mat{D}^{-1}\mat{D}^{-1}\mat{U}^T \mat{U}\mat{D}\mat{V}^T\\
&=\mat{V}\mat{V}^T
\end{align*}
Now, we know that $\left[\mat{V} \quad\frac{\one_{|\attr_i|}}{\sqrt{|\attr_i|}}\right]$ is an $|\attr_i|\times|\attr_i|$ orthogonal matrix, so
\begin{align*}
    \identity_{|\attr_i|} &=
    \left[\mat{V} \quad\frac{\one_{|\attr_i|}}{\sqrt{|\attr_i|}}\right] \left[\mat{V} \quad\frac{\one_{|\attr_i|}}{\sqrt{|\attr_i|}}\right]^T \\
    &= \left[\mat{V} \quad\frac{\one_{|\attr_i|}}{\sqrt{|\attr_i|}}\right] \left[
    \begin{matrix}
        \mat{V}^T\\ \quad\frac{\one_{|\attr_i|}^T}{\sqrt{|\attr_i|}}
    \end{matrix}
    \right]\\
    &= \mat{V}\mat{V}^T + \frac{1}{|\attr_i|}\one_{|\attr_i|}\one_{|\attr_i|}^T
\end{align*}
Combining both results, we get \\$\submat_{|\attr_i|}^T(\submat_{|\attr_i|}\submat_{|\attr_i|}^T)^{-1}\submat_{|\attr_i|}$ \\$= \identity_{|\attr_i|}- \frac{1}{|\attr_i|}\one_{|\attr_i|}\one^T_{|\attr_i|}$.

 \end{proofEnd}

\begin{theoremEnd}[category=base,proof end]{theorem}\label{thm:pcost}
The privacy cost of $\mech_{\margset}(\cdot; \sigma^2_{\margset})$ with noise parameter $\sigma^2_{\margset}$ is $\pcost(\mech_\margset(\cdot; \sigma^2_\margset))=\frac{1}{\sigma^2_{\margset}}\prod_{\attr_i\in\margset}\frac{|\attr_i|-1}{|\attr_i|}$ and the pseudocode  given in Algorithm \ref{alg:measure}  correctly implements $\mech_{\margset}$. The total privacy cost of releasing the outputs of $\mech_{\margset}$ for all $\margset\in\closure(\margworkload)$ is equal to\\ $\sum_{\margset\in\closure(\margworkload)}\pcost(\mech_\margset(\cdot; \sigma^2_\margset))$.
\end{theoremEnd}
\begin{proofEnd}
Without loss of generality (and to simplify notation), assume $\margset=\{\attr_1,\dots,\attr_\ell\}$ consists of the first $\ell$ attributes.

By definition, $\pcost(\mech_{\margset}(\cdot; \sigma^2_{\margset}))$ is the largest diagonal of $\frac{1}{\sigma^2}\resid_{\margset}^T\covar_{\margset}^{-1}\resid_{\margset}$. Thus we can write:

\begin{align}
\resid_{\margset} &= \left(\bigotimes_{i=1}^\ell \submat_{|\attr_i|}\right) \kron \left(\bigotimes_{j=\ell+1}^{\numattr}\one^T_{|\attr_j|}\right)\nonumber\\
\resid^T_{\margset} &= \left(\bigotimes_{i=1}^\ell \submat^T_{|\attr_i|}\right) \kron \left(\bigotimes_{j=\ell+1}^{\numattr}\one_{|\attr_j|}\right)\nonumber\\
\mat{H} &= \left(\bigotimes_{i=1}^\ell \submat_{|\attr_i|}\right) \kron \left(\bigotimes_{j=\ell+1}^{\numattr}\left[\begin{smallmatrix}1\end{smallmatrix}\right]\right)\nonumber\\&\quad\text{(rightmost krons use $1\times 1$ matrices)}\nonumber\\
\covar_{\margset}&=\mat{H}\mat{H}^T \nonumber\\ &= \left(\bigotimes_{i=1}^\ell (\submat_{|\attr_i|}\submat_{|\attr_i|}^T)\right) \kron \left(\bigotimes_{j=\ell+1}^{\numattr}\left[\begin{smallmatrix}1\end{smallmatrix}\right]\right)\nonumber\\
\covar^{-1}_{\margset}&= \left(\bigotimes_{i=1}^\ell (\submat_{|\attr_i|}\submat_{|\attr_i|}^T)^{-1}\right) \kron \left(\bigotimes_{j=\ell+1}^{\numattr}\left[\begin{smallmatrix}1\end{smallmatrix}\right]\right)\nonumber\\
\lefteqn{
\resid_{\margset}^T\covar_{\margset}^{-1}\resid_{\margset} }\nonumber\\&=\left(\bigotimes_{i=1}^\ell \submat_{|\attr_i|}^T(\submat_{|\attr_i|}\submat_{|\attr_i|}^T)^{-1}\submat_{|\attr_i|}\right) \nonumber\\&\qquad\kron \left(\bigotimes_{j=\ell+1}^{\numattr}\one_{|\attr_j|}\left[\begin{smallmatrix}1\end{smallmatrix}\right]\one_{|\attr_j|}^T\right)\label{eqn:pcost}
\end{align}
Now, by Lemma \ref{lem:pcostlemma}, 
\begin{align}
\lefteqn{\submat_{|\attr_i|}^T(\submat_{|\attr_i|}\submat_{|\attr_i|}^T)^{-1}\submat_{|\attr_i|} }\nonumber\\
&= \identity_{|\attr_i|}- \frac{1}{|\attr_i|}\one_{|\attr_i|}\one^T_{|\attr_i|}\label{eqn:pcost2}
\end{align}
Since its diagonals are $\frac{|\attr_i|-1}{|\attr_i|}$, then combined with Equation \ref{eqn:pcost} it proves the result for $\pcost(\mech_{\margset}(\cdot,\sigma^2_{\margset}))$. 

We next consider  the correctness of Algorithm \ref{alg:measure}. First, since the marginal on $\margset$ is $\marginal_{\margset}\datavec$, we need to show that for the matrix $\mat{H}$ defined in Line \ref{line:measureH} in Algorithm \ref{alg:measure}, $\mat{H}\marginal_{\margset}\datavec=\resid_{\margset}\datavec$.  Then we can write:
\begin{align*}
\resid_{\margset} &= \left(\bigotimes_{i=1}^\ell \submat_{|\attr_i|}\right) \kron \left(\bigotimes_{j=\ell+1}^{\numattr}\one^T_{|\attr_j|}\right)\\
\marginal_{\margset} &= \left(\bigotimes_{i=1}^\ell \identity_{|\attr_i|}\right) \kron \left(\bigotimes_{j=\ell+1}^{\numattr}\one^T_{|\attr_j|}\right)\\
&\quad\text{rightmost product is a matrix with 1 row}\\
\mat{H} &= \left(\bigotimes_{i=1}^\ell \submat_{|\attr_i|}\right) \kron \left[\begin{smallmatrix}1\end{smallmatrix}\right]\\&\quad\text{(rightmost term is a $1\times 1$ matrix)}\\
\mat{H}\marginal_{\margset} &= \left(\bigotimes_{i=1}^\ell \left(\submat_{|\attr_i|}\identity_{|\attr_i|}\right)\right) \kron \left(
\left[\begin{smallmatrix}1\end{smallmatrix}\right]
\left(\bigotimes_{j=\ell+1}^{\numattr}\one^T_{|\attr_j|}\right)\right)\\
&=\resid_{\margset}
\end{align*}
Next, we note that if $\vec{z}$ is distributed as $N(0,\mat{I}_m)$ (Line \ref{line:measureZ} in Algorithm \ref{alg:measure}) then $\sigma_{\margset}\mat{H}\vec{z}$ has the distribution $N(0,\sigma_\margset^2\mat{H}\mat{H}^T)=\covar_{\margset}$ and hence the algorithm is correct.

Composition follows from (1) the correspondence between privacy cost and the square of the Gaussian DP parameters (Definition 2), (2) the fact that composition in Gaussian DP follows from the summation of the squares of the Gaussian DP parameters \cite{fdp}, and (3) the fact that there is a base mechanism for each $\margset\in\closure(\margworkload)$.
\end{proofEnd}


\subsection{Reconstruction}
Next we explain how to reconstruct unbiased answers to marginal queries from the outputs of the base mechanisms and how to compute  (co)variances of the reconstructed marginals efficiently, without any heavy matrix operations (inversion, pseudo-inverses, etc.). Then, given the closed form expressions for marginals and privacy cost (Theorem \ref{thm:pcost}), we will be able to explain in Section \ref{sec:optselect} how to optimize the $\sigma^2_{\margset}$ parameters of the base mechanisms $\mech_{\margset}$ to optimize regular loss functions $\loss$.

Since the base mechanisms were built using a linearly independent basis, reconstruction is unique -- just efficiently invert the basis. Hence, unlike PGM and its extensions \cite{mckenna2019graphical,mckenna2021relaxed} our reconstruction algorithm does not need to solve an optimization problem and can  reconstruct each marginal independently, thus allowing marginals to be reconstructed in parallel, or as needed by users. The reconstructed marginals are consistent with each other (any two reconstructed marginals agree on their sub-marginals).
Just as the subtraction matrices $\submat_{k}$ were useful in constructing the base mechanisms $\mech_{\margset}$, their pseudo-inverses $\submat_{k}^\dagger$ are useful for reconstructing noisy marginals from the noisy answers of $\mech_{\margset}$. The pseudo-inverses have a closed form. For example $\submat_{4}=
\left[\begin{smallmatrix}
1  &-1  & \phantom{ -}0 &  \phantom{ -}0\\
 1 &  \phantom{ -}0 & -1  & \phantom{ -}0\\
 1 &  \phantom{ -}0 &  \phantom{ -}0 & -1\\
\end{smallmatrix}\right]$ and $\submat_{4}^\dagger=
\frac{1}{4}\left[\begin{smallmatrix}
\phantom{ -}1 & \phantom{ -}1 & \phantom{ -}1 \\
-3 & \phantom{ -}1 & \phantom{ -}1\\
\phantom{ -}1 & -3 & \phantom{ -}1\\
\phantom{ -}1 & \phantom{ -}1 & -3\\
\end{smallmatrix}\right]$. More generally, they are expressed as follows:

\begin{theoremEnd}[category=reconstruct]{lemma}\label{lem:psuedo} For any $\attr_i$, let $\ell=|\attr_i|$. The matrix $\submat_{\ell}$ has the following  block matrix, with dimensions $\ell\times (\ell-1)$, as its pseudo-inverse (and right inverse): 
$\submat_{\ell}^\dagger=\frac{1}{\ell}\left[\begin{smallmatrix}
\one^T_{\ell-1}\\
\one_{\ell-1}\one^T_{\ell-1}-\ell\identity_{\ell-1}
\end{smallmatrix}\right]$.
\end{theoremEnd}
\begin{proofEnd}
First, if a matrix has a right inverse then that is the pseudo-inverse. Hence we just need to show that $\submat_\ell\submat_\ell^\dagger=\identity_{\ell-1}$. 

Note that the $j^\text{th}$ row of $\submat_\ell$ has a 1 in position 1, -1 in position $j+1$, and is 0 everywhere else. 

Meanwhile, the $i^\text{th}$ column of our claimed representation of $\submat^\dagger_\ell$ has a $-(\ell-1)/\ell$ in position $i+1$ and $1/\ell$ everywhere else.

Hence if $j\neq i$ then the dot product between row $j$ of $\submat_\ell$ and column $i$ of $\submat^\dagger_\ell$ is 0 since the nonzero elements of the row from $\submat_\ell$ are being multiplied by $1/\ell$ and $1/\ell$.

If $i=j$ then the corresponding first elements that are multiplied are 1 and $1/\ell$ while the elements at position $i+1$ being multiplied are $-1$ and $-(\ell-1)/\ell$. Furthermore, $1(1/\ell) + (-1)(-(\ell-1)/\ell)=1$.
\end{proofEnd}

Each mechanism $\mech_{\margset}$, for $\margset\in\closure(\margworkload)$, has a noise scale parameter $\sigma^2_{\margset}$ and a noisy output that we denote by $\outp_{\margset}$. \underline{After} we have obtained the noisy outputs $\outp_{\margset}$ for all $\margset\in\closure(\margworkload)$, we can proceed with the reconstruction phase. The reconstruction of an unbiased noisy answer for any marginal on an attribute set $\margset\in\closure(\margworkload)$ is obtained using Algorithm \ref{alg:reconstruction}. We note that to reconstruct a marginal on attribute set $\margset$, one only needs to use the noisy answers $\outp_{\margset^\prime}$ for $\margset^\prime\in\closure(\margset)$. In other words, if we want to reconstruct a marginal on attribute set $\{\attr_1,\attr_2\}$, we only need the outputs of $\mech_{\emptyset}$, $\mech_{\{\attr_1\}}$, $\mech_{\{\attr_2\}}$, and $\mech_{\{\attr_1,\attr_2\}}$ no matter how many other attributes are in the data and no matter what other marginals are in the $\margworkload$. We emphasize again, the reconstruction phase \uline{does not run the base mechanisms anymore}, it is purely post processing.

\begin{algorithm}[h]
   \DontPrintSemicolon
   \KwIn{Noise scale parameters $\sigma^2_{\margset^\prime}$ and noisy answer vector $\outp_{\margset^\prime}$ of mechanism $\mech_{\margset^\prime}$ for every $\margset^\prime\in\closure(\margset)$.}
   \KwOut{$\vec{q}$ is output as an unbiased noisy estimate of $\marginal_{\margset}\datavec$.}
    $\vec{q} \gets \zero$\;
    \For{each $\margset' \in\closure(\margset)$}
    {
    $\mat{U}\gets $ $\mat{V}_1\kron\cdots\kron\mat{V}_{\numattr}$, where $\mat{V}_i=
    \begin{cases}
    \submat^{\dagger}_{|\attr_i|} & \text{ if } \attr_i\in\margset^\prime\\
    \frac{1}{|\attr_i|}\one_{|\attr_i|} & \text{ if }\attr_i \in \margset \setminus \margset^\prime\\
     [1] &\text{ if }\attr_i \notin \margset
    \end{cases}$\label{line:reconU}\;
    $\vec{q} \gets \vec{q} + \mat{U} \outp_{\margset^\prime}$\tcp{use  kron-product/vector multiplication from \cite{mckenna2018optimizing}}
    }
\Return $\vec{q}$\label{line:reconq}\;
\caption{Reconstruct Unbiased Answers to the Marginal on $\margset$}\label{alg:reconstruction}
\end{algorithm}

\begin{theoremEnd}[category=reconstruct,all end]{lemma}\label{lem:varlemma}For any attribute $\attr_i$, let $\ell=|\attr_i|$. Then we have
$\submat_{\ell}^{\dagger}(\submat_{\ell}\submat_{\ell}^T)\submat_{\ell}^{\dagger T} = \identity_{\ell}- \frac{1}{\ell}\one_{\ell}\one^T_{\ell}$
\end{theoremEnd}
\begin{proofEnd}
Because $\submat_{\ell}$ has linearly independent rows, the pseudo-inverse of it can be expressed as,
\begin{align*}
    \submat_{\ell}^{\dagger} = \submat_{\ell}^T (\submat_{\ell} \submat_{\ell}^T)^{-1} 
\end{align*}
From lemma \ref{lem:pcostlemma} we get,
\begin{align*}
    \submat_{\ell}^{\dagger}  \submat_{\ell} &= \submat_{\ell}^T (\submat_{\ell} \submat_{\ell}^T)^{-1} \submat_{\ell}\\
    &=\identity_{\ell} - \frac{1}{\ell} \one_{\ell} \one_{\ell}^T
\end{align*}
Therefore,
\begin{align*}
\lefteqn{
 \submat_{\ell}^{\dagger}(\submat_{\ell}\submat_{\ell}^T)\submat_{\ell}^{\dagger T}}\\
    =&  (\submat_{\ell}^{\dagger}\submat_{\ell})(\submat_{\ell}^{\dagger}\submat_{\ell})^T \\
    =& (\identity_{\ell} - \frac{1}{\ell} \one_{\ell} \one_{\ell}^T )(\identity_{\ell} - \frac{1}{\ell} \one_{\ell} \one_{\ell}^T) \\
    =& \identity_{\ell} - \frac{1}{\ell} \one_{\ell} \one_{\ell}^T - \frac{1}{\ell} \one_{\ell} \one_{\ell}^T + \frac{1}{\ell^2} \one_{\ell} (\ell)\one_{\ell}^T\\
    =&  \identity_{\ell} - \frac{1}{\ell} \one_{\ell} \one_{\ell}^T
\end{align*}
\end{proofEnd}

\begin{theoremEnd}[category=reconstruct, all end]{theorem}\label{thm:alg-recon}
 Let $\margset$ be a set of attributes and let $\marginal_{\margset}$ be the matrix representation of the marginal on $\margset$. Given the matrices $\resid_{\margset^\prime}$, for all $\margset^\prime\in \closure(\margset)$, we have
 $\marginal_{\margset} = \sum\limits_{\margset^{\prime} \in \closure(\margset)} \marginal_{\margset} \resid_{\margset^\prime}^{\dagger} \resid_{\margset^\prime} $.
\end{theoremEnd}
\begin{proofEnd}
\begin{align*}
\marginal_{\margset} &= \bigotimes_{i=1}^{\numattr} \mat{K}_i \quad\text{ where, for each $i$, }\\&\phantom{\bigotimes_{i=1}^{\numattr} \mat{K}_i \quad}\mat{K}_i = 
\begin{cases}
\identity_{|\attr_i|} & \text{ if } \attr_i \in \margset\\
 \one_{|\attr_i|}^T & \text{ if } \attr_i \notin \margset
\end{cases}\\
\resid_{\margset^{\prime}} &= \bigotimes_{i=1}^{\numattr} \mat{V}_i \quad \text{where, for each $i$, } \\&\phantom{\bigotimes_{i=1}^{\numattr} \mat{V}_i \quad}\mat{V}_i = 
\begin{cases}
    \submat_{|\attr_i|} &\text{ if} \attr_i \in \margset^{\prime}\\
    \one_{|\attr_i|}^T &\text{ if} \attr_i \notin \margset^{\prime} 
\end{cases}
\intertext{It is straightforward to verify that the following is a right inverse (and hence pseudo-inverse) of $\resid_{\margset^\prime}$}
\resid_{\margset^{\prime}}^{\dagger} &= \bigotimes_{i=1}^{\numattr} \mat{V}_i^{\dagger}\quad \text{where, for each $i$, } 
\\&\phantom{\bigotimes_{i=1}^{\numattr} \mat{V}_i^{\dagger}\quad\quad}\mat{V}_i^{\dagger} = 
\begin{cases}
    \submat_{|\attr_i|}^{\dagger} &\text{ if} \attr_i \in \margset^{\prime}\\
    \frac{1}{|\attr_i|}\one_{|\attr_i|} &\text{ if} \attr_i \notin \margset^{\prime} 
\end{cases}\\
\marginal_{\margset}\resid_{\margset^{\prime}}^{\dagger} \resid_{\margset^{\prime}} &= \bigotimes_{i=1}^{\numattr} \mat{K}_i\mat{V}_i^{\dagger} \mat{V}_i \quad \text{where, for each $i$, } 
\\\mat{K}_i\mat{V}_i^{\dagger} \mat{V}_i&= 
\begin{cases}
    \submat_{|\attr_i|}^{\dagger} \submat_{|\attr_i|} &\text{ if} \attr_i \in \margset^{\prime}\\
    \frac{1}{|\attr_i|} \one_{|\attr_i|} \one_{|\attr_i|}^T &\text{ if} \attr_i \in \margset / \margset^{\prime}\\
    \one_{|\attr_i|} ^T &\text{ if} \attr_i \notin \margset
\end{cases}\\
\end{align*}

Because $\submat_{|\attr_i|}$ has linearly independent rows, the pseudo-inverse of it can be expressed as,
\begin{align*}
    \submat_{|\attr_i|}^{\dagger} = \submat_{|\attr_i|}^T (\submat_{|\attr_i|} \submat_{|\attr_i|}^T)^{-1} 
\end{align*}
From lemma \ref{lem:pcostlemma} we get,
\begin{align*}
\lefteqn{    \submat_{|\attr_i|}^{\dagger}  \submat_{|\attr_i|}}\\ &= \submat_{|\attr_i|}^T (\submat_{|\attr_i|} \submat_{|\attr_i|}^T)^{-1} \submat_{|\attr_i|}\\
    &=\identity_{|\attr_i|} - \frac{1}{|\attr_i|} \one_{|\attr_i|} \one_{|\attr_i|}^T 
\end{align*}

Therefore,
\begin{align*}
\marginal_{\margset}\resid_{\margset^{\prime}}^{\dagger} \resid_{\margset^{\prime}} &= \bigotimes_{i=1}^{\numattr} \mat{T}_i \quad \text{where, for each $i$, } \\\mat{T}_i&= 
\begin{cases}
    \identity_{|\attr_i|} - \frac{\one_{|\attr_i|} \one_{|\attr_i|}^T}{|\attr_i|}   &\text{ if} \attr_i \in \margset^{\prime}\\
    \frac{1}{|\attr_i|} \one_{|\attr_i|} \one_{|\attr_i|}^T &\text{ if} \attr_i \in \margset / \margset^{\prime}\\
    \one_{|\attr_i|} ^T &\text{ if} \attr_i \notin \margset
\end{cases}
\end{align*}

Without loss of generality (and to simplify notation), assume $\margset=\{\attr_1,\dots,\attr_\ell\}$ consists of the first $\ell$ attributes, 
\begin{align*}
   \lefteqn{ \marginal_{\margset} =\left(\bigotimes_{i=1}^{\ell} \identity_{|\attr_i|}\right) \kron \left(\bigotimes_{i=\ell+1}^{\numattr}  \one_{|\attr_i|}^T\right) }\\
\lefteqn{    \sum_{\margset^{\prime} \in \closure(\margset)} \marginal_{\margset} \resid_{\margset^\prime}^{\dagger} \resid_{\margset^\prime} }\\&= \sum_{\margset^{\prime} \in \closure(\margset)} \left(\bigotimes_{i=1}^{\numattr} \mat{T}_i \right)\\
    &= \sum_{\margset^{\prime} \in \closure(\margset)} \left(\left(\bigotimes_{i=1}^{\ell} \mat{T}_i \right) \kron \left(\bigotimes_{i=\ell+1}^{\numattr}  \one_{|\attr_i|}^T\right) \right)\\ 
    &= \left(\sum_{\margset^{\prime} \in \closure(\margset)} \left(\bigotimes_{i=1}^{\ell} \mat{T}_i \right) \right) \kron \left(\bigotimes_{i=\ell+1}^{\numattr}  \one_{|\attr_i|}^T\right) \\ 
    & \text{where, for each $i\leq\ell$, } \\
    \mat{T}_i&= 
\begin{cases}
    \identity_{|\attr_i|} - \frac{1}{|\attr_i|} \one_{|\attr_i|} \one_{|\attr_i|}^T  &\text{ if} \attr_i \in \margset^{\prime}\\
    \frac{1}{|\attr_i|} \one_{|\attr_i|} \one_{|\attr_i|}^T &\text{ if} \attr_i \in \margset / \margset^{\prime}\\
\end{cases}
\end{align*}

Because of the distributive property of the Kronecker product,
\begin{align*}
\lefteqn{\bigotimes_{i=1}^{\ell} \identity_{|\attr_i|}}\\ &= \bigotimes_{i=1}^{\ell}
 \left( \left(\identity_{|\attr_i|} - \frac{1}{|\attr_i|} \one_{|\attr_i|} \one_{|\attr_i|}^T\right) + \frac{\one_{|\attr_i|} \one_{|\attr_i|}^T}{|\attr_i|}  \right) \\
 &= \sum_{\margset^{\prime} \in \closure(\margset)} \left(\bigotimes_{i=1}^{\ell} \mat{T}_i \right)
\end{align*}

Therefore, combining everything together,
\begin{align*}
   \lefteqn{ \sum_{\margset^{\prime} \in \closure(\margset)} \marginal_{\margset} \resid_{\margset^\prime}^{\dagger} \resid_{\margset^\prime} }\\&= \left(\sum_{\margset^{\prime} \in \closure(\margset)} \left(\bigotimes_{i=1}^{\ell} \mat{T}_i \right) \right) \kron \left(\bigotimes_{i=\ell+1}^{\numattr}  \one_{|\attr_i|}^T\right) \\
    &=\left(\bigotimes_{i=1}^{\ell} \identity_{|\attr_i|} \right) \kron \left(\bigotimes_{i=\ell+1}^{\numattr}  \one_{|\attr_i|}^T\right) \\
    &=\marginal_{\margset}
\end{align*}
\end{proofEnd}

\begin{theoremEnd}[category=reconstruct]{theorem}\label{thm:var}
Given a marginal workload $\margworkload$ and positive numbers $\sigma^2_{\margset}$ for each $\margset\in\closure(\margworkload)$, let $\mech$ be the mechanism that outputs $\{\mech_{\margset}(\datavec;\sigma^2_{\margset})~:~\margset\in\closure(\margworkload)\}$ and let  $\{\outp_{\margset}~:~\margset \in \closure(\margworkload) \}$ denote the privacy-preserving noisy answers (e.g., $\outp_{\margset}=\mech_{\margset}(\datavec,\sigma^2_\margset)$). Then for any marginal on an attribute set $\margset\in\closure(\margworkload)$, Algorithm  \ref{alg:reconstruction} returns the unique linear unbiased  estimate of $\marginal_{\margset}\datavec$ (i.e., answers to the marginal query) that can be computed from the noisy differentially private answers.

The variances $\varfun(\margset;\mech)$ of all the noisy cell counts of the marginal on $\margset$  is the vector whose components are all equal to 
$$\sum_{\margset' \subseteq \margset} \left(\sigma^2_{\margset'}\prod_{\attr_i\in\margset'}\frac{|\attr_i|-1}{|\attr_i|} *\prod_{\attr_j\in( \margset /\margset')}\frac{1}{|\attr_j|^2} \right).$$ 
%
\end{theoremEnd}
\begin{proofEnd}
We first verify the correctness and uniqueness of the reconstruction in Algorithm \ref{alg:reconstruction}.
Uniqueness follows from the fact that  the rows from all the matrices $\resid_{\margset}$ (for $\margset\in\closure(\margworkload)$) are linearly independent. 

Consider Line \ref{line:reconU} from Algorithm \ref{alg:reconstruction}. It uses a $\mat{U}$ matrix that depends on both the attributes $\margset$ of the marginal one wants to compute and a subset $\margset^\prime$ of it. So, for notational dependence, we write it as 
$\mat{U}_{\margset \gets \margset'}$. It is straightforward to verify that  $\mat{U}_{\margset \gets \margset'}= \marginal_{\margset} \resid_{\margset'}^{\dagger}$. From Theorem \ref{thm:alg-recon}, $\marginal_{\margset} \datavec = \sum_{\margset' \subseteq \margset} \marginal_{\margset} \resid_{\margset'}^{\dagger} \resid_{\margset'} \datavec = \sum_{\margset' \subseteq \margset} \mat{U}_{\margset \gets \margset'} \resid_{\margset'} \datavec$, and so Algorithm \ref{alg:reconstruction} is correct because each $\outp_{\margset^\prime}$ is an unbiased noisy version of $\resid_{\margset^\prime}\datavec$. 

Having established that the $\vec{q}$ returned by Line \ref{line:reconq} in Algorithm \ref{alg:reconstruction} is an unbiased estimate of the marginal query answer $\marginal_{\margset}\datavec$, the next step is to compute the covariance matrix $E[\vec{q}\vec{q}^T]$. 

\begin{align*}
E[\vec{q}\vec{q}^T] &= E\left[\sum_{\margset' \subseteq \margset} \mat{U}_{\margset \gets \margset'} \left(\outp_{\margset^\prime}\outp_{\margset^\prime}^T \right)\mat{U}_{\margset \gets \margset'}^T\right]\\
&=\sum_{\margset' \subseteq \margset} \mat{U}_{\margset \gets \margset'} \left(\sigma_{\margset'}^2  \covar_{\margset'} \right) \mat{U}_{\margset \gets \margset'}^T
\end{align*}


Without loss of generality (and to simplify notation), assume $\margset=\{\attr_1,\dots,\attr_\ell\}$ consists of the first $\ell$ attributes, $\margset'=\{\attr_1,\dots,\attr_t\}$ consists of the first $t \leq \ell$ attributes, then $ \margset / \margset'=\{\attr_{t+1},\dots,\attr_{\ell}\}$.

By definition, $\varfun(A; \mech)$ is the  diagonal of $E[\vec{q}\vec{q}^T]=\sum_{\margset' \in \closure(\margset)} {\sigma_{\margset'}^2} \mat{U}_{\margset \gets \margset'}\covar_{\margset'}\mat{U}_{\margset \gets \margset'
}^{T}$. Thus we can write:

{\small
\begin{align}
\marginal_{\margset} &= \left(\bigotimes_{i=1}^t \identity_{|\attr_i|}\right) \kron \left(\bigotimes_{j=t+1}^{\ell}\identity_{|\attr_j|}\right) \kron \left(\bigotimes_{k=\ell+1}^{\numattr}\one^T_{|\attr_k|}\right)\nonumber\\
\resid_{\margset'} &= \left(\bigotimes_{i=1}^t \submat_{|\attr_i|}\right) \kron \left(\bigotimes_{j=t+1}^{\ell}\one^T_{|\attr_j|}\right) \kron \left(\bigotimes_{k=\ell+1}^{\numattr}\one^T_{|\attr_k|}\right)\nonumber\\
\lefteqn{\resid_{\margset'}^{\dagger}}\nonumber\\ &\hspace{-0.5cm}= \left(\bigotimes_{i=1}^t \submat^{\dagger}_{|\attr_i|}\right) \kron \left(\bigotimes_{j=t+1}^{\ell}\frac{\one_{|\attr_j|}}{|\attr_j|}\right) \kron \left(\bigotimes_{k=\ell+1}^{\numattr}\frac{\one_{|\attr_k|}}{\attr_k}\right)\nonumber
\end{align}}
\begin{align}
\lefteqn{
 \mat{U}_{\margset \gets \margset'} = \marginal_{\margset} \resid_{\margset'}^{\dagger} }\nonumber\\
 &= \left(\bigotimes_{i=1}^t \submat^{\dagger}_{|\attr_i|}\right) \kron \left(\bigotimes_{j=t+1}^{\ell}\frac{1}{|\attr_j|}\one_{|\attr_j|}\right) \kron \left(\bigotimes_{k=\ell+1}^{\numattr} [1]\right)\nonumber \\
\lefteqn{  \mat{U}_{\margset \gets \margset'}^T }\nonumber\\& = \left(\bigotimes_{i=1}^t \submat^{\dagger T}_{|\attr_i|}\right) \kron \left(\bigotimes_{j=t+1}^{\ell}\frac{1}{|\attr_j|}\one^T_{|\attr_j|}\right) \kron \left(\bigotimes_{k=\ell+1}^{\numattr} [1]\right)\nonumber\\
%
\lefteqn{\covar_{\margset'}}\nonumber\\&= 
\left(\bigotimes_{i=1}^t \submat_{|\attr_i|} \submat_{|\attr_i|}^T\right) \kron \left(\bigotimes_{j=t+1}^{\ell}\left[\begin{smallmatrix}1\end{smallmatrix}\right]\right) \kron \left(\bigotimes_{k=\ell+1}^{\numattr}\left[\begin{smallmatrix}1\end{smallmatrix}\right]\right)\nonumber\\
\lefteqn{ \mat{U}_{\margset \gets \margset'}\covar_{\margset'}\mat{U}_{\margset \gets \margset'
}^{T}}\nonumber\\&= \left(\bigotimes_{i=1}^t \submat_{|\attr_i|}^{\dagger}\submat_{|\attr_i|}\submat_{|\attr_i|}^T \submat_{|\attr_i|}^{\dagger T}\right) \nonumber\\
&\kron \left(\bigotimes_{j=t+1}^{\ell} 
\frac{1}{|\attr_j|^2} \one_{|\attr_j|}\left[\begin{smallmatrix}1\end{smallmatrix}\right] \one_{|\attr_j|}^T \right) \kron \left(\bigotimes_{k=\ell+1}^{\numattr}\left[\begin{smallmatrix}1\end{smallmatrix}\right]\right)
\label{eqn:var}
\end{align}
Now, by Lemma \ref{lem:varlemma}, 
\begin{align}
\lefteqn{\submat_{|\attr_i|}^{\dagger}\submat_{|\attr_i|}\submat_{|\attr_i|}^T \submat_{|\attr_i|}^{\dagger T}}\\
&= \identity_{|\attr_i|}- \frac{1}{|\attr_i|}\one_{|\attr_i|}\one^T_{|\attr_i|}\label{eqn:pcost3}
\end{align}
So the diagonals of $\mat{U}_{\margset \gets \margset'}\covar_{\margset'}\mat{U}_{\margset \gets \margset'
}^{T}$ can be computed by multiplying $\frac{|\attr_i|-1}{|\attr_i|}$ for each $\attr_i\in\margset^\prime$ and $1/|\attr_j|$ for each $\attr_j \in \margset\setminus\margset^\prime$. Meanwhile, the off diagonals are all the same and can be computed by multiplying $\frac{-1}{|\attr_i|}$ for each $\attr_i\in\margset^\prime$ and $\frac{1}{|\attr_j|^2}$ for each 
$\attr_j \in \margset\setminus\margset^\prime$.

Computing the variance  of the marginal query answer is therefore the summation of these quantities for all $\margset^\prime\subseteq \margset$ and is what the theorem states.


\end{proofEnd}

\subsection{Optimizing the Base Mechanism Selection}\label{sec:optselect}

We now consider how to find the optimal Gaussian linear mechanism $\mech^*$ that solves the optimization problems in Equations \ref{eqn:privacyconstrained} or \ref{eqn:accuracyconstrained}. Given a workload on marginals $\margworkload$, the optimization  involves $\varfun(\margset;\mech^*)$ for $\margset\in\margworkload$ (the variance of the  marginal answers reconstructed from the output of $\mech^*$) and $\pcost(\mech^*)$, from which the privacy parameters of different flavors of differential privacy can be computed. 

By Theorem \ref{thm:optimal},  $\mech^*$ releases $\mech_{\margset}(\datavec;\sigma^2_{\margset})$ for each $\margset\in\closure(\margworkload)$ for appropriately chosen values of $\sigma^2_{\margset}$. The privacy cost $\pcost(\mech^*)$ is the sum of the privacy costs of the $\mech_{\margset}$. Theorem \ref{thm:pcost} therefore shows that $\pcost(\mech^*)$ is a positive linear combination of the values $1/\sigma^2_{\margset}$ for $\margset\in\closure(\margworkload)$ and is therefore convex in the $\sigma^2_{\margset}$ values. Meanwhile, Theorem \ref{thm:var} shows how to represent, for  each $\margset\in\closure(\margworkload)$, the quantity $\varfun(\margset;\mech^*)$ as a positive linear combination of $\sigma^2_{\margset^\prime}$ for $\margset^\prime\in\closure(\margset)\subseteq\closure(\margworkload)$. Therefore, the loss function $\loss$ is also convex in the $\sigma^2_{\margset}$ values.

Thus the optimization problems in Equations \ref{eqn:privacyconstrained} and \ref{eqn:accuracyconstrained} can be written as minimizing a convex function of the $\sigma^2_{\margset}$ subject to convex constraints. In fact, in Equation \ref{eqn:accuracyconstrained}, the constraints are linear when the optimization variables represent the $\sigma^2_{\margset}$ and in Equation \ref{eqn:privacyconstrained} the constraints are linear when the optimization variables represent the $1/\sigma^2_{\margset}$.  Furthermore, when the loss function is the weighted sum of variances of the marginal cells,  the solution can be obtained in closed form (see supplementary material).
Otherwise, we use 
CVXPY/ECOS \cite{diamond2016cvxpy,domahidi2013ecos}
for solving these convex optimization problems.

\subsection{Computational Complexity}
The time complexity of the steps of our framework is provided in the following theorem. It can be expressed in terms of the sizes of the marginals the user is asking for.  Crucially, it does \emph{not} depend on the universe size $|\attr_1|\times\cdots\times|\attr_\numattr|$, which accounts for the scalability.
\begin{theoremEnd}[category=complexity]{theorem}
Let $\numattr$ be the total number of attributes. Let $\cellcount(\margset)$ denote the number of cells in the marginal on attribute set $\margset$. Then:
\begin{enumerate}[leftmargin=*,parsep=0pt]
\item Expressing the privacy cost of the optimal mechanism $\mech^*$ as a linear combination of the $1/\sigma^2_{\margset}$  values takes $O(\sum_{\margset\in\margworkload} \cellcount(\margset))$  total time.
\item Expressing \underline{all} of the $\varfun(\margset;\mech^*)$, for $\margset\in\margworkload$, as a linear combinations of the $\sigma^2_{\margset}$ values can be done in $O(\sum_{\margset\in\margworkload} \cellcount(\margset))$ total time.
\item Computing \underline{all} the noisy outputs of the optimal mechanism (i.e., $\mech_{\margset}(\datavec;\sigma^2_{\margset})$ for $\margset\in\closure(\margworkload)$) takes $O\left(\numattr \sum_{\margset\in\margworkload}\prod_{\attr_i\in\margset}(|\attr_i|+1)\right)$  time  after the true answers have been precomputed (Line \ref{line:eval} in Algorithm \ref{alg:measure}). Note  the total number of cells of the marginals contained in  $\margworkload$ is \\ $O\left(\sum_{\margset\in\margworkload}\prod_{\attr_i\in\margset}|\attr_i|\right)$. 
\item Reconstructing  marginals for \underline{all}  $\margset\in\margworkload$ takes total time $O(\sum_{\margset\in\margworkload}|\margset|\cellcount(\margset)^2)$ 
\item Computing the variance of the cells for all of the marginals  $\margset\in\margworkload$ can be done in  \\ $O(\sum_{\margset\in\margworkload} \cellcount(\margset))$ total time.
\end{enumerate}
\end{theoremEnd}
\begin{proofEnd}
First we establish that \\$|\closure(\margworkload)|\leq\sum_{\margset\in\margworkload}\cellcount(\margset)$. Given an set $\margset\in\margworkload$, we note that it has $2^{|\margset|}$ subsets, so that $|\closure(\margset)|=2^{|\margset|}$. However, $\cellcount(\margset)$ is at least $2^{|\margset|}$ (because each attribute has at least 2 attribute values). We also note that $\closure(\margworkload)=\bigcup\limits_{\margset\in\margworkload}\closure(\margset)$. Hence
\begin{align*}
|\closure(\margworkload)|&\leq \sum_{\margset\in\margworkload} |\closure(\margset)|\\
&= \sum_{\margset\in\margworkload}\cellcount(\margset)
\end{align*}

To analyze the time complexity of symbolically representing the privacy cost, as a linear combination of the $1/\sigma^2_{\margset}$ values (for all $\margset\in\closure(\margworkload)$) we note that 
the coefficient of $1/\sigma^2_{\margset}$ is $\prod\limits_{\attr_i\in\margset}\frac{|\attr_i|-1}{|\attr_i|}$. Thus computing the coefficient $1/\sigma^2_{\emptyset}$ takes $O(1)$ time. Then, computing the coefficient of $1/\sigma^2_{\{\attr_i\}}$ can be computed from the coefficient of $1/\sigma^2_{\emptyset}$ in $O(1)$ additional time. Thus, we if go level by level, first computing the coefficients of $1/\sigma^2_{\margset}$ with $|\margset|=1$ then for $|\margset|=2$, etc. then computing the coefficient for each new $\margset$ takes incremental $O(1)$ time. Thus the overall time is $O(|\closure(\margworkload)|)$ and therefore is $O(\sum_{\margset\in\margworkload}\cellcount(\margset))$. 

\vspace{0.5cm}
Let $\ncells=\sum_{\margset\in\margworkload}\cellcount(\margset)$.
To express the variance symbolically as a linear function of the $\sigma^2_{\margset}$ values via Theorem \ref{thm:var}, we note from the previous part that computing $\prod\limits_{\attr_i\in\margset^\prime}\frac{|\attr_i|-1}{|\attr_i|}$ for all $\margset^\prime\in\closure(\margworkload)$ can be done in total $O(\ncells)$ time. Similarly, computing $\prod\limits_{\attr_i\in{\margset^\prime}}\frac{1}{|\attr_i|^2}$ for all $\margset^\prime\in\closure(\margworkload)$ also take total $O(\ncells)$ time. Once this is pre-computed, then for any $\margset^\prime\subseteq\margset\in\closure(\margworkload)$, the product 
$\prod_{\attr_i\in\margset'}\frac{|\attr_i|-1}{|\attr_i|} *\prod_{\attr_j\in( \margset /\margset')}\frac{1}{|\attr_j|^2} $ can be computed in $O(1)$ time since $\margset\setminus\margset^\prime\in\closure(\margworkload)$.
Now, $\varfun(\margset; \mech^*)$ equals
$$\sum\limits_{\margset^\prime\subseteq \margset}\sigma^2_{\margset^\prime}\prod\limits_{\attr_i\in\margset'}\frac{|\attr_i|-1}{|\attr_i|} *\prod\limits_{\attr_j\in( \margset /\margset')}\frac{1}{|\attr_j|^2}. $$ This is a linear combination of $2^{|\margset|}$ terms (one term for each variable $\sigma^2_{\margset^\prime}$ for $\margset^\prime\subseteq\margset$). Each term is computed in $O(1)$ time after the precomputation phase. Thus the symbolic representation of $\varfun(\margset;\mech^*)$ takes $O(2^{|\margset|})$ time (which is at most the number of cells in the marginal on $\margset$) time after precomputation. Thus computing $\varfun(\margset;\mech^*)$ for all $\margset\in\margworkload$ can be done in total $O(\ncells)$ time after precomputation, but precomputation also takes $O(\ncells)$ time. Thus the overall total time is $O(\ncells)$.

\vspace{0.5cm} We next analyze the time it takes to generate noisy answers once the true answers have been  precomputed (Line \ref{line:eval} in Algorithm \ref{alg:measure}). This involves (1) computing the product $\mat{H}\vec{v}$ in the algorithm, (2) generating one Gaussian random variable for each column of $\mat{H}$  and (3) computing$\mat{H}\vec{z}$. Now, the first and third steps take the same amount of time. The second step  generates one Gaussian for each row of $\mat{H}$ and hence, for each $\mech_\margset$ takes time $\Pi_{\attr_i\in\margset} (|\attr_i|-1)$. 

For the first step, the fast kronecker-product multiplication algorithm (Algorithm 1 of \cite{mckenna2021hdmm}) has the following complexity. Given a kronecker product of $\ell$ matrices of sizes  $(m_1-1)\times m_1, \dots, (m_\ell-1)\times m_\ell$ and a vector with $m_1\times\cdots\times m_\ell$ components, their algorithm has $\ell$ iterations. In iteration $i$, the $i^\text{th}$ matrix (with size  $m_{i-1}\times m_i$) is multiplied by a matrix with shape $(m_i, \prod_{j=1}^{i-1} m_j * \prod^\ell_{j=i+1}(m_j-1))$. In our case, each $m_i$ is a subtraction matrix with two nonzero elements in each row. Thus, in each iteration, the product makes $2\prod_{j=1}^{i-1} m_j * \prod^\ell_{j=i}(m_j-1)$ scalar multiplication operations. There are $\ell$ iterations, so the multiplication algorithm uses $O(\ell\prod_{i=1}^\ell m_i)$ multiplications.

Now, to run algorithm $\mech_{\margset}$, the number of kron products $\ell$ is $|\margset|$ and each  $m_i$ is  $|\attr_i|$ for $\attr_i\in\margset$. Hence the running time of $\mech_{\margset}$ is $O(|\margset|\prod_{\attr_i\in\margset}|\attr_i|)$ which is at most $|\margset|$ times the number of cells in the marginal on $\margset$. Note that the constant in the big-O notation is bounded across all $\margset$. Next, when adding up the complexity across all $\margset^\prime\in\closure(\margset)$, we can replace  $|\margset^\prime|$ with $|\margset|$, and then the summation looks like the product $\prod\limits_{\attr_i\in\margset}(|\attr_i|+1)$ when this product is expanded. Hence the time to run all $\marginal_{\margset^\prime}$ for all $\margset^\prime\in \closure(\margset)$ is $O(|\margset|\prod\limits_{\attr_i\in\margset}(|\attr_i|+1))$. Adding up over all $\margset\in\margworkload$ gets the results.

\vspace{0.5cm}
Next we consider the reconstruction phase. Using the same analysis of the fast kron-product vector multiplication, we see that in each iteration of Algorithm \ref{alg:reconstruction}, there is a kron product vector multiplication. Using similar reasoning as for the previous item, each such multiplication takes $O(|\margset|\prod_{\attr_i\in\margset})|\attr_i|=O(|\margset|\cellcount(\margset))$ time. The number of iterations in the algorithm is $2^{|\margset|} \leq \cellcount(\margset)$. Thus the overall runtime is $O(\sum_{\margset\in\margworkload}|\margset|\cellcount(\margset)^2)$.

\vspace{0.5cm} Finally, the variance computation is no harder than expressing the $\varfun(\margset;\mech^*)$ as linear combinations of the optimization variables and we have shown this to be $O(\ncells)$.
\end{proofEnd}
To get a sense of these numbers, consider a dataset with 20 attributes, each having 3 possible values. If the workload consists of all 3-way marginals, there are 1,140 marginals each with 27 cells so $\ncells=30,780$. The quantity inside the big-O for the selection step is $1,459,200$ (roughly the number of scalar multiplications needed). These are all easily manageable on modern computers even without GPUs. Our experiments, under more challenging conditions, run in seconds.

\section{A Numerically Secure Implementation with Discrete Gaussian Noise}\label{sec:dgm}

The basic ResidualPlanner algorithm described so far uses correlated Gaussian noise to protect privacy.
In practice, the naive use of floating point computation and floating point randomness could result in exploitable
vulnerabilities due to rounding errors and gaps in the possible values of the least significant bits \cite{widespreadsens,MironovBits}.
While it is possible to provide floating point Gaussian sampling algorithms that are secure for differential
privacy \cite{haney2022precisionbasedattacksintervalrefining}, it is highly recommended to avoid floating point altogether \cite{widespreadsens,MironovBits}. That is, differentially private platforms are encouraged to use integer or fixed-point computation to avoid exploitable bugs in noise generation and query processing. For such systems, securely implemented discrete Laplace and discrete Gaussian noise implementations are available \cite{discgauss,tdahdsr}.

In this section, we show how ResidualPlanner can be modified to work with discrete Gaussian noise instead of continuous Gaussian noise. The challenge is non-trivial: there is no known secure algorithm for sampling \textbf{correlated} discrete Gaussian noise. Furthermore, simply replacing the continuous noise in Line \ref{line:measureZ} of the Algorithm \ref{alg:measure} with independent discrete Gaussian noise can result in substantially degraded privacy guarantees, as demonstrated in the following example.
\begin{example}
Suppose $\sigma_{\margset}=1$ in Algorithm \ref{alg:measure} and suppose the vector of independent standard Gaussian noise $\vec{z}$ in Line \ref{line:measureZ} were replaced by a vector of independent discrete Gaussian random variables (also with mean 0 and scale 1). Then the resulting measurement mechanism $\mech_{\margset}(\datavec; 1)$ is equivalent to computing an exact answer $\vec{v}$ to a marginal query, adding the noise $\vec{z}$, then postprocessing by multiplication with the matrix $\mat{H}$ defined in Algorithm \ref{alg:measure} (i.e., it is $\mat{H}(\vec{v}+\vec{z})$). This is clearly an application of the discrete Gaussian mechanism to a sensitivity 1 query (a marginal) followed by postprocessing and therefore it satisfies $\rho$-zCDP with $\rho=1/2$ \cite{discgauss}.

However, it is possible to do better in the continuous case because of some special mathematical properties related to the interaction of continuous Gaussian noise and linear postprocessing. Specifically, in the continuous noise case, the exact distribution of $\mat{H}(\vec{v}+\vec{z})$ can be computed in closed form because it is a continuous Gaussian (i.e., $\mat{H}\vec{v}+N(\vec{0}, \mat{H}\mat{H}^T)$), and so has a closed form covariance matrix $\mat{H}\mat{H}^T$ that can be analyzed for a tighter privacy bound as in Definition \ref{def:lgm} (in the discrete case, even the addition of two independent discrete Gaussians is no longer a discrete \emph{Gaussian}). If the marginal is a $k$-way marginal on $k$ binary attributes, by Theorem \ref{thm:pcost}, the privacy cost calculation is $\pcost=1/2^k$ and therefore $\rho$-zCDP is satisfied with $\rho=\frac{1}{2}2^{-k}$ (Definition \ref{def:lgm}).

Comparing the discrete calculation ($\rho=1/2$) to the continuous calculation ($\rho=\frac{1}{2}2^{-k}$) shows that naively replacing continuous Gaussian noise with discrete Gaussian noise can cause a blow-up in the privacy parameters of mechanism $\mech_{\margset}$ by $2^k$, where $k$ is the number of attributes featured in a marginal.
\end{example}

We next explain how we convert the continuous-noise mechanism to a discrete-noise mechanism without blowing up the privacy budget (resulting in the mechanism shown in Algorithm \ref{alg:discrete}). Although the algorithm we work out is specific to  ResidualPlanner, it may be possible to adapt the techniques to other settings.

The first step is to recall that $\mech_{\margset}(\datavec; \sigma^2_{\margset}) = \resid_{\margset}\datavec + N(\vec{0}, \sigma^2_{\margset}\covar_\margset)$, 
its privacy cost matrix (Definition \ref{def:lgm})  is  $\frac{1}{\sigma_{\margset}^2}\resid_{\margset}^T\covar^{-1}\resid_{\margset}$ and $\pcost(\mech_\margset(\cdot; \sigma^2_\margset))$ is the largest diagonal of that matrix. 

The second step is to round \textbf{up} $\sigma_{\margset}$ to some rational number $\overline{\sigma}_{\margset}\equiv s/t$, where $s$ and $t$ have no common factors. For example, if $\sigma_{\margset}=\sqrt{2}\approx 1.414$, we can set $s=71$ and $t=50$ so that $\overline{\sigma}_{\margset}=s/t=1.42$.  The privacy cost matrix of $\mech(\cdot; \overline{\sigma}^2_\margset)$ is $\sigma^2_{\margset}/\overline{\sigma}^2_\margset$ times the privacy cost matrix of $\mech(\cdot; \sigma^2_\margset)$ and hence its privacy cost is smaller by that factor. Thus the conversion from floating point to arithmetic based on integers will result in a tiny increase in privacy and a corresponding tiny decrease in utility.

The third step is to create an intermediate continuous noise mechanism $\mech^\prime_{inter}(\datavec)=\Xi \datavec + N(0, \gamma^2\identity)$ that is equivalent to $\mech(\datavec; \overline{\sigma}_\margset^2)$. That is, there exists a matrix $\mat{Y}$ such that the output distribution of $\mech_{inter}^\prime(\datavec)$ is the same as $\mat{Y}\mech_\margset(\datavec; \overline{\sigma}_\margset^2)$ (i.e., running $\mech_\margset$ and multiplying the answer by $\mat{Y}$) and the output distribution of $\mech_\margset(\datavec; \overline{\sigma}^2_{\margset})$ is equal to $\mat{Y}^\dagger \mech_{inter}^\prime(\datavec)$. Since each mechanism can be obtained by postprocessing the other one, this means that $\mech_{inter}^\prime$ and $\mech$ have the same exact privacy and utility properties. However, $\mech^\prime_{inter}(\datavec)=\Xi \datavec + N(0, \gamma^2\identity)$ has additional desirable characteristics: it uses independent (not correlated) noise, the matrix $\Xi$ has integer entries and $\gamma$ is a rational number. This means that the independent continuous noise in $\mech^\prime_{inter}$ can be replaced by independent discrete noise with the same scale parameter $\gamma$ and achieving the same privacy for $\rho$-zCDP. Calling this discrete noise mechanism $\mech^\prime$, the overall algorithm will first run $\mech^\prime$, then multiply the result by $\mat{Y}^\dagger$ to reconstruct the answer to the original mechanism $\mech_{\margset}$.


Now, the requirement  $\mech^\prime_{inter}(\datavec)=\Xi \datavec + N(\vec{0}, \gamma^2\identity)$ is equivalent to first running $\mech_{\margset}(\datavec; \overline{\sigma}^2_\margset)=\resid_{\margset}\datavec + N(\vec{0}, \overline{\sigma}^2_\margset \covar_{\margset})$ and then multiplying by $\mat{Y}$ means that the following conditions must be satisfied:
\begin{align*}
    \Xi &= \mat{Y}\resid_\margset \quad \text{(the mean of $\mech^\prime_{inter}(\datavec)$)}\\
    \sigma_\margset^2\mat{Y}\covar_\margset \mat{Y}^T &= \gamma^2 \identity \quad\text{(its noise covariance)}
\end{align*}

We achieve these requirements with the following settings (we will prove correctness in Theorem \ref{thm:discrete}):
\begin{align}
\overline{\sigma}&\equiv s/t\label{eqn:discst}\\
\mat{Y} &\gets \bigotimes\limits_{\attr_i\in\margset}|\attr_i|* \submat_{|\attr_i|}^\dagger \label{eqn:discy}\\ 
\Xi &\gets \mat{Y}\resid_\margset \label{eqn:discxi}\\ 
\gamma^2 &\gets \frac{s^2}{t^2} * \prod\limits_{\attr_i\in\margset}|\attr_i|^2 \label{eqn:discj}
\end{align}
\begin{example}
    Suppose attribute $\attr_1$ has 4 possible values and $\margset=\{\attr_1\}$. Consider the continuous noise mechanism $\mech_\margset(\datavec; \sigma_{\margset}) = \resid_\margset\datavec + N(\vec{0}, \sigma^2_{\margset}\covar_\margset^T)$ where
    \begin{align*}
        \sigma_\margset &= 2/3, &
        \resid_{\margset}
        &= 
\left[
\begin{array}{cccc}
1 & -1 & \phantom{-}0 & \phantom{-}0 \\
1 & \phantom{-}0 & -1 & \phantom{-}0 \\
1 & \phantom{-}0 & \phantom{-}0 & -1 \\
\end{array}
\right]\\
\end{align*}
Plugging these values into Equations \ref{eqn:discst}, \ref{eqn:discy}, \ref{eqn:discxi}, and \ref{eqn:discj}, we get:
\begin{align*}
    \overline{\sigma} &= 2/3\\
    \mat{Y} &=  4 * \resid_{\margset}^\dagger =  \begin{bmatrix}
        \phantom{-}1 & \phantom{-}1 & \phantom{-}1 \\
        -3 & \phantom{-}1 & \phantom{-}1 \\
        \phantom{-}1 & -3 & \phantom{-}1 \\
        \phantom{-}1 & \phantom{-}1 & -3
    \end{bmatrix} \\
    \Xi &= \mat{Y} \resid_{\margset} =
    \begin{bmatrix}
        \phantom{-}3 & -1 & -1 & -1 \\
        -1 & \phantom{-}3 & -1 & -1 \\
        -1 & -1 & \phantom{-}3 & -1 \\
        -1 & -1 & -1 & \phantom{-}3
    \end{bmatrix} \\
    \gamma^2 &= \frac{4}{9}*16 = 64/9 
\end{align*}
The resulting mechanism $\mech^\prime(\datavec)=\Xi\datavec + N(\vec{0}, \frac{64}{9}\identity)$ is adding discrete Gaussian noise with scale $\gamma=\sqrt{64/9}=8/3$ to the vector of integers $\Xi\datavec$ (while the intermediate mechanism $\mech^\prime_{inter}$ uses continuous Gaussian noise instead). The squared $L_2$ sensitivity of $\Xi$ is $9+1+1+1=12$ and the $\rho$-zCDP privacy parameter is the $L_2$ sensitivity squared divided by $2\gamma^2$, so that $\rho=\frac{12}{2(64/9)}=27/32$, which is the same as $\rho$-zCDP parameter for $\mech_{\margset}$. Furthermore, it is easy to check numerically that $\mech_{\margset}$ is identical to first running $\mech^\prime_{inter}$ on the input data and multiplying the result by $\mat{Y}^\dagger$ (the pseudo-inverse of $\mat{Y}$) while $\mech^\prime_{inter}$ is the same as first running $\mech_{\margset}$ and then multiplying the result by $\mat{Y}$.
\end{example}
The general algorithm that replaces a continuous noise mechanism $\mech_{\margset}(\datavec; \overline{\sigma}^2)$ with a discrete noise algorithm  is shown in Algorithm \ref{alg:discrete}. First it creates $\mech^\prime$ as in the above discussion, then multiplies its output by $\mat{Y}^\dagger$ so that it is answering the same query as $\mech_\margset$. The pseudocode (after algebraic simplification of $\mat{Y}\resid_\margset$) is shown in Algorithm \ref{alg:discrete}. Note that the Kronecker products are not expanded. Instead, the algorithm uses fast Kronecker-vector multiplication \cite{mckenna2018optimizing}.
\begin{algorithm}[h]
   \DontPrintSemicolon
   $\overline{\sigma}_{\margset} \gets s/t $ where $s$ and $t$ are integers and $s/t\geq \sigma_{\margset}$\;
    $\vec{v} \gets \marginal_{\margset}\datavec$\tcp{Evaluate the true marginal}
    $\mat{Y}^\dagger \gets \bigotimes\limits_{\attr_i\in\margset} \frac{1}{|\attr_i|}\submat_{|\attr_i|}$\;
    $\mat{H} \gets \bigotimes\limits_{\attr_i\in\margset}\left(|\attr_i|\identity_{|\attr_i|} - \one_{|\attr_i|}\one_{|\attr_i|}^T \right)$\;\tcc{Note: $\mat{H}\vec{v}=\Xi\datavec$}
    $\gamma^2 \gets \frac{s^2}{t^2} * \prod\limits_{\attr_i\in\margset}|\attr_i|^2$\;
    $\vec{z} \gets$ vector of independent discrete Gaussian noise with scale $\gamma$.\;
    \Return $\mat{Y}^\dagger(\mat{H}\vec{v} + \vec{z})$
\caption{Replacement for $\mech_\margset(\datavec;\sigma_{\margset}^2)$ using discrete Gaussian noise.}\label{alg:discrete}
\end{algorithm}

\subsection{Correctness}

We next discuss the correctness of Algorithm \ref{alg:discrete}. We show correctness in a two-stage manner.
First, we show that if Algorithm \ref{alg:discrete} uses continuous noise then it would be the same as Algorithm \ref{alg:measure} (using noise parameter $\overline{\sigma}$). That is, for each input, both algorithms would have the same output distribution and so they are functionally equivalent. However, the difference is that discrete noise can be directly swapped into Algorithm \ref{alg:discrete} to maintain the same privacy guarantees, which is the second part of the proof of correctness. Here, the privacy guarantees are in terms of $\rho$-zCDP \cite{zcdp}, as this was the privacy definition used by Canonne et al. \cite{discgauss} to prove the privacy properties of discrete Gaussian noise. This two-stage correctness result is stated in the following theorem.

\begin{theoremEnd}[category=discrete,all end]{lemma}\label{lem:subpinvsub}
  For all $i$, \\$\submat_{|\attr_i|}^\dagger\submat_{|\attr_i|}=\identity_{|\attr_i|} - \frac{1}{|\attr_i|}\one_{|\attr_i|}\one_{|\attr_i|}^T$.
\end{theoremEnd}
\begin{proofEnd}
Recall that $\submat_{|\attr_i|}$ and $\submat^\dagger_{|\attr_i|}$ can be represented as block matrices as follows:
\begin{align*}
\submat_{|\attr_i|} &= \left[\begin{smallmatrix}
    \one_{|\attr_i|-1},\quad -\identity_{|\attr_i|-1}
\end{smallmatrix}\right]\\
    \submat_{|\attr_i|}^\dagger&=\frac{1}{|\attr_i|}\left[\begin{smallmatrix}
\one^T_{|\attr_i|-1}\\
\one_{|\attr_i|-1}\one^T_{|\attr_i|-1}-|\attr_i|\identity_{|\attr_i|-1}
\end{smallmatrix}\right]
\end{align*}
\begin{align*}
\lefteqn{\submat_{|\attr_i|}^\dagger \submat_{|\attr_i|}}\\
&=\frac{1}{|\attr_i|}
\left[\begin{smallmatrix} 
|\attr_i|-1, & - \one^T_{|\attr_i|-1}\\
-\one_{|\attr_i|-1},\quad &-\one_{|\attr_i|-1}\one^T_{|\attr_i|-1}+|\attr_i|\identity_{|\attr_i|-1}
\end{smallmatrix}\right]\\
&=\identity_{|\attr_i|} - \frac{1}{|\attr_i|}
\left[\begin{smallmatrix} 
1, &  \one^T_{|\attr_i|-1}\\
\one_{|\attr_i|-1},\quad &\one_{|\attr_i|-1}\one^T_{|\attr_i|-1}
\end{smallmatrix}\right]\\
&=\identity_{|\attr_i|} - \frac{1}{|\attr_i|}\one_{|\attr_i|}\one_{|\attr_i|}^T
\end{align*}\qed
\end{proofEnd}

\begin{theoremEnd}[category=discrete]{theorem}\label{thm:discrete}
    Let $\mech_{disc}$ denote the mechanism in Algorithm \ref{alg:discrete}. Let $\mech_{cont}$ denote the version of this algorithm when continuous Gaussian noise is used instead of discrete Gaussian noise. Then the output distribution of $\mech_{cont}$ is exactly the same as $\mech_\margset(\cdot; \overline{\sigma}^2)$. Furthermore,  If $\mech_{cont}$ satisfies zCDP with privacy parameter $\rho$, then the discrete noise version $\mech_{disc}$ satisfies zCDP with the same privacy parameter.
\end{theoremEnd}
\begin{proofEnd}
Without loss of generality, let $\margset=\{\attr_1,\dots, \attr_\ell\}$ for some $\ell$ (that is, the attributes in $\margset$ are the first $\ell$ attributes). Recall that $\resid_{\margset}$, $\mat{Y}$, and the marginal query $\marginal_\margset$ can be represented as:
\begin{align*}
    \resid_{\margset} &= \left(\bigotimes_{i=1}^\ell \submat_{|\attr_i|}\right) \kron \left(\bigotimes_{j=\ell+1}^{\numattr}\one^T_{|\attr_j|}\right)\nonumber\\
    \mat{Y} &= \bigotimes\limits_{\attr_i\in\margset}|\attr_i|* \submat_{|\attr_i|}^\dagger \\ 
    &=  \left(\bigotimes_{i=1}^\ell |\attr_i| *\submat_{|\attr_i|}^{\dagger}\right) \kron \left(\bigotimes_{j=\ell+1}^{\numattr}\left[\begin{smallmatrix}1\end{smallmatrix}\right]\right)\\
    \marginal_{\margset} &= \left(\bigotimes_{i=1}^\ell \identity_{|\attr_i|}\right) \kron \left(\bigotimes_{j=\ell+1}^{\numattr}\one^T_{|\attr_j|}\right)\\
\end{align*}
Therefore
\small{
\begin{align*}
\mat{Y}\resid_\margset &=\left(\bigotimes_{i=1}^\ell |\attr_i| \submat_{|\attr_i|}^\dagger\submat_{|\attr_i|}\right) \kron \left(\bigotimes_{j=\ell+1}^{\numattr}\one^T_{|\attr_j|}\right)
\end{align*}}
and
\begin{align*}
\mat{Y}\resid_\margset\datavec &= \left(\bigotimes_{i=1}^\ell |\attr_i| \submat_{|\attr_i|}^\dagger\submat_{|\attr_i|}\right) \marginal_\margset\datavec\\
&=\left(\bigotimes_{i=1}^\ell \left(|\attr_i|\identity_{|\attr_i|} - \one_{|\attr_i|}\one_{|\attr_i|}^T \right)\right) \marginal_\margset\datavec\\
&\text{By Lemma \ref{lem:subpinvsub}}\\
&=\mat{H}\marginal_{\margset}\datavec=\mat{H}\vec{v} \quad\text{in the notation of Algorithm \ref{alg:discrete}}\\
\end{align*}
Now, when Algorithm \ref{alg:discrete} uses continuous noise, it is expressible as $\mat{Y}^\dagger(\mat{H}\marginal_{\margset}\datavec +N(\vec{0}, \gamma^2\identity))$. We need to show that it is identical to $\mech_{\margset}(\datavec; \overline{\sigma}_\margset^2)=\resid_\margset\datavec + N(\vec{0}, \overline{\sigma}_\margset^2\covar_{\margset})$. This is equivalent to showing that $\mat{Y}^\dagger\mat{H}\marginal_\margset=\resid_\margset$ and $\mat{Y}^\dagger \gamma^2 (\mat{Y}^\dagger)=\overline{\sigma}^2_\margset\covar_\margset$.

Note that we have already shown that for any $\datavec$, we have $\mat{H}\marginal_\margset\datavec=\mat{Y}\resid_\margset\datavec$ and so for any $\datavec$:
\begin{align*}
    \mat{Y}^\dagger\mat{H}\marginal_\margset\datavec &=\left(\bigotimes\limits_{i=1}^\ell \frac{1}{|\attr_i|}\submat_{|\attr_i|}\right)\mat{H}\marginal_\margset\datavec\\
    &=\left(\bigotimes\limits_{i=1}^\ell \submat_{|\attr_i|}\right)\marginal_\margset\datavec\\
    &=\resid_{\margset}\datavec
\end{align*}
where the second equality is a consequence of the equality $\submat_{|\attr_i|}|\attr_i|(\identity_{|\attr_i|} - \one_{|\attr_i|}\one_{|\attr_i|}^T)=|\attr_i|\submat_{|\attr_i|}$ and so $\mat{Y}^\dagger\mat{H}\marginal_\margset=\resid_\margset$. Next,
\begin{align*}
    \lefteqn{\mat{Y}^\dagger \gamma^2 (\mat{Y}^\dagger) }\\&= \frac{s^2}{t^2} \prod\limits_{i=1}^\ell{|\attr_i|^2}\bigotimes\limits_{i=1}^\ell \frac{1}{|\attr_i|}\submat_{|\attr_i|}\bigotimes\limits_{i=1}^\ell \frac{1}{|\attr_i|}\submat_{|\attr_i|}^T\\
    &=\overline{\sigma}^2_\margset \prod\limits_{i=1}^\ell{|\attr_i|^2}\bigotimes\limits_{i=1}^\ell \frac{1}{|\attr_i|^2}\submat_{|\attr_i|}\submat_{|\attr_i|}^T\\
    &=\overline{\sigma}^2_\margset \covar_\margset
\end{align*}

So, we have shown that under continuous noise, Algorithm \ref{alg:discrete} is equivalent to $\mech_\margset(\cdot; \overline{\sigma}^2_\margset)$.

Next, we prove the privacy properties under discrete noise. First note that $\mat{H}\marginal_\margset$ is a matrix with integer-valued entries, so it is appropriate to add discrete Gaussian noise (whose support is the set of integers) to $\mat{H}\marginal_\margset$. Since the noise is independent with scale $\gamma$, the algorithm satisfies $\rho$-zCDP with $\rho=\frac{\Delta^2_2(\mat{H}\marginal_\margset)}{2\gamma^2}$ \cite{discgauss}, where $\Delta^2_2(\mat{H}\marginal_\margset)$ is the square of the $L_2$ sensitivity of multiplication by $\mat{H}\marginal_\margset$, which is the same as the largest squared $L_2$ norm of any column of $\mat{H}\marginal_\margset$.

Now, since the expression for $\marginal_\margset$ is a kronecker product of the identity matrices and row vectors full of ones, the largest squared $L_2$ norm of any column of $\mat{H}\marginal_\margset$ is the same as the largest squared $L_2$ norm of any column of $\mat{H}$. Thus, 
\begin{align*}
    \Delta^2_2(\mat{H}\marginal_\margset)&=\prod\limits_{i=1}^\ell \left((|\attr_i|-1)^2 + |\attr_i|-1 \right)\\
    &= \prod\limits_{i=1}^\ell |\attr_i|(|\attr_i|-1)\\
\end{align*}
thus
\begin{align*}
    \rho = \frac{\Delta^2_2(\mat{H}\marginal_\margset)}{2\gamma^2}
    =\frac{1}{2 \frac{s^2}{t^2}}\prod\limits_{i=1}^\ell \frac{|\attr_i|-1}{|\attr_i|}=\frac{1}{2}\frac{1}{\overline{\sigma}^2_\margset}\prod\limits_{i=1}^\ell \frac{|\attr_i|-1}{|\attr_i|}
\end{align*}
This matches the $\rho$-zCDP value of $\mech_\margset(\cdot; \overline{\sigma}^2_\margset)$, which is equal to $\frac{1}{2}\pcost(\mech_\margset(\cdot; \overline{\sigma}^2_\margset))$, where $\pcost(\mech_\margset(\cdot; \overline{\sigma}^2_\margset))$  is given by Theorem \ref{thm:pcost}. \qed
\end{proofEnd}

\subsection{Utility and Scalability.}
We note that the changes needed to support discrete Gaussian noise only affect the measurement phase, while the selection and reconstruction phases do not change at all. Hence, utility and running time are only affected by the variance of the discrete Gaussian noise, the number of discrete Gaussian random variables that are needed, and the cost of generating a discrete Gaussian random variable. The latter concern is orthogonal to our work.

Comparing Algorithms \ref{alg:measure} and \ref{alg:discrete} for the continuous and discrete Gaussians, respectively, we see that they require the same number of random variables. For a set $\margset$ of attributes, the number of random variables needed is the product of the domains of the attributes in $\margset$. Hence, the increase in running time only depends on the cost of generating a discrete Gaussian random variable vs. a continuous one -- this is an engineering issue that is external to ResidualPlanner. We note that all of the mechanisms used in the measurement phase can be run in parallel, hence reducing wall clock time.

In terms of utility, the variance of a discrete Gaussian with parameter $\sigma^2$ is at most $\sigma^2$ \cite{discgauss}. Thus, the variance of any linear combination of independent discrete Gaussian random variables 
is at most the variance of the same linear combination of continuous Gaussians. The only possible loss in utility
arises in Line 1 of Algorithm \ref{alg:discrete} where the optimal choice of $\sigma^2$ needs to be replaced by
a fixed precision upper bound $\overline{\sigma}^2$. This can be controlled by the user. For example, $\sigma=1/3$ can be rounded up in its 4th decimal digit to $\overline{\sigma}=0.3334$ for a negligible percentage increase in variance (in this case, a $0.04\%$ increase).
Hence, there is no meaningful change in utility either.

\section{Using ResidualPlanner to Explore Consequences of Objective Functions}\label{sec:fair}


One of the benefits of the scalability of ResidualPlanner is that it allows data curators to
explore the consequences of optimizing marginals according to different objective functions.
For example, one of the historically most popular objective functions used for optimizing query workload
error is the weighted sum of variances of the query workload answers \cite{LMHMR15}. Given a mechanism $\mech$, let 
$\avgv_\mech(\margset)$ be the average of the cell variances when $\mech$ (e.g., ResidualPlanner) is used to 
produce the answer to the marginal on $\margset$ and let $\ncells(\margset)$ be the number of cells in
the marginal on $\margset$. Let $\text{Imp}_{\margset}$ be a user-defined importance weight associated with the marginal on $\margset$. Then the weighted sum of variances over a workload can be expressed as:
    $\sum_{\margset\in\margworkload} \text{Imp}_{\margset}*\avgv_\mech(\margset)$.

With ResidualPlanner, we can provide a \emph{closed form} expression for the optimal weighted sum of variances for any  marginal workload for a given privacy budget. This enables data curators to examine per-marginal errors to determine if any marginals are being treated unfairly (e.g., have higher variances) compared to other marginals, especially ones that have the same importance weights. In other words, it can be used to quickly identify unintended consequences of the weighted sum of variances objective function. This is an important feature for data curators who need to balance the utility needs of different end-users.

We next derive a closed form expression  for the ResidualPlanner parameters $\sigma^2_{\margset}$ that optimize such objective functions (Section \ref{subsec:closedvar}). Plugging those values into Theorem \ref{thm:var} then gives the formula for the variance of each marginal cell. Then, in Section \ref{subsec:numerical}, we use these results to provide an example analysis of marginal fairness.

\subsection{A Closed Form Expression for Optimizing the Weighted Sum of Variances}\label{subsec:closedvar}
Recall from Theorem \ref{thm:pcost} that the privacy cost is a linear combination of the $1/\sigma^2_{\margset}$ values. Specifically, when we define $p_\margset=\prod_{\attr_i\in\margset}\frac{|\attr_i|-1}{|\attr_i|}$ then 
the privacy cost of ResidualPlanner is $\sum_{\margset\in\closure(\margworkload)} p_\margset/\sigma^2_\margset$.

We can use Theorem \ref{thm:var} to write the expression for the weighted sum of variance:
\begin{align*}
    \sum_{\margset\in\margworkload}\hspace{-0.4cm}\text{Imp}_\margset\hspace{-0.15cm}\sum_{\margset' \subseteq \margset}\hspace{-0.05cm}\left(\sigma^2_{\margset'}\hspace{-0.25cm}\prod_{\attr_i\in\margset'}\hspace{-0.2cm}\frac{|\attr_i|-1}{|\attr_i|} \hspace{-0.2cm}\prod_{\attr_j\in\margset /\margset'}\frac{1}{|\attr_j|^2} \right)
\end{align*}
which is a linear combination of the $\sigma^2_\margset$ values. For any $\margset^\prime\in\closure(\margworkload)$, define
\begin{align*}
    v_{\margset'}=\hspace{-0.2cm}\sum_{\substack{\margset\in\margworkload\\\margset\supset \margset^\prime}}\hspace{-0.4cm}\text{Imp}_\margset\left(\prod_{\attr_i\in\margset'}\frac{|\attr_i|-1}{|\attr_i|} \hspace{-0.2cm}\prod_{\attr_j\in\margset /\margset'}\frac{1}{|\attr_j|^2} \right)
\end{align*}
then the weighted sum of variances simplifies to the expression $\sum_{\margset\in\closure(\margworkload)} v_\margset\sigma^2_\margset$.
 Thus, minimizing the weighted sum of reconstructed marginal variances subject to the privacy cost being $\leq c$ can be formulated as the following problem:
\begin{align}
\substack{\argmin\\{\sigma^2_{\margset}:~\margset\in\closure(\margworkload)}} & \sum_{\margset\in\closure(\margworkload) } v_{\margset} \sigma^2_{\margset} \label{eqn:totalvar}\\
     s.t. & \sum_{\margset \in\closure(\margworkload)}\frac{p_{\margset}}{\sigma^2_{\margset}}\leq c\nonumber
\end{align}
 The closed form solution is given by the following lemma.

\begin{theoremEnd}[category=fair,proof here]{lemma}\label{lemma:wsum}
Given the optimization problem in Equation \ref{eqn:totalvar},
the optimal objective function value is $T =  \left( \sum_{\margset} \sqrt{v_{\margset}p_{\margset}}\right)^2/c$ and the optimal value of each noise scale parameter is $\sigma^2_{\margset} = \sqrt{T*p_{\margset} / (c v_{\margset} )}$.
\end{theoremEnd}
\begin{proofEnd}
Clearly, for the optimal solution, the inequality constraint must be tight (i.e., $=c$) because if it is not tight, we can lower variance while increasing privacy cost by dividing each $\sigma^2_{\margset}$ by a number $>1$. Thus we just need to solve the problem subject to $\sum_{\margset} p_{\margset}/\sigma^2_{\margset}=c$.

    From the Cauchy-Schwarz inequality,
    \begin{align*}
      \sum_{\margset } v_{\margset} \sigma^2_{\margset} & =  \left(\sum_{\margset } v_{\margset} \sigma^2_{\margset} \right)
      \left(\sum_{\margset } 
 \frac{p_{\margset}}{\sigma^2_{\margset}} \right)/c \\
& \geq  \left( \sum_{\margset} \sqrt{v_{\margset}p_{\margset}}\right)^2/c = T
    \end{align*}
    Equality holds when 
    $\frac{v_{\margset}}{p_{\margset}} \sigma^4_{\margset} =t$ for all $\margset$ (for some constant $t$).  Since $c=\sum_{\margset } 
 \frac{p_{\margset}}{\sigma^2_{\margset}} =\sum_{\margset } 
 \sqrt{v_{\margset}p_{\margset}/{t}}$, then we must have $t = T/c$. Plugging this into the definition of $t$, we get  $\sigma^2_{\margset} = \sqrt{Tp_{\margset} / (c v_{\margset})}$.
\end{proofEnd}


Thus, if the loss function is the weighted sum of variances, ResidualPlanner does not need any optimization steps. The selection of the noise scales and the computation of reconstructed marginal variances is a direct computation via formulas.

\subsection{Numerical Explorations of Cell Fairness}\label{subsec:numerical}

As an example application of these results, we study how optimizing the weighted sum of variances, as is common in the matrix mechanism literature, affects the variances of individual marginals in the workload (e.g., are any marginals neglected and tend to have higher variances?). For this study, we use the Adult dataset \cite{Dua:2019}, which contains 14 attributes with different domain sizes. The domain sizes are: $100$, $100$, $100$, $99$, $85$, $42$, $16$, $15$, $9$, $7$, $6$, $5$, $2$, $2$. Thus, some 1-way marginals can have 2 cells while others can have up to 100 cells, some 2-way marginals can have as few as $2*2=4$ cells while others can have up to $100*100=10,000$ cells.

The workload of interest here is the set of all marginals with up to 3-attributes. This includes the 0-way marginal (total count), the 14 one-way marginals, the 91 two-way marginals, and the 364 three-way marginals.

For the optimization, we require the privacy cost to be $1$ and we consider the following 3 different weighting schemes for the weighted sum of variances loss function:
\begin{itemize}
    \item \textbf{Equi-weighted} optimization: each marginal receives the same importance weight (i.e., Imp$_{\margset}=1$).
    \item \textbf{Cell-size weighted} optimization: the weight of each marginal is the number of cells it contains (Imp$_{\margset}=\ncells(\margset)$). This is equivalent to the sum of the variances of each cell of each marginal.
    \item\textbf{Square-root weighted} optimization: the weight of each marginal is the square root of the number of cells it contains (Imp$_{\margset}=\sqrt{\ncells(\margset)}$). This is one way to interpolate between the equi-weighted and cell-size weighted schemes.
\end{itemize}

\begin{figure}[h]
\centering
\includegraphics[width=0.45\textwidth]{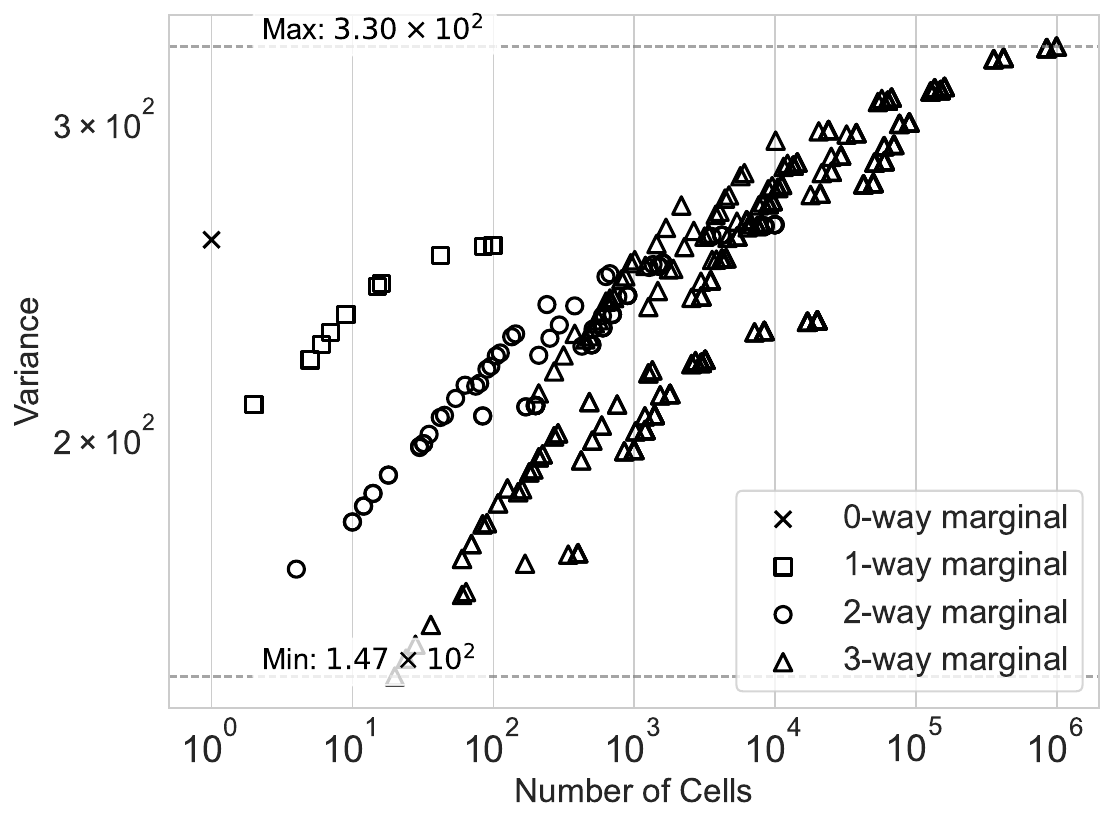}
\caption{Variance of a marginal cell (y-axis) vs. the number of cells in that marginal (x-axis) when optimizing for the equi-weighted objective function. There are 4 clear bands in this figure: the single 0-way marginal at the top, followed by the band of 1-way marginals (squares), followed by the band of 2-way (circle) and 3-way (triangle) marginals. The latter two bands eventually merge.}
\label{fig:unweighted-comb-log}
\end{figure}

\begin{figure}[h]
\centering
\includegraphics[width=0.45\textwidth]{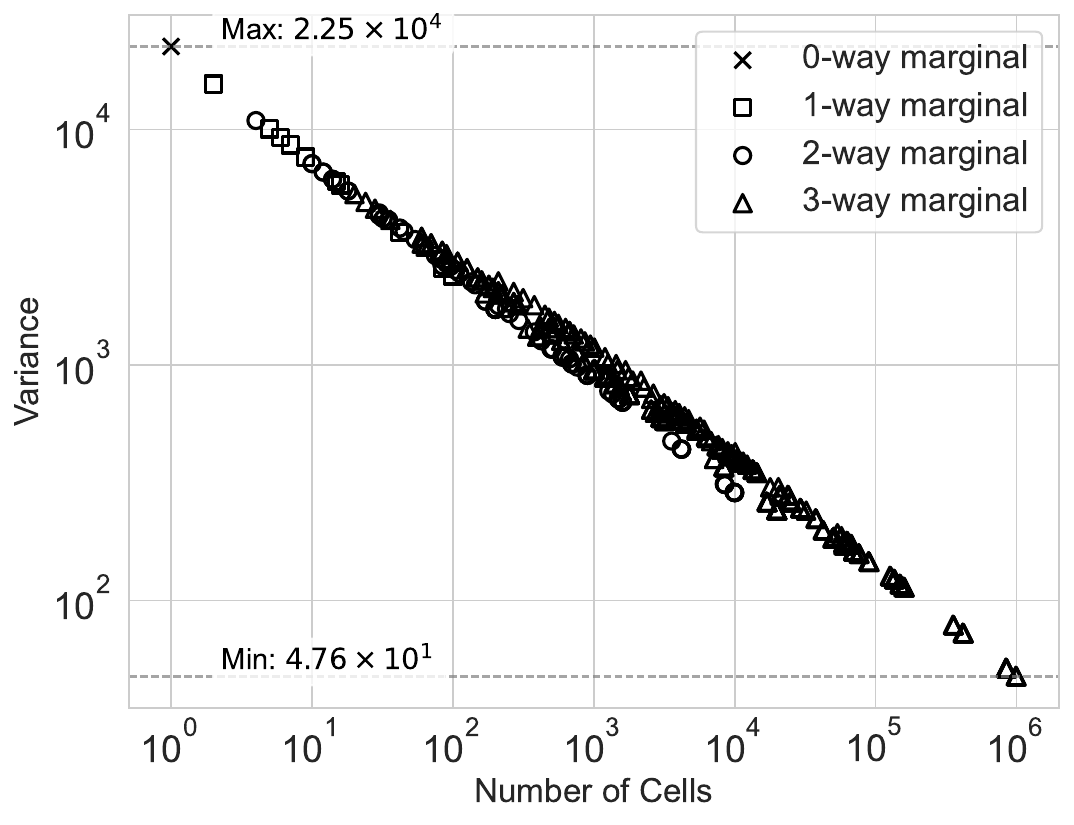}
\caption{Variance of a marginal cell (y-axis) vs. the number of cells in that marginal (x-axis) when optimizing for the cell-size weighted objective function. Aside from the 0-way marginal at the top left, all of the bands now overlap and the bottom right is dominated by some of the 3-way marginals.}
\label{fig:orig-comb-log}
\end{figure}

\begin{figure}[h]
\centering
\includegraphics[width=0.45\textwidth]{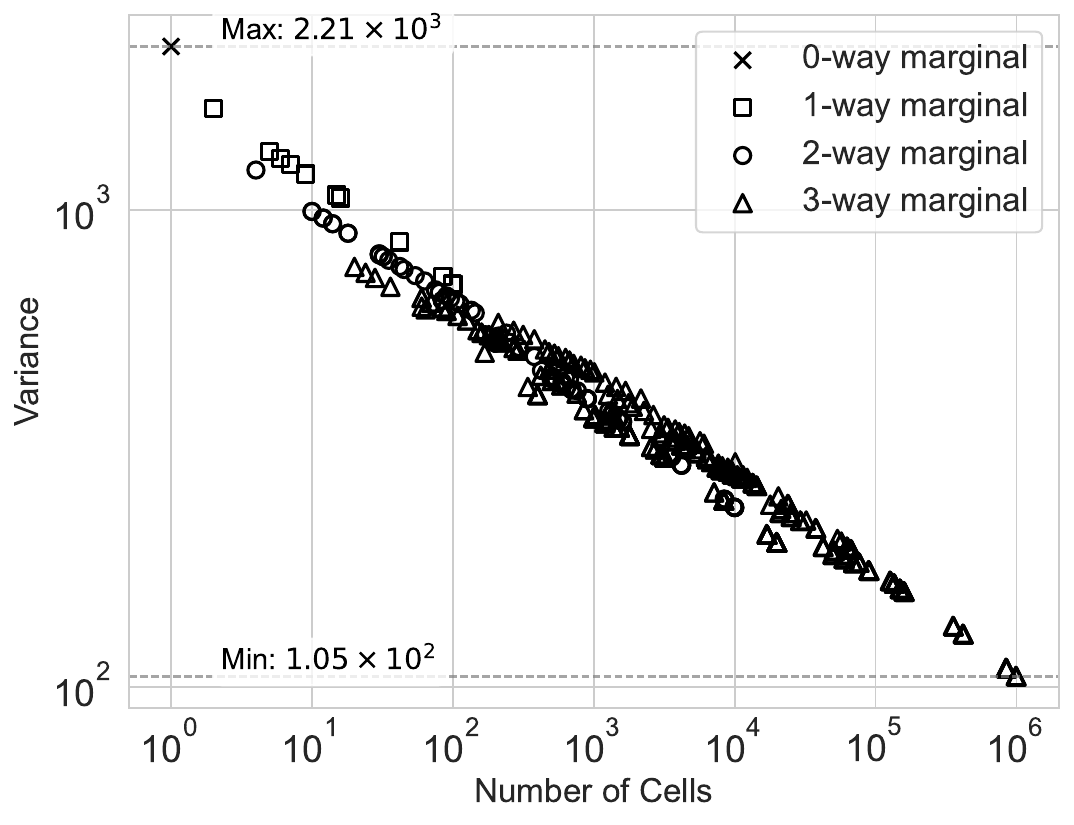}
\caption{Variance of a marginal cell (y-axis) vs. the number of cells in that marginal (x-axis) when optimizing for the square root weighted objective function. Aside from the 0-way marginal at the top left, all of the bands now overlap and the bottom right is dominated by some of the 3-way marginals.}
\label{fig:sqrt-comb-log}
\end{figure}

Figure \ref{fig:unweighted-comb-log} plots the cell variances for each of the marginals when the equi-weighted optimization scheme is used. The points are separated by whether they come from the 0-way, 1-way, 2-way, or 3-way marginals. Even though each marginal receives the same weight, we see that marginals with more cells can have over 2 times as much  noise per cell than the smaller marginals.

On the other hand, Figure \ref{fig:orig-comb-log} plots cell variances when the cell-size weighted optimization is used. This is the most common formulation \cite{LMHMR15}. We see that now the marginals with large cells are disproportionally favored, and achieve a variance per cell that is several orders of magnitudes smaller than for the smaller marginals. 

When we use square root weighting (Figure \ref{fig:sqrt-comb-log}) to try to lessen the weighting given to large marginals, we see that larger marginals are still heavily prioritized, but this time by approximately 1 order of magnitude.

Overall, we see that the proper choice of weighting schemes is extremely important when minimizing the weighted sum of variance objective function, with the equi-weighting providing the most reasonable results. To get finer control over per-cell variances, we advise considering the maximum weighted variance objective discussed in Section \ref{sec:residloss}, which more commonly appears in the theory literature. ResidualPlanner can provide optimal solutions for this loss function as well, although it runs slower because it is a more difficult optimization problem.

\begin{remark}
    In this section, we focused on mean squared error analysis because it is data independent, and hence the findings can generalize across datasets. Analysis of \emph{relative} errors, on the other hand, would be data-dependent, but data curators can use these results to estimate them. Specifically, the data curators can compute the data independent mean-squared errors for different marginals. They can use their domain knowledge to estimate the sizes of the true counts in the marginals (this way they don't peek at the data when making algorithm design choices). The ratio between root-mean-squared error and estimated counts would be a measure of relative error.
\end{remark}

\section{\newrp: From Marginals to More Expressive Queries}\label{sec:extension}
In this section, we show how the ideas that make ResidualPlanner optimal and scalable for marginals can be extended to a much larger class of queries, such as range queries, or range queries mixed with marginals. We call this extension \newrp. Although not necessarily optimal anymore, it is still highly scalable and accurate, and consistently outperforms the next best scalable approach (HDMM).

\subsection{A More General Class of Queries}\label{subsec:moregeneral}
In a \emph{marginals} workload, every query has the form $\mathbf{V}_1\kron\cdots\kron \mathbf{V}_{\numattr}$, where $\mathbf{V}_i$ is either the identity matrix (when we want attribute $\attr_i$ in the marginal) or the row vector $\one^T_{|\attr_i|}$ containing all ones (when we do not want attribute $\attr_i$ in the marginal). 
The \emph{\newrp Workload} is a generalization defined as follows:
\begin{enumerate}[leftmargin=*]
    \item Each attribute $\attr_i$ is associated with a basic matrix $\attwork_i$ (e.g., $\attwork_i$ could be the identity, or encode all ranges  or prefix-sums \footnote{Prefix-sums  on an ordinal attribute like Age have the form ``How many people have age $\leq x$?''} on $\attr_i$). The only restriction on $\attwork_i$ is that the vector $\one^T_{|\attr_i|}$ must be in the row space of  $\attwork_i$ (this is true for ranges, prefix-sums, and the identity).
    \item Every query $\marginal_{\margset}$ in the workload $\margworkload$ has the form  $\marginal_{\margset}=\mathbf{V}_1\kron\cdots\kron \mathbf{V}_{\numattr}$,  where $\mathbf{V}_i=\attwork_i$ when $i\in\margset$ and $\mathbf{V}_i=\one^T_{|\attr_i|}$ when $i\notin\margset$.
\end{enumerate}
This generalization, which we call \emph{generalized marginals},  can represent a much richer class of queries. For example, let $\attr_i=\text{Age}, \attr_2=\text{Height}, \attr_3=\text{Gender}, \attr_4=\text{Occupation}$. Now, let $\attwork_{agerange}$, encoding all possible ranges on Age, be the matrix associated with $\attr_1$. Let $\attwork_{heightrange}$, encoding all possible ranges on Height, be the matrix associated with $\attr_2$. Let $\identity_{|\attr_3|}$ and $\identity_{|\attr_4|}$ be the (identity) queries associated with attributes $\attr_3$ and $\attr_4$, respectively.  Now a data analyst can express a more complex workload that contains queries such as:
\begin{itemize}[leftmargin=*]
    \item 1-dimensional range query on attribute \textit{Age}: $\marginal_{\{\text{Age}\}}=\attwork_{agerange}\kron \one^T_{|\attr_2|}\kron \one^T_{|\attr_3|}\kron \one^T_{|\attr_4|}$. This query matrix uses the range query workload on the first attribute (Age) and marginalizes out the other attributes (using the $\one^T$ vector). Intuitively, this is a range query marginal on the first attribute.
    \item 2-D range queries on \textit{Age}, \textit{Height}: $\marginal_{\{\text{Age, Height}\}}=\attwork_{agerange}\kron \attwork_{heighrange}\kron \one^T_{|\attr_3|}\kron \one^T_{|\attr_4|}$. This query matrix uses the range query workloads on Age and Height and marginalizes out the rest. Hence, the queries represented by the rows of this query matrix are like a two-way range marginals that can answer questions like ``how many people have age $\in[a,b]$ and height in $[c,d]$?''
    \item Range queries on Age for Females, and separately  range queries on Age for Males (this is a type of hybrid marginal/range query where we have equality on the categorical attributes and ranges on the numerical ones). It is represented as $\marginal_{\{\text{Age, Gender}\}}=\attwork_{agerange}\kron \one^T_{|\attr_2|}\kron \identity_{|\attr_3|}\kron \one^T_{|\attr_4|}$. Here, attributes $\attr_2$ and $\attr_4$ are marginalized out using the $\one^T$ vector, while the first attribute (Age) is given the range query workload and the third one (Gender) is given the identity workload. Hence each query in this query matrix answers a question of the form ``How many people with gender X have ages in the range $[a, b]$?''
    \item 2-way marginal on the attributes \textit{Gender}, \textit{Occupation}: $\marginal_{\{\text{Gender, Occupation}\}}=\one^T_{|\attr_1|}\kron \one^T_{|\attr_2|} \kron \identity_{|\attr_3|}\kron \identity_{|\attr_4|}$. Here the first two attributes are marginalized out and the last two (Gender and Occupation) are each given the identity workload. Hence this represents an ordinary 2-way marginal on Gender and Occupation.
\end{itemize}





Although $\attwork_i$ can be arbitrary, we expect the most common use-cases will set $\attwork_i$ to be the identity for categorical attributes and ranges or prefix-sums for numerical attributes. As a reference example, if a numerical attribute can take 3 values, the prefix-sum $\attwork_{prefix}$ and ranges $\attwork_{range}$ matrices look like:
{\small
\begin{align*}
    \attwork_{prefix} &=
\begin{bmatrix}
1      & 0      & 0 \\[6pt]
1      & 1      & 0 \\[6pt]
1      & 1      & 1  \\[6pt]
\end{bmatrix}
&\attwork_{range} &=
\begin{bmatrix}
1      & 0      & 0 \\[6pt]
0      & 1      & 0 \\[6pt]
0      & 0      & 1  \\[6pt]
1      & 1      & 0 \\[6pt]
0      & 1      & 1 \\[6pt]
1      & 1      & 1  \\[6pt]
\end{bmatrix}
\end{align*}
}

\subsection{Constructing Residual Mechanisms for \newrp}
The closure is again defined as: $\closure(\margworkload)=\{\margset^\prime~:~\margset^\prime\subseteq \margset \text{ for some $\margset\in\margworkload$}\}$.
\newrp will re-define the concept of residual mechanism $\resid_{\margset}$ for  $\margset\in\closure(\margworkload)$.

The first step is to define the analogue of the subtraction matrix.
Recall that every attribute $\attr_i$ has an associated matrix $\attwork_i$. A \emph{strategy replacement} \cite{LMHMR15}  for $\attwork_i$ is a matrix $\strat_i$  such that $ \attwork_i=\attwork_i \strat_i^\dagger \strat_i$ (that is, each row of $\attwork_i$ is in the row space of $\strat_i$) and for which the noisy answers $\strat_i\datavec+N(\mat{0},\identity)$ have low privacy cost. For example, for the ranges matrix $\attwork_{range}$, possible strategy replacements are $\attwork_{range}$ itself, or the prefix-sum matrix $\attwork_{prefix}$, or some more exotic matrix \cite{YYZH16,xiao2020optimizing,mckenna2021hdmm} (e.g., those works all have code for creating strategy replacements for small matrices $\attwork_i$). Strategy replacement matrices $\strat_i$  typically contain the same information as $\attwork_i$ but have lower privacy costs \cite{LMHMR15} and hence are widely used for intermediate calculations (i.e., noisy answers to the linear queries in $\strat_i$ can be used to reconstruct answers to $\attwork_i$ with lower variance for a given privacy budget than by directly getting noisy answers to the linear queries in $\attwork_i$).
Given a strategy replacement $\strat_i$  as input, \newrp creates the corresponding subtraction matrix $\submat_i$ and \ covariance matrix it needs $\strathalf_i\strathalf_i^T$ by using Algorithm \ref{alg:subtraction} (the covariance is either the same as in ResidualPlanner or is the identity matrix). The goal of the algorithm is to make the columns of $\submat_i$ orthogonal to the vector $\one_{|\attr_i|}$, since that vector is a replacement for $\submat_i$ when $\attr_i$ is not in a query $\marginal_{\margset}$ (i.e., when $\attr_i\notin \margset$). We use the orthogonalization procedure from \cite{commonmech}, which, when applied to $\one_{|\attr_i|}$, results in the Algorithm \ref{alg:subtraction}.

\begin{algorithm}[h]
   \DontPrintSemicolon
   \KwIn{Matrix $\strat_i \in \mathbb{R}^{m\times n}$, where $n=|\attr_i|$}
   \If{$\strat_i$ is the identity matrix}{
        $\submat_i =$ the subtraction matrix from Section \ref{subsec:margbase}\;
        $\strathalf_i = \submat_i$\;
   } 
   \Else{
    $\mat{P}_1 = \strat_i - \strat_i\one_n\one_n^T/n$\;
    Use Cholesky decomposition to represent $\mat{P}_1^T\mat{P}_1$ as $\mat{L}\mat{L}^T$, where $\mat{L}$ is lower triangular.\;
    $\mat{P}_2 \gets $ linearly independent columns of $\mat{L}$ \; 
    $r\gets$ \# of columns of $\mat{P}_2$.\;
    $\submat_i=\mat{P}_2^T,$ $\quad \strathalf_i=\identity_{r}$\;
   }
    \Return $\submat_i, \strathalf_i$

\caption{Constructing a Subtraction matrix $\submat_i$ and covariance $\strathalf_i\strathalf_i^T$ from $\strat_i$} \label{alg:subtraction}
\end{algorithm}

With these subtraction matrices, the definition of the residual basis and base mechanisms follows the pattern of ResidualPlanner. For any subset $\margset$ of attributes,
$\resid_{\margset}=\mat{V}_1\kron\cdots\kron\mat{V}_{\numattr}$, where $\mat{V}_i=\one^T_{|\attr_i|}$ if $\attr_i\notin \margset$ and $\mat{V}_i=\submat_{i}$ if $\attr_i\in\margset$. 
Similarly, $\covar_{\margset}=\bigotimes\limits_{\attr_i\in\margset} \strathalf_i\strathalf_i^T$.
The base mechanisms are defined as:
\begin{align}
 \mech_{\margset}(\datavec; \sigma_{\margset}^2)&\equiv \resid_{\margset}\datavec + N(\zero,\sigma_{\margset}^2\covar_{\margset})\label{eqn:baseextension}
 \end{align}

\begin{theoremEnd}[category=extension,proof end]{lemma}\label{lem:semipos}
Let $\submat_i$ be a subtraction matrix produced by Algorithm \ref{alg:subtraction}. Then 
$\submat_i\one_{|\attr_i|}=\vec{0}$ 
and the residual matrices are mutually orthogonal. That is, if $\margset\neq\margset^\prime$ then $\resid_{\margset}\resid_{\margset^\prime}^T=\mat{0}$. Also,  the rows from  the residual matrices in $\closure(\margworkload)$ are linearly independent and the space spanned by them contains the rows of the workload queries.
\end{theoremEnd}
\begin{proofEnd}

    We note that $\mat{P}_1\one_{|\attr_i|}=\mat{0}$, so $\one_{|\attr_i|}^T\mat{P}_1^T\mat{P}\one_{|\attr_i|}=0$ so $\one_{|\attr_i|}^T\mat{L}\mat{L}^T\one_{|\attr_i|}=0$ so $\one_{|\attr_i|}^T\mat{L}=\mat{0}$ so $\one_{|\attr_i|}^T\mat{P}_2=\mat{0}$ so $\submat_i \one_{|\attr_i|}=\mat{0}$ and $\one_{|\attr_i|}^T\submat_i^T=\mat{0}$.

    To prove pairwise orthogonality of residual matrices, 
    we write $\resid_\margset=\mat{V}_1\kron\cdots\kron\mat{V}_\numattr$ and $\resid_\margset=\mat{V}^\prime_1\kron\cdots\kron\mat{V}^\prime_\numattr$.
    Thus $\resid_{\margset}\resid_{\margset^\prime}^T = (\mat{V}_1(\mat{V}^\prime_1)^T)\kron\cdots\kron (\mat{V}_\numattr(\mat{V}^\prime_\numattr)^T)$.
    
    Since $\margset\neq \margset^\prime$ then some $i$ is in one of them but not the other, so that either $\mat{V}_i=\one_{|\attr_i|}^T$ and $\mat{V}_i^\prime=\submat_i$, or $\mat{V}_i=\submat_i$ and $\mat{V}^\prime_i=\one_{|\attr_i|}^T$. Therefore $\mat{V}_i(\mat{V}^\prime_i)^T$ is either $\one_{|\attr_i|}\submat_i^T=\mat{0}$ or $\submat_i\one_{|\attr_i|}=\mat{0}$. In both cases, this means  $\resid_{\margset}\resid_{\margset^\prime}^T$ is a kron product of terms, one of which is $\mat{0}$, and so $\resid_{\margset}\resid_{\margset^\prime}^T=\mat{0}$.

    Given the mutual orthogonality, linear independence follows if the rows within a residual matrix are linearly independent. The rows of $\submat_i$ are linearly independent by construction. 
    Hence each residual matrix is a kron product of matrices with linearly independent rows and hence the residual matrices have linearly independent rows.

    Next we note that the row space of $\strat_i$ is spanned by the row space of $\mat{P}_1$ and $\one_{|\attr_i|}^T$, which is the same as the rowspace spanned by the rows of $\mat{P}_1^T\mat{P}_1$ and $\one_{|\attr_i|}^T$, which is the same as the rowspace spanned by the rows of $\mat{L}\mat{L}^T$ and $\one_{|\attr_i|}^T$, which is the same as the row space spanned by $\mat{P}_2\mat{P}_2^T$ (because $\mat{P}_2$ extracts the linearly independent columns of $\mat{L}$) and $\one_{|\attr_i|}^T$, and that is exactly the same as the row space of $\submat_i^T\submat_i$ and $\one_{|\attr_i|}^T$, and is therefore the same as the row space spanned by the rows of $\submat_i$ and $\one_{|\attr_i|}^T$.
    Thus the rows of $\attwork_i$ are spanned by $\submat_i$ and $\one_{|\attr_i|}^T$.

    Next, using this result, we see that a query \\$\marginal_{\margset}=\Kron_{i=1}^\numattr \left(\begin{cases}
     \attwork_i& \text{ if }  i\in\margset\\
     \one_{|\attr_i|}^T &\text{ otherwise}
     \end{cases}\right)$
     has all of its rows in the row space of
     $\marginal_{\margset}=\Kron_{i=1}^\numattr \left(\begin{cases}
     \left[\begin{smallmatrix}\one_{|\attr_i|^T}\\\submat_i\end{smallmatrix}\right]& \text{ if }  i\in\margset\\
     \one_{|\attr_i|}^T &\text{ otherwise}
     \end{cases}\right)$
     and the rows of that matrix are all of the rows in all the matrices $\resid_{\margset^\prime}$ for $\margset^\prime\in\closure(\margset)$. Now since the closure of a workload is the union of closures of the  individual queries, the rowspace spanned by the rows of the residual matrices in the closure contain all of the rows of the workload queries.
\end{proofEnd}

\subsection{ResidualPlanner+ Measurement Phase}\label{sec:rpp:measure}
The updated base mechanism pseudocode is shown in Algorithm \ref{alg:measureplus}.
The following theorem provides the privacy cost.

\begin{algorithm}[t]
   \DontPrintSemicolon
    $\vec{v} \gets $ compute marginal  on $\margset$ from $\datavec$\tcp{equals $\marginal_{\margset}\datavec$} \label{line:eval2}
    $\mat{H}_1\gets \bigotimes\limits_{\attr_i\in\margset} \submat_{i}$\tcp{ implicit representation}\label{line:measureHPlus}
    $\mat{H}_2\gets \bigotimes\limits_{\attr_i\in\margset} \strathalf_{i}$\tcp{ implicit representation}
    $m\gets $ number of columns of $\mat{H}_2$\;
    $\vec{z}\gets N(\zero, \identity_{m})$\tcp{ independent noise}\label{line:measureZPlus}
    \Return $\mat{H}_1\vec{v} + \sigma_{\margset}\mat{H}_2\vec{z}$\tcp{use  kron-product/vector multiplication from \cite{mckenna2018optimizing}}
\caption{ResidualPlanner+ measurement: $\mech_{\margset}(\datavec; \sigma^2_{\margset})\equiv\resid_{\margset}\datavec + N(\zero,\sigma^2_{\margset}\covar_{\margset})$}\label{alg:measureplus}
\end{algorithm}


\begin{theoremEnd}[category=extension,all end]{lemma}\label{lem:pcostlemmaextension}
If $\strat_i$ has full column rank and $\submat_i$ is constructed from $\strat_i$ using Algorithm \ref{alg:subtraction}, then $\submat_i$ has linearly independent rows and  $$\submat_{i}^T(\submat_{i}\submat_{i}^T)^{-1}\submat_{i}= \identity_{|\attr_i|}- \frac{1}{|\attr_i|}\one_{|\attr_i|}\one^T_{|\attr_i|}.$$
\end{theoremEnd}
\begin{proofEnd}
First, note that since $\strat_i$ has full column rank, then $\strat_i\vec{x}\neq \vec{0}$ for any $\vec{x}$ orthogonal to $\one_{|\attr_i|}$. The matrix $\mat{P}_1$ in Algorithm \ref{alg:subtraction} therefore has rank $|\attr_i|-1$ since $\one_{|\attr_i|}$ is in its null space, but no other vector orthogonal to $\one_{|\attr_i|}$ is in the null space (because of the previously mentioned fact about $\strat_i$). It then follows that the ranks of $\mat{L}, \mat{P}_2$ and $\submat_i$ from Algorithm \ref{alg:subtraction} all have rank $|\attr_i|-1$. 

We note therefore that $\submat_{i}$ has size $|\attr_i|-1 \times |\attr_i|$, rank $|\attr_i|-1$ and its rows are orthogonal to $\one_{|\attr_i|}$ (as a consequence of Lemma \ref{lem:semipos}). Hence its rows are linearly independent. Using the SVD decomposition, express $\submat_{i}=\mat{U}\mat{D}\mat{V}^T$, where $\mat{U}$ is an $|\attr_i|-1\times |\attr_i|-1$ orthogonal matrix, $\mat{D}$ is a $|\attr_i|-1\times |\attr_i|-1$ diagonal matrix, and $\mat{V}$ is an $|\attr_i|\times |\attr_i|-1$ matrix with orthogonal columns.

We note that $\mat{D}$ is invertible because the rank of $\submat_i$ is $|\attr_i|-1$ and the columns of $\mat{V}$ must be orthogonal to $\one_{|\attr_i|}$ because $\submat_{i}\one_{|\attr_i|}=\vec{0}$.
Then 
\begin{align*}
\lefteqn{    \submat_{i}^T(\submat_{i}\submat_{i}^T)^{-1}\submat_{i}}\\
&= \mat{V}\mat{D}\mat{U}^T (\mat{U}\mat{D}\mat{V}^T \mat{V}\mat{D}\mat{U}^T)^{-1} \mat{U}\mat{D}\mat{V}^T\\
&= \mat{V}\mat{D}\mat{U}^T (\mat{U}\mat{D}\mat{D}\mat{U}^T)^{-1} \mat{U}\mat{D}\mat{V}^T\\
&= \mat{V}\mat{D}\mat{U}^T \mat{U}\mat{D}^{-1}\mat{D}^{-1}\mat{U}^T \mat{U}\mat{D}\mat{V}^T\\
&=\mat{V}\mat{V}^T
\end{align*}
Now, we know that $\left[\mat{V} \quad\frac{\one_{|\attr_i|}}{\sqrt{|\attr_i|}}\right]$ is an $|\attr_i|\times|\attr_i|$ orthogonal matrix, so
\begin{align*}
    \identity_{|\attr_i|} &=
    \left[\mat{V} \quad\frac{\one_{|\attr_i|}}{\sqrt{|\attr_i|}}\right] \left[\mat{V} \quad\frac{\one_{|\attr_i|}}{\sqrt{|\attr_i|}}\right]^T \\
    &= \left[\mat{V} \quad\frac{\one_{|\attr_i|}}{\sqrt{|\attr_i|}}\right] \left[
    \begin{matrix}
        \mat{V}^T\\ \quad\frac{\one_{|\attr_i|}^T}{\sqrt{|\attr_i|}}
    \end{matrix}
    \right]\\
    &= \mat{V}\mat{V}^T + \frac{1}{|\attr_i|}\one_{|\attr_i|}\one_{|\attr_i|}^T
\end{align*}
Combining both results, we get \\$\submat_{i}^T(\submat_{i}\submat_{i}^T)^{-1}\submat_{i}$ \\$= \identity_{|\attr_i|}- \frac{1}{|\attr_i|}\one_{|\attr_i|}\one^T_{|\attr_i|}$.

 \end{proofEnd}

\begin{theoremEnd}[category=extension,proof end]{theorem}\label{thm:pcostextension}
For each $i$, 
let $\beta_i$ be the  largest diagonal element of $\submat_{i}^T(\strathalf_{i}\strathalf_{i}^T)^{-1}\submat_{i}$.
Then the privacy cost of the \newrp base mechanism $\mech_{\margset}$ having noise parameter $\sigma^2_{\margset}$ is $\frac{1}{\sigma^2_{\margset}}\prod_{\attr_i\in\margset}\beta_i$ and the evaluation of $\mech_{\margset}$ given in Algorithm \ref{alg:measureplus} is correct.
\end{theoremEnd}
\begin{proofEnd}
Without loss of generality (and to simplify notation), assume $\margset=\{\attr_1,\dots,\attr_\ell\}$ consists of the first $\ell$ attributes.

By definition, $\pcost(\mech_{\margset}(\cdot; \sigma^2_{\margset}))$ is the largest diagonal of $\frac{1}{\sigma^2}\resid_{\margset}^T\covar_{\margset}^{-1}\resid_{\margset}$. Thus we can write:

\begin{align}
\resid_{\margset} &= \left(\bigotimes_{i=1}^\ell \submat_{i}\right) \kron \left(\bigotimes_{j=\ell+1}^{\numattr}\one^T_{|\attr_j|}\right)\nonumber\\
\resid^T_{\margset} &= \left(\bigotimes_{i=1}^\ell \submat^T_{i}\right) \kron \left(\bigotimes_{j=\ell+1}^{\numattr}\one_{|\attr_j|}\right)\nonumber\\
\mat{H}_1 &= \left(\bigotimes_{i=1}^\ell \submat_{i}\right) \kron \left(\bigotimes_{j=\ell+1}^{\numattr}\left[\begin{smallmatrix}1\end{smallmatrix}\right]\right)\nonumber\\&\quad\text{(rightmost krons use $1\times 1$ matrices)}\nonumber\\
\covar_{\margset}&=\mat{H}_2\mat{H}_2^T \nonumber\\ &= \left(\bigotimes_{i=1}^\ell (\strathalf_{i}\strathalf_{i}^T)\right) \kron \left(\bigotimes_{j=\ell+1}^{\numattr}\left[\begin{smallmatrix}1\end{smallmatrix}\right]\right)\nonumber\\
\covar^{-1}_{\margset}&= \left(\bigotimes_{i=1}^\ell (\strathalf{i}\strathalf_{i}^T)^{-1}\right) \kron \left(\bigotimes_{j=\ell+1}^{\numattr}\left[\begin{smallmatrix}1\end{smallmatrix}\right]\right)\nonumber\\
\lefteqn{
\resid_{\margset}^T\covar_{\margset}^{-1}\resid_{\margset} }\nonumber\\&=\left(\bigotimes_{i=1}^\ell \submat_{i}^T(\strathalf{i}\strathalf_{i}^T)^{-1}\submat_{i}\right) \nonumber\\&\qquad\kron \left(\bigotimes_{j=\ell+1}^{\numattr}\one_{|\attr_j|}\left[\begin{smallmatrix}1\end{smallmatrix}\right]\one_{|\attr_j|}^T\right)\label{eqn:pcostextension}
\end{align}
and so the privacy cost is the product of the largest diagonals of $\submat_{i}^T(\strathalf_{i}\strathalf_{i}^T)^{-1}\submat_{i}$ for $i\in\margset$. 
This proves the result for $\pcost(\mech_{\margset}(\cdot,\sigma^2_{\margset}))$. 


We next consider  the correctness of Algorithm \ref{alg:measureplus}. First, since the marginal on $\margset=\{\attr_1,\dots, \attr_\ell\}$ is $$\left(\Kron_{i=1}^\ell \identity_{|\attr_i|}\right)\kron\left(\Kron_{i=\ell+1}^\numattr \one_{|\attr_i|}^T\right)\datavec,$$ we need to show that for the matrix $\mat{H}_1$,
$$\mat{H}_1\left(\Kron_{i=1}^\ell \identity_{|\attr_i|}\right)\kron\left(\Kron_{i=\ell+1}^\numattr \one_{|\attr_i|}^T\right)\datavec=\resid_{\margset}\datavec.$$
Then we can write:
\begin{align*}
\resid_{\margset} &= \left(\bigotimes_{i=1}^\ell \submat_{i}\right) \kron \left(\bigotimes_{j=\ell+1}^{\numattr}\one^T_{|\attr_j|}\right)\\
\widetilde{\marginal}_{\margset} &\equiv \left(\bigotimes_{i=1}^\ell \identity_{|\attr_i|}\right) \kron \left(\bigotimes_{j=\ell+1}^{\numattr}\one^T_{|\attr_j|}\right)\\
&\quad\text{rightmost product is a matrix with 1 row}\\
\mat{H}_1 &= \left(\bigotimes_{i=1}^\ell \submat_{i}\right) \kron \left[\begin{smallmatrix}1\end{smallmatrix}\right]\\&\quad\text{(rightmost term is a $1\times 1$ matrix)}\\
\mat{H}_1\widetilde{\marginal}_{\margset} &= \left(\bigotimes_{i=1}^\ell \left(\submat_{i}\identity_{|\attr_i|}\right)\right) \kron \left(
\left[\begin{smallmatrix}1\end{smallmatrix}\right]
\left(\bigotimes_{j=\ell+1}^{\numattr}\one^T_{|\attr_j|}\right)\right)\\
&=\resid_{\margset}
\end{align*}
Next, we note that if $\vec{z}$ is distributed as $N(0,\mat{I}_m)$  in Algorithm \ref{alg:measureplus} then $\sigma_{\margset}\mat{H}_2\vec{z}$ has the distribution $N(0,\sigma_\margset^2\mat{H}_2\mat{H}_2^T)=\covar_{\margset}$ and hence the algorithm is correct.
\end{proofEnd}

\subsection{The Reconstruction Phase}\label{sec:rpp:reconstruct}
Recall that each attribute $\attr_i$ now has an associated base matrix $\attwork_i$ that tells us what to do with that attribute when it is part of the query (i.e., a query is defined as $\marginal_{\margset}=\mathbf{V}_1\kron\cdots\kron \mathbf{V}_{\numattr}$,  where $\mathbf{V}_i=\attwork_i$ when $i\in\margset$ and $\mathbf{V}_i=\one^T_{|\attr_i|}$ when $i\notin\margset$.). Recall, also, that in the matrix mechanism paradigm, intermediate computations  are performed with carefully designed strategy matrices $\strat_i$ in place of the $\attwork_i$ \cite{LMHMR15}. Both the $\attwork_i$ and $\strat_i$ are inputs to \newrp (if $\strat_i$ is missing, we set $\strat_i=\attwork_i$). This makes the reconstruction slightly more complex than before. The reconstruction procedure is shown in Algorithm \ref{alg:reconstructionextension}.

\begin{algorithm}[h]
   \DontPrintSemicolon
   \KwIn{Desired query $\marginal_{\widetilde{\margset}}$. Noise scale parameters $\sigma^2_{\margset^\prime}$, noisy answer vector $\outp_{\margset^\prime}$ of mechanism $\mech_{\margset^\prime}$ for every $\margset^\prime\in\closure(\widetilde{\margset})$, and  basic attribute matrices $\attwork_i$ for $i=1,\dots, \numattr$.}
   \KwOut{$\vec{q}$ is output as an unbiased noisy estimate of $\marginal_{\widetilde{\margset}}\datavec$.}
    $\vec{q} \gets \zero$\;
    \For{each $\margset' \in\closure(\widetilde{\margset})$}
    {
    $\mat{U}\gets $ $\mat{V}_1\kron\cdots\kron\mat{V}_{\numattr}$, where $\mat{V}_i=
    \begin{cases}
    \submat^{\dagger}_{i} & \text{ if } \attr_i\in\margset^\prime\\
    \frac{1}{|\attr_i|}\one_{|\attr_i|} & \text{ if }\attr_i \in \widetilde{\margset} \setminus \margset^\prime\\
     [1] &\text{ if }\attr_i \notin \widetilde{\margset}
    \end{cases}$\label{line:reconUextension}\;
    $\vec{q} \gets \vec{q} + \mat{U} \outp_{\margset^\prime}$\tcp{use  kron-product/vector multiplication from \cite{mckenna2018optimizing}}
    }
    $\widehat{\mat{W}} \gets \Kron_{i\in\widetilde{\margset}} \attwork_i$ \;
    answer $\gets \widehat{\mat{W}}\vec{q}$ \tcp{use  kron-product/vector multiplication from \cite{mckenna2018optimizing}}
\Return answer\;
\caption{\newrp Reconstruction Phase}\label{alg:reconstructionextension}
\end{algorithm}

The following theorem shows that the reconstruction algorithm is correct and shows how to
compute the covariance matrix of the reconstructed query answers and the sum of the cell variances
within any query supported by the workload. As with ResidualPlanner, we see that the computation scales
with the size of the query answer being reconstructed, rather than with the overall dataset domain size as with prior work.

\begin{theoremEnd}[category=extension,proof end]{theorem}\label{thm:varextension}
Let $\attwork_1,\dots,\attwork_\numattr$ be the base matrices for the attributes $\attr_1,\dots, \attr_{\numattr}$.
Let $\margworkload$ be  a \newrp workload. Given positive numbers $\sigma^2_{\margset}$ for  each $\margset\in\closure(\margworkload)$, let $\mech$ be the mechanism that runs all the  $\mech_{\margset}(\datavec;\sigma^2_{\margset})$ for $\margset\in\closure(\margworkload)$. Let  $\{\outp_{\margset}~:~\margset \in \closure(\margworkload) \}$ denote the privacy-preserving noisy answers (e.g., $\outp_{\margset}=\mech_{\margset}(\datavec,\sigma^2_\margset)$). Then for any attribute set $\widetilde{\margset}\in\closure(\margworkload)$, Algorithm  \ref{alg:reconstructionextension} returns the unique linear unbiased  estimate of $\marginal_{\widetilde{\margset}}\datavec$. 

The covariance matrix for the reconstructed answer to the query $\marginal_{\widetilde{\margset}}$ is equal to:
$\sum_{\margset\subseteq\widetilde{\margset}}\sigma^2_{\margset}
{\Kron}_{i=1}^\numattr \mat{\Psi}_{\margset,i}\mat{\Psi}_{\margset,i}^T$, where 
$\mat{\Psi}_{\margset,i} = 1$ if $\attr_i\not\in \widetilde{\margset}$; 
and $\mat{\Psi}_{\margset,i} = \attwork_i\frac{\one_{|\attr_i|}}{|\attr_i|}$ if $\attr_i\in\widetilde{\margset}\setminus\margset$; and 
$\mat{\Psi}_{\margset,i} = \attwork_i\submat_i^\dagger\strathalf_i$  if $\attr_i\in\margset$.

The trace of this covariance matrix, which is the same as the sum of the squares of the cell variances in the reconstructed answer to $\marginal_{\widetilde{\margset}}$, is equal to
$$\sum_{\margset\subseteq\widetilde{\margset}} \sigma^2_\margset \prod_{\attr_i\in \margset}
||\attwork_i\submat_i^\dagger\strathalf_i||^2_F *
\prod_{\attr_j\in\widetilde{\margset}\setminus\margset} \frac{||\attwork_j \one_{|\attr_j|}||^2_2}{|\attr_j|^2}$$
x\end{theoremEnd}
\begin{proofEnd}$~$\\

Pick $\widetilde{\margset}\in\closure(\margworkload)$.

Let $\resid_{all}$ be the matrix one obtains by vertically stacking all the $\resid_{\margset}$ for $\margset\in\closure(\margworkload)$ and let $\outp_{all}$ be the vector of all of the noisy answers stacked in the same way. By Lemma \ref{lem:semipos}, the rows of $\resid_{all}$ are linearly independent and its rowspace contains all the rows of the query matrices $\marginal_\margset$ for $\margset\in\closure(\margworkload)$. Also, $\outp_{all}$ equals $\resid_{all}\datavec +\vec{z}$, where $\vec{z}$ is a vector of correlated Gaussian noise. 

Thus, the query answer $\marginal_{\widetilde{\margset}}\datavec$  has a unique unbiased linear estimator which equals:
\begin{align}
    \marginal_{\widetilde{\margset}}\resid_{all}^\dagger \outp_{all}\label{eqn:extensionsanswer}
\end{align}
and the covariance matrix of this estimate is
\begin{align}
    \marginal_{\widetilde{\margset}}\resid_{all}^\dagger E[\vec{z}\vec{z}^T] (\resid_{all}^\dagger)^T    \marginal_{\widetilde{\margset}}^T\label{eqn:extensioncovariance}
\end{align}

We first note that since the rows of $\resid_{all}$ are linearly independent, then $\resid_{all}\resid_{all}^T$ is invertible and so $\resid^\dagger_{all}=\resid_{all}^T(\resid_{all}\resid_{all}^T)^{-1}$. Next, by Lemma \ref{lem:semipos}, we know that $\resid_\margset\resid_{\margset^\prime}=\mat{0}$ whenever $\margset\neq \margset^\prime$. That means that $\resid_{all}\resid_{all}^T$ is a block-diagonal matrix where the blocks are $\resid_\margset\resid_{\margset}$ and appear in the same order as the stacking of $\resid_{all}$. Thus $\resid^\dagger_{all}$ is the matrix obtained by \emph{horizontally} stacking $\resid_{\margset}^T(\resid_{\margset}\resid_{\margset}^T)^{-1}$ for all $\margset$ in the same order, which is the same as stacking $\resid^\dagger_\margset$ horizontally (since $\resid^\dagger_\margset=\resid_{\margset}^T(\resid_{\margset}\resid_{\margset}^T)^{-1}$). 

Thus we have:
\begin{align*}
    \resid^\dagger_{all}\outp_{all} &= \sum_{\margset} \resid_{\margset}^T(\resid_{\margset}\resid_{\margset}^T)^{-1}\outp_{\margset}\\
    &=\sum_{\margset}\resid^\dagger_\margset\outp_{\margset}\\
   \marginal_{\widetilde{\margset}}\resid_{all}^\dagger \outp_{all} 
   &= \sum_{\margset}\marginal_{\widetilde{\margset}}\resid_{\margset}^T(\resid_{\margset}\resid_{\margset}^T)^{-1}\outp_{\margset}\\
    &= \sum_{\margset\subseteq \widetilde{\margset}}\marginal_{\widetilde{\margset}}\resid_{\margset}^T(\resid_{\margset}\resid_{\margset}^T)^{-1}\outp_{\margset}\\
    &= \sum_{\margset\subseteq\widetilde{\margset}}\marginal_{\widetilde{\margset}}\resid^\dagger_\margset\outp_{\margset}
\end{align*}
The second-to-last equality follows because for any $\margset\not\subseteq\widetilde{\margset}$ there is an attribute $\attr_\ell\in \margset$ with $\attr_\ell\notin\widetilde{\margset}$; therefore the kron representation of $\marginal_{\widetilde{\margset}}$ has a $\one^T_{|\attr_\ell|}$ in position $\ell$ while that kron representation of $\resid^\dagger_\margset =\resid_{\margset}^T(\resid_{\margset}\resid_{\margset}^T)^{-1}$ has $\submat_\ell^T(\submat_\ell\submat_\ell^T)^{-1}$ in position $\ell$. Thus their product is $\mat{0}$ by Lemma \ref{lem:semipos}.

We also note that $\one_{|\attr_i|}(\one_{|\attr_i|}^T\one_{|\attr_i|})^{-1}=\frac{1}{|\attr_i|}\one_{|\attr_i|}$. Thus it follows that:
\begin{align*}
\lefteqn{    \marginal_{\widetilde{\margset}}\resid_{all}^\dagger \outp_{all} }\\
    &= \sum_{\margset\subseteq\widetilde{\margset}}\marginal_{\widetilde{\margset}}\resid^\dagger_\margset\outp_{\margset}\\
    &=\sum_{\margset\subseteq\widetilde{\margset}} 
    \left(\Kron_{i=1}^\numattr\left.
    \begin{cases}
        \attwork_i \submat_i^\dagger & \text{if } \attr_i\in \margset\\
        \attwork_i \frac{\one_{|\attr_i|}}{|\attr_i|} & \text{if }\attr_i\in \widetilde{\margset}\setminus\margset\\
        \one_{|\attr_i|}^T \frac{\one_{|\attr_i|}}{|\attr_i|}  & \text{if }\attr_i \notin\widetilde{\margset}
    \end{cases}\right\}\right)\outp_{\margset}
    \\
    &=\sum_{\margset\subseteq\widetilde{\margset}} 
    \left(\Kron_{i=1}^\numattr
    \left.\begin{cases}
        \attwork_i \submat_i^\dagger & \text{if } \attr_i\in \margset\\
        \attwork_i \frac{\one_{|\attr_i|}}{|\attr_i|} & \text{if }\attr_i\in \widetilde{\margset}\setminus\margset\\
        [1]  & \text{if }\attr_i \notin\widetilde{\margset}
    \end{cases}\right\}\right)\outp_{\margset}
    \\
\end{align*}
Which is exactly what Algorithm \ref{alg:reconstructionextension} computes.

We next consider the covariance. Starting from Equation \ref{eqn:extensioncovariance}, we note that $E[\vec{z}\vec{z}^T]$ is a block diagonal matrix whose diagonals blocks are $\sigma^2_{\margset}\Kron_{\attr_i\in\margset}\strathalf_i\strathalf_i^T$ for each $\margset\in\closure(\margworkload)$, appearing in the same order as the stacking in $\resid_{all}$. Recall that  $\resid^\dagger_{all}$ is the matrix obtained by \emph{horizontally} stacking $\resid_{\margset}^T(\resid_{\margset}\resid_{\margset}^T)^{-1}$ (which equals $\resid_{\margset}^\dagger$) for all $\margset$.
Thus
{\scriptsize
\begin{align*}
\lefteqn{    \resid_{all}^\dagger E[\vec{z}\vec{z}^T] (\resid_{all}^\dagger)^T}  \\
&=\left(\sum_{\margset}\resid_{\margset}^\dagger   \left(\sigma^2_{\margset} \Kron_{\attr_i\in\margset}\strathalf_i\strathalf_i^T\right)
  (\resid_{\margset}^\dagger)^T\right)\\
&=\sum_{\margset}\sigma^2_{\margset}\underset{i=1}{\overset{\numattr}{\Kron}}\left.
\begin{cases}
    \submat_i^\dagger \strathalf_i\strathalf_i^T(\submat_i^\dagger)^T & \text{if }\attr_i\in\margset\\
    (\one^T_{|\attr_i|})^\dagger [1] [1]^T ((\one^T_{|\attr_i|})^\dagger)^T & \text{if }\attr_i\notin\margset
\end{cases}
\right\}\\
&=\sum_{\margset}\sigma^2_{\margset}\underset{i=1}{\overset{\numattr}{\Kron}}\left.
\begin{cases}
    \submat_i^\dagger \strathalf_i\strathalf_i^T(\submat_i^\dagger)^T & \text{if }\attr_i\in\margset\\
    \frac{\one_{|\attr_i|}}{|\attr_i|} [1] [1]^T \frac{\one^T_{|\attr_i|}}{|\attr_i|} & \text{if }\attr_i\notin\margset
\end{cases}
\right\}\\
\lefteqn{\marginal_{\widetilde{\margset}}\resid_{all}^\dagger E[\vec{z}\vec{z}^T] (\resid_{all}^\dagger)^T    \marginal_{\widetilde{\margset}}^T}\\
&=\sum_{\margset}\marginal_{\widetilde{\margset}}\sigma^2_{\margset}\overset{\numattr}{\underset{i=1}{\Kron}}\left.
\begin{cases}
    \submat_i^\dagger \strathalf_i\strathalf_i^T(\submat_i^\dagger)^T & \text{if }\attr_i\in\margset\\
    \frac{\one_{|\attr_i|}}{|\attr_i|} [1] [1]^T \frac{\one^T_{|\attr_i|}}{|\attr_i|} & \text{if }\attr_i\notin\margset
\end{cases}
\right\}\marginal_{\widetilde{\margset}}^T\\
&=\sum_{\margset\subseteq\widetilde{\margset}}\marginal_{\widetilde{\margset}}\sigma^2_{\margset}\overset{\numattr}{\underset{i=1}{\Kron}}\left.
\begin{cases}
    \submat_i^\dagger \strathalf_i\strathalf_i^T(\submat_i^\dagger)^T & \text{if }\attr_i\in\margset\\
    \frac{\one_{|\attr_i|}}{|\attr_i|} [1] [1]^T \frac{\one^T_{|\attr_i|}}{|\attr_i|} & \text{if }\attr_i\notin\margset
\end{cases}
\right\}\marginal_{\widetilde{\margset}}^T\\
\end{align*}
}
the last equality (allowing us to only consider subsets of $\widetilde{\margset}$) comes from the same discussion that allowed us to consider only those subsets when proving the correctness of the reconstruction algorithm above.
Continuing the derivation,
{\scriptsize
\begin{align*}
    \lefteqn{\marginal_{\widetilde{\margset}}\resid_{all}^\dagger E[\vec{z}\vec{z}^T] (\resid_{all}^\dagger)^T    \marginal_{\widetilde{\margset}}^T}\\
    &=\sum_{\margset\subseteq\widetilde{\margset}}\marginal_{\widetilde{\margset}}\sigma^2_{\margset}\overset{\numattr}{\underset{i=1}{\Kron}}\left.
\begin{cases}
    \submat_i^\dagger \strathalf_i\strathalf_i^T(\submat_i^\dagger)^T & \text{if }\attr_i\in\margset\\
    \frac{\one_{|\attr_i|}}{|\attr_i|} [1] [1]^T \frac{\one^T_{|\attr_i|}}{|\attr_i|} & \text{if }\attr_i\notin\margset
\end{cases}
\right\}\marginal_{\widetilde{\margset}}^T\\
&=\sum_{\margset\subseteq\widetilde{\margset}}\sigma^2_{\margset}
\overset{\numattr}{\underset{i=1}{\Kron}}\left.
\begin{cases}
     \attwork_i\submat_i^\dagger\strathalf_i\strathalf_i^T(\submat_i^\dagger)^T\attwork_i^T & \hspace{-0.2cm}\text{if }\attr_i\in\margset\\
    \attwork_i\frac{\one_{|\attr_i|}}{|\attr_i|}  \frac{\one^T_{|\attr_i|}}{|\attr_i|}\attwork_i^T & \hspace{-0.2cm}\text{if }\attr_i\in\widetilde{\margset}\setminus\margset\\
    [1] & \hspace{-0.2cm}\text{if }\attr_i\notin \widetilde{\margset}
\end{cases}
\right\}
\end{align*}
}
This is the covariance matrix of the the reconstructed query answers, and so the variances are on the diagonals. Since the trace operator (sum of the diagonals) commutes with addition over matrices (but max over diagonals does not), one can get an expression that is linear in the $\sigma^2_{\margset}$ values. The trace of a kron product of matrices is the product of their traces, hence:
\begin{align*}
     \lefteqn{\text{trace}\left(   \marginal_{\widetilde{\margset}}\resid_{all}^\dagger E[\vec{z}\vec{z}^T] (\resid_{all}^\dagger)^T    \marginal_{\widetilde{\margset}}^T\right)}\\
     &=
     \sum_{\margset\subseteq\widetilde{\margset}}\sigma^2_{\margset}
\overset{\numattr}{\underset{i=1}{\prod}}\left.
\begin{cases}
     ||\attwork_i\submat_i^\dagger\strathalf_i||_F^2 & \text{if }\attr_i\in\margset\\
    \frac{1}{|\attr_i|}^2||\attwork_i\one_{|\attr_i|}||_F^2   & \text{if }\attr_i\in\widetilde{\margset}\setminus\margset\\
    1 & \text{if }\attr_i\notin \widetilde{\margset}
\end{cases}
\right\}
\end{align*}
\end{proofEnd}

\subsection{Optimizing the choice of $\sigma_{\margset}^2$.}\label{sec:optextension}

 Each \newrp base mechanism $\mech_{\margset}$ has a single tuning parameter $\sigma_{\margset}^2$. The selection step of \newrp involves optimizing the choice of the $\sigma^2_{\margset}$ for $\margset\in\closure(\margworkload)$ to minimize a loss function $\loss$ of the workload query variances subject to an upper bound on the privacy cost. Since \newrp provides closed form and data independent expressions of the privacy cost and reconstructed query variances in terms of the $\sigma_{\margset}^2$ tuning parameters, any loss function can be used.

However, some loss functions $\loss$ are easier to work with as they result in a convex optimization problem with respect to the tuning parameters, thus allowing the use of off-the-shelf optimizers. Given a query $\marginal_{\widetilde{\margset}}$, let $\sov(\marginal_{\widetilde{\margset}})$ denote the sum of the variances of the reconstructed answers to $\marginal_{\widetilde{\margset}}$ (recall that each query produces a vector of answers, just like with marginal queries). Theorem \ref{thm:varextension} shows that $\sov(\marginal_{\widetilde{\margset}})$ is a linear function of the $\sigma_{\margset}^2$ tuning parameters for $\margset\subseteq {\widetilde{\margset}}$ and Theorem \ref{thm:pcostextension} shows that the privacy cost is a linear function of the $1/\sigma_{\margset}^2$, for $\margset\in\closure(\margworkload)$, with nonnegative coefficients. Therefore when $\loss$ is a convex function of the collection of $\sov(\marginal_{\margset})$ values for $\margset\in\margworkload$, then minimizing $\loss$ subject to an upper bound on the privacy cost is a convex optimization problem with convex constraints.

\section{Experiments for ResidualPlanner}\label{sec:experiments}
In this section, we evaluate accuracy and scalability for marginal workloads and compare \newrp (which is identical to ResidualPlanner on marginals)
against HDMM \cite{mckenna2021hdmm}, including variations of HDMM with faster reconstruction phases \cite{mckenna2019graphical}. More complex workloads, like mixtures of marginals and prefix-sum queries, are evaluated in Section \ref{sec:exprange}. 

The hardware used was an Ubuntu 22.04.2 server with 12 Intel(R) Core(TM) i7-8700 CPU @ 3.20GHz processors and 32GB of DDR4 RAM. We use 3 real datasets to evaluate accuracy and 1 synthetic dataset to evaluate scalability. The real datasets are (1) the Adult dataset \cite{Dua:2019} with 14 attributes, each having domain sizes $100,100,100,99,85,42,16,15,9,7,6,5,2,2$, respectively, resulting in a record domain size of
$ 6.41 * 10^{17}$; (2) the CPS dataset \cite{cps2023link} with 5 attributes, each having domain size $100, 50, 7, 4, 2$, respectively, resulting in a record domain size of $2.8 * 10^{5}$; (3) the Loans dataset \cite{kaggleloans} with 12 attributes, each having domain size $101, 101, 101, 101, 3, 8,
    36, 6, 51, 4, 5, 15$, respectively, resulting in a record domain size of $8.25* 10^{15}$. The synthetic dataset is called Synth-$n^d$. Here $d$ refers to the number of attributes (we experiment from $d=2$ to $d=100$) and $n$ is the domain size of each attribute. The running times of the algorithms only depend on $n$ and $d$ and not on the records in the synthetic data.
For all experiments, we set the privacy cost $\pcost$ to 1, so all mechanisms being compared satisfy $0.5$-zCDP and 1-Gaussian DP.

\subsection{Scalability of the Selection Phase}
\begin{figure}[h]
\centering
\includegraphics[width=0.45\textwidth]{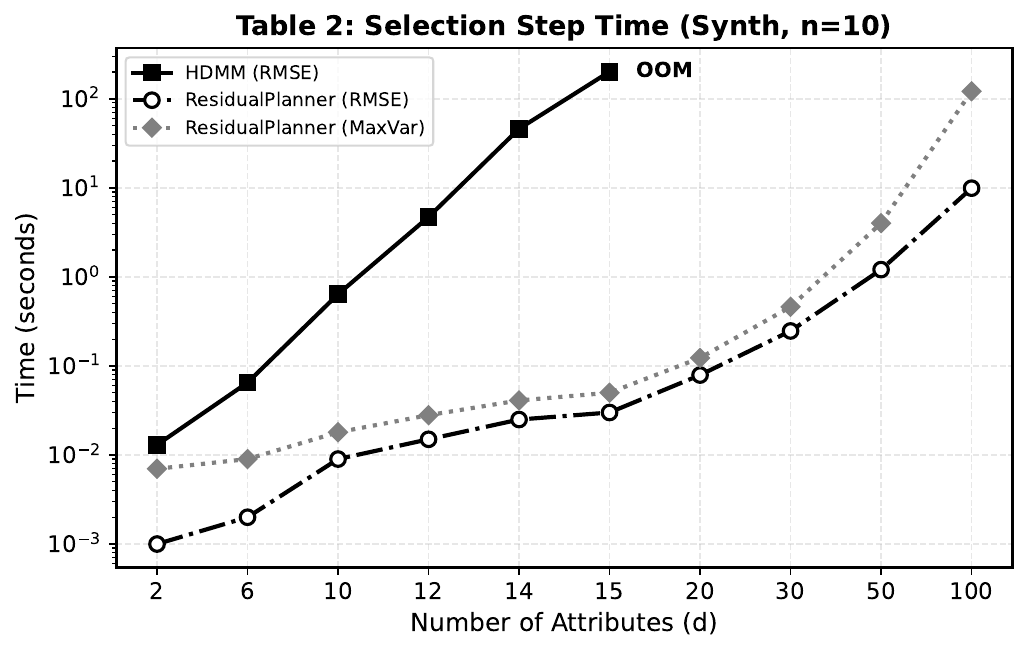}
\caption{Graphical depiction of Table~\ref{tab:timeselect}. 
Selection step running time (seconds) on $\text{Synth-}n^d$ with $n=10$ 
and varying $d$. HDMM runs out of memory at $d=20$, while ResidualPlanner 
scales to $d=100$ for both RMSE and max variance objectives. }
\label{fig:figtab2}
\end{figure}

We first consider how long each method takes to perform the selection phase (i.e., determine what needs noisy answers and how much noise to use). HDMM can only optimize total variance, which is equivalent to root mean squared error. For ResidualPlanner we consider both RMSE and max variance as objectives (the latter is a harder to solve problem). Each algorithm is run 5 times and the average is taken. Table \ref{tab:timeselect} shows the numerical running time results with Figure \ref{fig:figtab2} plotting them graphically for easier comparisons. Accuracy results will be presented later.

\begin{table*}[h!]
\centering
\caption{\textbf{Time for Selection Step in seconds} on Synth$-n^d$ dataset. $n=10$ and the number of attributes $d$ varies. The workload consists of all marginals on $\leq 3$ attributes each. Times for HDMM are reported with $\pm 2$ standard deviations.}
\label{tab:timeselect}
\begin{tabular}{|c|c|c|c|}\hline
$d$ & \multicolumn{1}{|p{0.25\textwidth}|}{\centering HDMM\newline RMSE Objective} & \multicolumn{1}{|p{0.25\textwidth}|}{\centering ResidualPlanner\newline RMSE Objective} & \multicolumn{1}{|p{0.25\textwidth}|}{\centering ResidualPlanner\newline Max Variance Objective}\\\hline
2  & $0.013 \pm 0.003$ & $0.001 \pm 0.0008$ & $0.007 \pm 0.001$\\
6 & $0.065 \pm 0.012$& $0.002 \pm 0.0008$& $0.009 \pm 0.001$ \\
10 & $0.639 \pm 0.059$ & $0.009 \pm 0.001$ & $0.018 \pm 0.001$\\
12 & $4.702 \pm 0.315$& $0.015 \pm 0.001$ & $0.028 \pm 0.001$\\
14 & $46.054 \pm 12.735$& $0.025 \pm 0.002$& $0.041 \pm 0.001$\\
15 & $201.485 \pm 13.697$& $0.030 \pm 0.017$ & $0.050 \pm 0.001$ \\
20 & Out of memory & $0.079 \pm 0.017$ & $0.123 \pm 0.023$\\
30 & Out of memory & $0.247 \pm 0.019$& $0.461 \pm 0.024$\\
50 & Out of memory & $1.207 \pm 0.047$& $4.011 \pm 0.112$\\
100 & Out of memory & $9.913 \pm 0.246 $& $121.224 \pm 3.008$\\
\hline
\end{tabular}
\end{table*}

As we can see, optimizing for max variance is more difficult than for RMSE, but ResidualPlanner does it quickly even for data settings too big for  HDMM. The runtime of HDMM increases rapidly, while even for the extreme end of our experiments, ResidualPlanner needs just a few minutes.

\subsection{Scalability of the Reconstruction Phase}
\begin{figure}[h]
\centering
\includegraphics[width=0.45\textwidth]{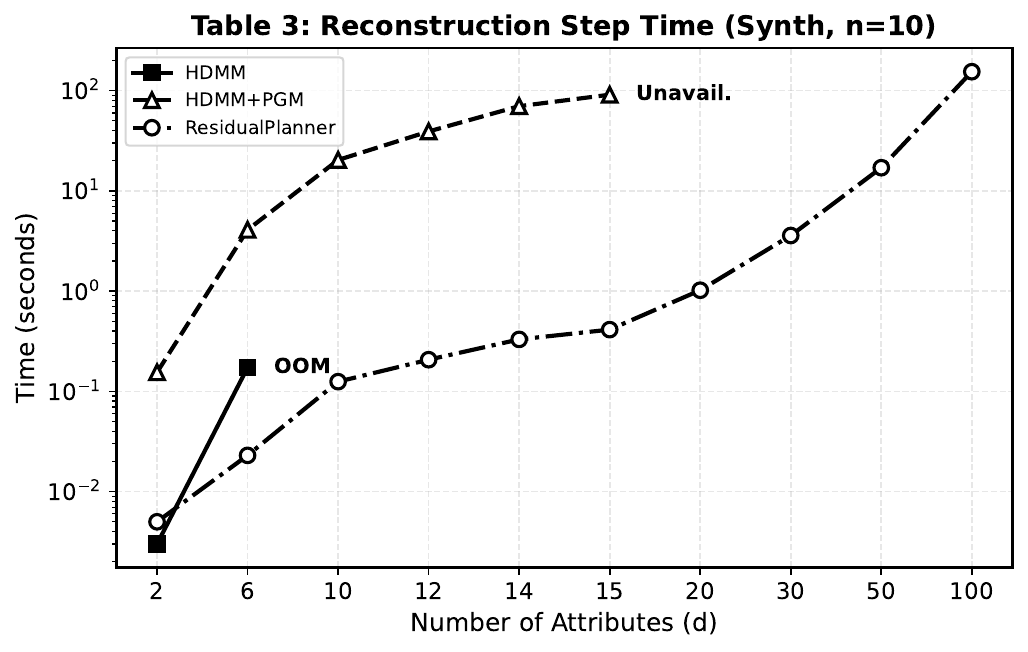}
\caption{Graphical depiction of Table~\ref{tab:time-reconstruct-phase}. 
Reconstruction step running time (seconds) on $\text{Synth-}n^d$ with $n=10$ 
and varying $d$. HDMM fails for $d=10$ (out of memory), HDMM+PGM becomes unavailable at $d=20$ (the selection step failed), 
while ResidualPlanner handles $d=100$ efficiently. }
\label{fig:figtab3}
\end{figure}

We next evaluate the scalability of the reconstruction phase under the same settings. The reconstruction speed for ResidualPlanner does not depend on the objective of the selection phase. Here we compare against HDMM \cite{mckenna2021hdmm} and a version of HDMM with improved reconstruction scalability called HDMM+PGM \cite{mckenna2021hdmm,mckenna2019graphical} (the PGM settings used 50 iterations of  its Local-Inference estimator, as the default 1000 was too slow). Since HDMM cannot perform the selection phase after a certain point, reconstruction results also become unavailable.  Table \ref{tab:time-reconstruct-phase} provides the numerical running time
results and Figure \ref{fig:figtab3} plots the numbers for easier comparison. They show that ResidualPlanner is clearly faster.

\begin{table*}[h!]
\centering
\caption{\textbf{Time for Reconstruction Step in seconds} on Synth$-n^d$ dataset. $n=10$ and the number of attributes $d$ varies. The workload consists of all marginals on $\leq 3$ attributes each. Times are reported with $\pm 2$ standard deviations. Reconstruction can only be performed if the select step completed.}
\label{tab:time-reconstruct-phase}
\begin{tabular}{|c|c|c|c|}\hline
$d$ & \multicolumn{1}{|p{0.25\textwidth}|}{\centering HDMM} & \multicolumn{1}{|p{0.25\textwidth}|}{\centering HDMM + PGM} & \multicolumn{1}{|p{0.19\textwidth}|}{\centering ResidualPlanner}\\\hline
2  & $0.003 \pm 0.0006$ & $0.155 \pm 0.011$  & $0.005 \pm 0.003$ \\
6 &  $0.173 \pm 0.011$ & $4.088 \pm 0.233$  & $0.023 \pm 0.004$\\
10 &  Out of memory in reconstruction& $20.340 \pm 2.264$  & $0.125 \pm 0.032$\\
12 &  Out of memory  in reconstruction& $39.162 \pm 1.739$  & $0.207 \pm 0.004$\\
14 &  Out of memory  in reconstruction& $69.975 \pm 4.037$  & $0.330 \pm 0.006$\\
15 &  Out of memory  in reconstruction& $91.101 \pm 7.621$  & $0.413 \pm 0.006$\\
20 & Unavailable (select step failed) & Unavailable (select step failed) & $1.021 \pm 0.011$\\
30 & Unavailable (select step failed)& Unavailable (select step failed)& $3.587 \pm 0.053$  \\
50 & Unavailable (select step failed)& Unavailable (select step failed)& $17.029 \pm 0.212$\\
100 & Unavailable (select step failed)& Unavailable (select step failed)& $154.538 \pm 15.045 $\\
\hline
\end{tabular}
\end{table*}

\subsection{Accuracy Comparisons}
Since ResidualPlanner is optimal, the purpose of the accuracy comparisons is a sanity check. For RMSE, we compare the quality of  ResidualPlanner to the theoretically optimal lower bound known as the SVD bound \cite{li2013optimal} (they match, as shown in Table \ref{tab:accrmse}). We note ResidualPlanner can provide solutions even when the SVD bound is infeasible to compute. Then we compare ResidualPlanner to HDMM when the user is interested in the maximum variance objective. This just shows that it is important to optimize for the user's objective function and that the optimal solution for RMSE (the only objective HDMM can optimize) is not a good general-purpose approximation for other objectives (as shown in Table \ref{tab:accmax}). Additional comparisons are provided in the supplementary material.

\begin{table*}[h!]
\centering
\caption{RMSE Comparisons to the theoretical lower bound SVD Bound \cite{li2013optimal}}
\label{tab:accrmse}
\begin{tabular}{|c|c|c|| c|c|| c|c|}
\cline{2-7}
\multicolumn{1}{c}{} &
\multicolumn{2}{|c||}{Adult Dataset}&
\multicolumn{2}{c||}{CPS Dataset}&
\multicolumn{2}{c|}{Loans Dataset}
\\\hline
Workload
 & \algoname & SVDB
 & \algoname & SVDB
 & \algoname & SVDB
 \\
\hline
1-way Marginals & 3.047 & 3.047 & 1.744 &1.744 &2.875 &2.875 \\
2-way Marginals & 6.359 & 6.359 &2.035 &2.035 &5.634 &5.634 \\
3-way Marginals & 10.515 & 10.515 &2.048 &2.048 &8.702 &8.702 \\
$\leq 3$-way Marginals & 10.665 &10.665 &2.276 &2.276 & 8.876 & 8.876\\
\hline
\end{tabular}
\end{table*}

\begin{table*}[h!]
\centering
\caption{Max Variance Comparisons with ResidualPlanner and HDMM (showing that being restricted to optimizing only RMSE is not a good approximation of Max Variance optimization).}
\label{tab:accmax}
\begin{tabular}{|c|c|c|| c|c|| c|c|}
\cline{2-7}
\multicolumn{1}{c}{} &
\multicolumn{2}{|c||}{Adult Dataset}&
\multicolumn{2}{c||}{CPS Dataset}&
\multicolumn{2}{c|}{Loans Dataset}
\\\hline
Workload
 & \algoname & HDMM
 & \algoname & HDMM
 & \algoname & HDMM
 \\
\hline
1-way Marginals  & 12.047 &41.772  & 4.346 & 13.672 & 10.640 &33.256\\
2-way Marginals  &67.802 &599.843 &7.897 &47.741 &52.217 &437.478 \\
3-way Marginals  &236.843 &5675.238 & 7.706  &71.549 &156.638 &3095.997 \\
$\leq 3$-way Marginals  &253.605 & 6677.253  &13.216 &415.073 &180.817 &4317.709\\
\hline
\end{tabular}
\end{table*}


\section{Experiments for \newrp}\label{sec:exprange}

We next evaluate \newrp by comparing its performance to HDMM for queries that mix marginals with prefix-sums. 
That is, the base matrices $\attwork_i$ for categorical attributes are the identity matrices and for the numerical attributes the base matrices are prefix-sum matrices. 
Thus, for example, a two-way marginal on (age, race) would provide prefix sums on age for each race (and hence also provide range queries on age for each race). We also experiment with synthetic data for which the numeric domain size of each attribute is small enough to directly represent the complete set of possible range queries. The datasets we use have the following properties.
\begin{enumerate}
    \item The Adult dataset has 5 numerical attributes and 9 categorical attributes.
    \item The CPS dataset has 2 numerical attributes and 3 categorical attributes.
    \item The Loans dataset has 4 numerical attributes and 8 categorical attributes.
    \item The Synthetic-$n^d$ datasets are designed with $d$ numerical attributes, each attribute having $n=10$ possible values. For these datasets, the base matrices are \emph{range queries} (instead of prefix-sums) because the number of possible ranges is tractable. 
\end{enumerate}

For all experiments, we set the privacy cost to be 1. We use the 1-dimensional optimizer included with HDMM to obtain a replacement strategy matrix $\strat_i$ for each base matrix $\attwork_i$ after projecting out the $\one$ vector from each workload query. As HDMM provides several different template strategies to optimize a workload, we used DefaultKron (with approx=True), DefaultUnionKron (with approx=True) and Marginals (with approx=True) and used the best-performing template.

\subsection{Scalability of the Selection Phase}

\begin{table*}[h!]
\centering
\caption{\textbf{Time for Selection Step in seconds} on Synth$-n^d$ dataset for allrange query. $n=10$ and the number of attributes $d$ varies. The workload consists of all range queries on $\leq 3$ attributes. Times for HDMM are reported with $\pm 2$ standard deviations.}
\label{tab:seleRange}
\begin{tabular}{|c|c|c|c|}\hline
$d$ & \multicolumn{1}{|p{0.25\textwidth}|}{\centering HDMM\newline RMSE Objective} & \multicolumn{1}{|p{0.25\textwidth}|}{\centering \newrp \newline RMSE Objective} & \multicolumn{1}{|p{0.25\textwidth}|}{\centering \newrp \newline Max Variance Objective}\\\hline
2  & 0.095$ \pm $ 0.001 & $0.016 \pm 0.001$  & $ 1.455 \pm 0.0158  $ \\
6 & 0.33$  \pm $ 0.001& $0.07 \pm 0.011$  &  $4.528 \pm 0.171$  \\
10 & 0.73 $ \pm $ 0.001 & $0.164 \pm 0.004$  & $9.994 \pm 0.599$ \\
12 & 1.087$ \pm $ 0.001& $0.202 \pm 0.008$  & $15.181 \pm 0.244$ \\
14 & 1.60$ \pm $ 0.001& $0.253 \pm 0.013$  & $16.168\pm3.381$ \\
15 & 1.91 $ \pm $ 0.001 & $0.281 \pm 0.002$  & $18.617\pm3.079$ \\
20 & Out of memory & $0.474 \pm 0.005$  & $57.918\pm3.024$ \\
30 & Out of memory & $1.210 \pm 0.0343$  & $172.041\pm 2.551$ \\
100 & Out of memory & $38.005 \pm 0.168$  & Out of memory \\
\hline
\end{tabular}
\end{table*}

\begin{table*}[h!]
\centering
\caption{\textbf{Time for Reconstruction Step in seconds} on Synth$-n^d$ dataset. $n=10$ and the number of attributes $d$ varies. The workload consists of all range queries on $\leq 3$ attributes. Times are reported with $\pm 2$ standard deviations. Reconstruction can only be performed if the select step completed.}
\label{tab:time-reconstruct-phase-range}
\begin{tabular}{|c|c|c|c|}\hline
$d$ & \multicolumn{1}{|p{0.25\textwidth}|}{\centering HDMM} & \multicolumn{1}{|p{0.25\textwidth}|}{\centering HDMM + PGM} & \multicolumn{1}{|p{0.19\textwidth}|}{\centering \newrp}\\\hline
2  & 0.0026 $ \pm $ 0.001 & 0.14$\pm$ 0.005  & $0.0007\pm  0.0001$ \\
6 & 0.24 $ \pm $ 0.001 & 3.17 $\pm $ 0.069   & $0.0062 \pm 0.0002$ \\
10 &  Out of memory in reconstruction & 19.58 $\pm$ 0.32  & $ 0.0297 \pm 0.0009$ \\
12 &  Out of memory  in reconstruction& 38.32 $\pm $ 0.49  & $ 0.0512\pm  0.002$\\
14 &  Out of memory  in reconstruction& 68.10$\pm $0.49  & $ 0.0816 \pm 0.00352$ \\
15 &  Out of memory  in reconstruction& 87.68$\pm $0.78  & $0.103\pm 0.003$ \\
20 & Unavailable (select step failed) & Unavailable (select step failed)& $0.247 \pm0.02$  \\
30 & Unavailable (select step failed)& Unavailable (select step failed)& $ 0.8455 \pm 0.0296$ \\
50 & Unavailable (select step failed)& Unavailable (select step failed)& $3.987 \pm  0.177$ \\
100 & Unavailable (select step failed)& Unavailable (select step failed)& Out of memory \\
\hline
\end{tabular}
\end{table*}

\begin{table*}[h!] 
  \centering       
  \caption{RMSE Comparisons for prefix-sum queries with \newrp and HDMM.}
  \label{tab:rmseprefix_Rp+}
  \begin{tabular}{|c|c|c|| c|c|| c|c|}
  \cline{2-7}                         
  \multicolumn{1}{c}{} &  
  \multicolumn{2}{|c||}{Adult Dataset}&       
  \multicolumn{2}{c||}{CPS Dataset}&        
  \multicolumn{2}{c|}{Loans Dataset}            
  \\\hline          
  Workload        
   & \newrp & HDMM           
   & \newrp & HDMM           
   & \newrp & HDMM                   
   \\             
  \hline         
  1-way Prefix  & 5.114 & 5.648  & 3.181 &  3.347 & 4.728  &5.127\\     
  2-way Prefix  & 17.632 & 21.111 & 6.357 &  6.625 &14.913 &17.149\\                   
  3-way Prefix  & 47.193 & 60.025 & 8.124 & 8.634 &  36.108 &42.938\\   
  $\leq 3$-way Prefix  & 48.903 & 61.613 & 8.392 & 8.623 & 36.651 &44.338\\ 
  \hline          
  \end{tabular}      
  \end{table*}

\begin{table*}[h!]
  \centering
  \caption{Max Variance Comparisons for prefix-sum queries with \newrp and HDMM.}          
  \label{tab:maxvarprefix}
  \begin{tabular}{|c|c|c|| c|c|| c|c|}
  \cline{2-7}     
  \multicolumn{1}{c}{} &          
  \multicolumn{2}{|c||}{Adult Dataset}&          
  \multicolumn{2}{c||}{CPS Dataset}&            
  \multicolumn{2}{c|}{Loans Dataset}               
  \\\hline         
  Workload
   & \newrp & HDMM        
   & \newrp & HDMM               
   & \newrp & HDMM
   \\               
  \hline      
  1-way Prefix  & 16.247 & 105.440  & 7.158 &  32.788 & 14.631  &87.830\\
  2-way Prefix  & 88.718 & 922.546 & 24.193 &  160.173 &66.074 &631.476\\
  3-way Prefix  & 139.103 & 1958.304 & 12.814 & 57.015 &  90.632 &981.259\\ 
  $\leq 3$-way Prefix  & 165.942 & 1813.099 & 28.526 & 81.030 & 124.318 &1261.880\\
  \hline        
  \end{tabular}     
  \end{table*}  
  
\begin{figure}[h]
\centering
\includegraphics[width=0.45\textwidth]{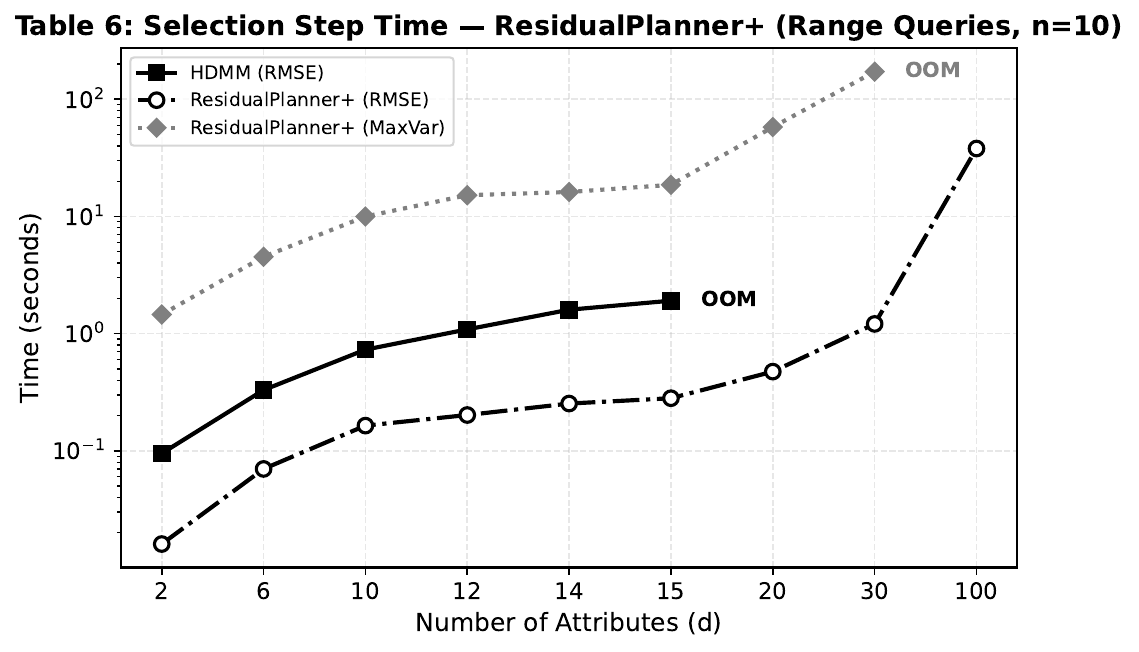}
\caption{Graphical depiction of Table~\ref{tab:seleRange}. 
Selection step running time (seconds) for range queries on $\text{Synth-}n^d$ 
with $n=10$ and varying $d$. HDMM runs out of memory at $d=20$, while 
ResidualPlanner+ scales to $d=100$ for RMSE optimization.}
\label{fig:figtab6}
\end{figure}

We use the Synthetic-$n^d$ datasets to test scalability as the number of attributes $d$ increases and present the results in Table \ref{tab:seleRange} and the graphical version in Figure \ref{fig:figtab6}. The workload is the set of all range queries on $\leq 3$ attributes (this includes range queries on individual attributes, 2-d range queries on all pairs of attributes, and 3-d range queries on all triples of attributes).
We evaluate the time it takes \newrp to optimize for the sum of variances (referred to as RMSE in the table) and the time it takes \newrp to optimize for the max weighted variance. We compare it to the time taken by HDMM for the sum of variances only, as it cannot optimize for the max weighted variance.

 Table \ref{tab:seleRange} provides timing results for the selection phase on Synth-$n_d$ datasets with fixed domain size n = 10 while varying the number of attributes d. The results indicate that \newrp is considerably faster and easily handles dimensionalities that cause HDMM to run out of memory. 

Optimization for maximum variance, a feature HDMM does not support, presents greater computational demands. \newrp manages dimensions up to d = 30 within reasonable timeframes (under 3 minutes). Max variance is a much more difficult optimization task and the limitation becomes apparent at d = 100, where \newrp also runs out of memory.
These findings confirm that \newrp extends the scalability benefits to range queries that were previously demonstrated for marginals.

\subsection{Scalability of Reconstruction Phase}

\begin{figure}[h]
\centering
\includegraphics[width=0.45\textwidth]{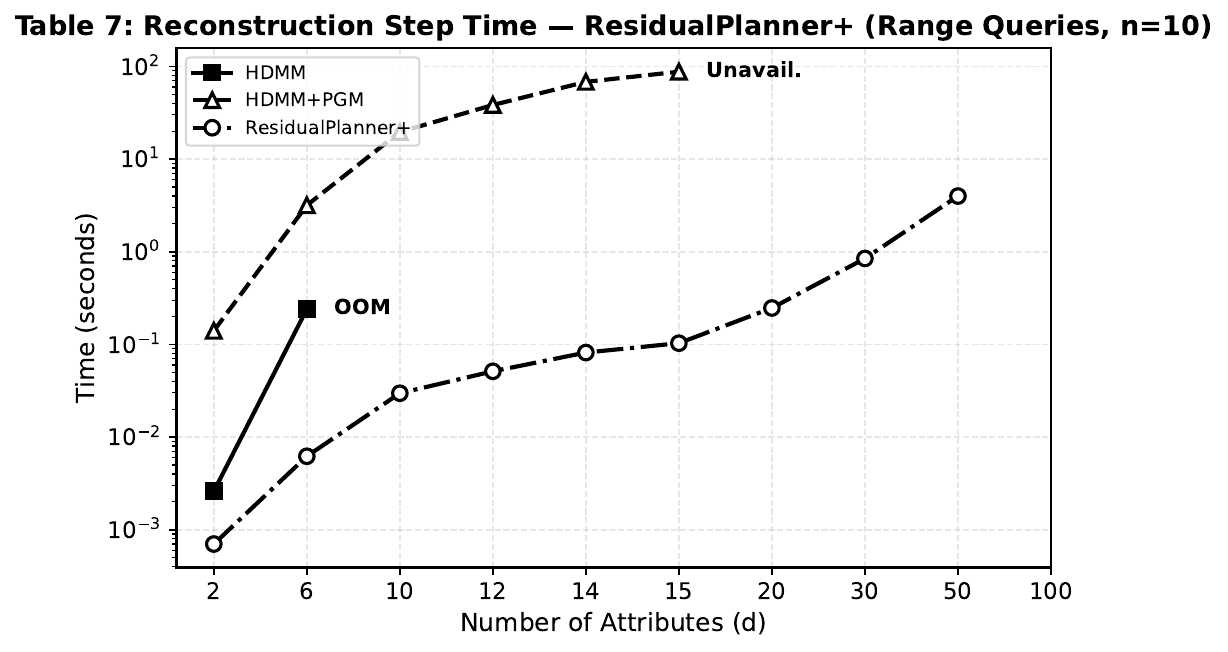}
\caption{Graphical depiction of Table~\ref{tab:time-reconstruct-phase-range}. 
Reconstruction step running time (seconds) for range queries on $\text{Synth-}n^d$ 
with $n=10$ and varying $d$. HDMM fails at $d=10$, HDMM+PGM becomes unavailable 
at $d=20$ (selection step failed), while ResidualPlanner+ handles $d=50$ in under 4 seconds.}
\label{fig:tab7}
\end{figure}

We next evaluate the efficiency of the reconstruction phase. Using the same experimental settings as the selection phase evaluation, we measured reconstruction time for range queries as $d$ increases.
Table \ref{tab:time-reconstruct-phase-range} (and plotted version in Figure \ref{fig:tab7})
demonstrates that \newrp significantly outperforms both the standard HDMM and the optimized HDMM+PGM  \cite{mckenna2021hdmm,mckenna2019graphical} implementation for reconstruction. 
HDMM fails for datasets with as few as 10 attributes. HDMM+PGM fails when the number of attributes is 20, while \newrp can handle 50 attributes in under 4 seconds.

\subsection{ Accuracy Comparisons }
We next evaluate \newrp for the generalized marginal workloads involving prefix queries on numerical attributes using the Adult, CPS and Loan datasets. 
Table \ref{tab:rmseprefix_Rp+} presents RMSE comparisons between \newrp and HDMM. \newrp consistently outperforms HDMM across all workloads and datasets. For 1-way queries (marginals on categorical and prefix-sums on numerical attributes), \newrp outperforms HDMM by 10-15 percent, and the advantage becomes more pronounced with higher-dimensional queries.

When optimizing for maximum variance (Table \ref{tab:maxvarprefix}), the improvements are even more substantial because \newrp has the capability for optimizing this objective function while HDMM does not. 
\newrp's ability to handle different query types and objective functions both efficiently and with high accuracy makes it suitable for real-world  scenarios where understanding distributions across numerical ranges is crucial.
These results demonstrate that ResidualPlanner's design extends successfully to queries beyond standard marginals, maintaining its dual advantages of improved accuracy and scalability across different query types.

\subsection{HDMM/ResidualPlanner Accuracy Crossover.}
\label{sec:kron-discussion}

Feedback from this paper's reviewers helped identify situations where HDMM outperforms \newrp and an interesting accuracy cross-over point phenomenon. Namely,
when a dataset has $d$ attributes and the workload consists of $k$-way queries (e.g., all $k$-dimensional prefix sums), then \newrp performs better when $k$ is 
small compared to $d$ and HDMM performs better when $k$ is close to $d$.



At one extreme, $k=d$, so that the entire workload is represented as a Kronecker product (rather than the union of Kronecker products). This is a setting for which HDMM is optimal \cite{mckenna2021hdmm}. Table \ref{tab:single-kron-range} shows the comparison between HDMM and \newrp for the range query case and Table \ref{tab:single-kron-prefix} shows the comparison
for prefix sums. The variable $n$ represents the domain size of each attribute. As we can see, HDMM outperforms \newrp, and the gap grows with $n$.


\begin{table}[h!]
\centering
\caption{RMSE of the Kronecker product range query workload. In this schema there are $d$ attributes and the domain size of each attribute is $n$. The workload is the set of $d$-dimensional axis-aligned range queries. HDMM is optimal for this setting.}
\label{tab:single-kron-range}
\begin{tabular}{r r r}
\hline
$n$ & HDMM & RP+ \\
\hline
\multicolumn{3}{c}{\textit{d = 3}} \\
\hline
2  & \textbf{1.39} & \textbf{1.39} \\
4  & \textbf{2.11} & 2.25 \\
8  & \textbf{3.34} & 3.64 \\
16 & \textbf{5.29} & 5.77 \\
32 & \textbf{8.18} & 8.87 \\
64 & \textbf{12.20} & 13.18 \\
\hline
\multicolumn{3}{c}{\textit{d = 4}} \\
\hline
2  & \textbf{1.55} & \textbf{1.55} \\
4  & \textbf{2.70} & 2.95 \\
8  & \textbf{5.00} & 5.59 \\
16 & \textbf{9.23} & 10.34 \\
32 & \textbf{16.48} & 18.37 \\
64 & \textbf{28.09} & 31.13 \\
\hline
\multicolumn{3}{c}{\textit{d = 5}} \\
\hline
2  & \textbf{1.73} & \textbf{1.73} \\
4  & \textbf{3.47} & 3.86 \\
8  & \textbf{7.47} & 8.60 \\
16 & \textbf{16.08} & 18.54 \\
32 & \textbf{33.21} & 38.03 \\
64 & \textbf{64.67} & 73.54 \\
\hline
\end{tabular}
\end{table}

\begin{table}[h!]
\centering
\caption{RMSE of the Kronecker product prefix-sum workload. In this schema there are $d$ attributes and the domain size of each attribute is $n$. The workload is the set of $d$-dimensional axis-aligned prefix-sum queries. HDMM is optimal for this setting.}
\label{tab:single-kron-prefix}
\begin{tabular}{r r r}
\hline
$n$ & HDMM & RP+ \\
\hline
\multicolumn{3}{c}{\textit{d = 3}} \\
\hline
2  & \textbf{1.50} & \textbf{1.50} \\
4  & \textbf{2.25} & 2.54 \\
8  & \textbf{3.34} & 4.01 \\
16 & \textbf{4.82} & 5.89 \\
32 & \textbf{6.77} & 8.23 \\
64 & \textbf{9.26} & 11.09 \\
\hline
\multicolumn{3}{c}{\textit{d = 4}} \\
\hline
2  & \textbf{1.71} & \textbf{1.71} \\
4  & \textbf{2.95} & 3.47 \\
8  & \textbf{4.99} & 6.37 \\
16 & \textbf{8.15} & 10.64 \\
32 & \textbf{12.82} & 16.62 \\
64 & \textbf{19.44} & 24.73 \\
\hline
\multicolumn{3}{c}{\textit{d = 5}} \\
\hline
2  & \textbf{1.96} & \textbf{1.96} \\
4  & \textbf{3.87} & 4.74 \\
8  & \textbf{7.45} & 10.12 \\
16 & \textbf{13.76} & 19.23 \\
32 & \textbf{24.25} & 33.56 \\
64 & \textbf{40.83} & 55.16 \\
\hline
\end{tabular}
\end{table}

The difference in performance arises from \newrp's strategy of making subtraction matrices orthogonal to the $\one_{\attr_i}$ queries -- this breaks optimality but improves scalability (as this is the feature that makes the original ResidualPlanner scalable for marginals).

We next present experiments where $k$ varies. For this setting, sometimes the HDMM OPT$_\otimes$ template outpeformed the HDMM OPT$_{+}$ template, so we include results from both.
Table~\ref{tab:sharing} considers $k$-way prefix sum queries over datasets with $d=5$ attributes, with each attribute having domain size $n=10$. \newrp has the best performance for 1-way and 2-way prefix queries, while HDMM performs better for 3-way, 4-way, and 5-way. Here, $k=3$ was the accuracy crossover point.

%

\begin{table}[h!]
\centering
\caption{RMSE on $k$-way prefix-sum workloads over  datasets with $d=5$ dimensions with each attribute having domain size $n=10$. \#marg is the number of $k$-way views (number of ways of choosing $k$ attributes out of $d$)}
\label{tab:sharing}
\begin{tabular}{r r r r r}
\hline
Workload & \#marg & RP+ & OPT$_\otimes$ & OPT$_{+}$ \\
\hline
1-way  & 5  & \textbf{2.94} & 3.59 & 3.48 \\
2-way  & 10 & \textbf{5.84} & 6.32 & 7.66 \\
3-way  & 10 & 9.00 & \textbf{8.44} & 11.92 \\
4-way  & 5  & 11.48 & \textbf{9.40} & 13.11 \\
5-way  & 1  & 12.58 & \textbf{9.13} & --- \\
\hline
\end{tabular}
\end{table}



Table \ref{tab:sharing-10-10} increases $d$ to 10 and again considers $k$-way prefix-sum queries for $k=1,\dots, 5$. In this case, the crossover point happens later, with \newrp performing better for 1-way, 2-way, and 3-way prefix-sum queries, while HDMM is best for $4$-way and above.

\begin{table}[h!]
\centering
\caption{RMSE on $k$-way prefix-sum workloads over  datasets with $d=10$ dimensions with each attribute having domain size $n=10$. \#marg is the number of $k$-way views (number of ways of choosing $k$ attributes out of $d$)}
\label{tab:sharing-10-10}
\begin{tabular}{r r r r r}
\hline
Workload & \#marg & RP+ & OPT$_\otimes$ & OPT$_{+}$ \\

\hline
1-way & 10   & \textbf{3.91}  & 5.11  & 4.92 \\
2-way & 45   & \textbf{10.51} & 12.90 & 16.24 \\
3-way & 120  & \textbf{22.50} & 24.82 & 41.28 \\
4-way & 210  & 40.68 & \textbf{39.95} & 84.98 \\
5-way & 252  & 64.26 & \textbf{56.28} & 144.86 \\
\hline
\end{tabular}
\end{table}

For Table \ref{tab:crossover}, we considered $d=20$ attributes, with each attribute having a domain size of $n=20$. In this setting, \newrp performed better from 1-way up to 6-way prefix-sum queries and HDMM took over for 7-way.

\begin{table}[h!]
\centering
\caption{RMSE on $k$-way prefix-sum workloads over  datasets with $d=20$ dimensions with each attribute having domain size $n=20$. \#marg is the number of $k$-way views (number of ways of choosing $k$ attributes out of $d$)}
\label{tab:crossover}
\begin{tabular}{r r r r r}
\hline
Workload & \#marg & RP+ & OPT$_\otimes$ & OPT$_{+}$ \\

\hline
1-way & 20      & \textbf{6.31}    & 8.95    & 7.84 \\
2-way & 190     & \textbf{27.64}   & 39.51   & 42.40 \\
3-way & 1140    & \textbf{97.23}   & 132.39  & 182.18 \\
4-way & 4845    & \textbf{291.01}  & 369.15  & 658.76 \\
5-way & 15504   & \textbf{765.37}  & 894.65  & 2066.90 \\
6-way & 38760   & \textbf{1805.21} & 1933.59 & --- \\
7-way & 77520   & 3872.58 & \textbf{3790.64} & --- \\
\hline
\end{tabular}
\end{table}

%

\paragraph{Implications for real datasets.}
%

These observations help explain why \newrp outperformed HDMM on real datasets (Adult, CPS, Loans). While their dimensions ranged from 5 - 14, the workloads consisted of 1-way, 2-way, and 3-way queries. We believe that in practice, most datasets and data publishing applications will have a similar pattern -- a moderate to large amount of dimensions, but the queries of interest are low-dimensional.

\section{Conclusion and Future Work.}\label{sec:conc}

In this paper, we introduced ResidualPlanner, a matrix mechanism that is scalable and optimal for marginals under Gaussian noise, for a large class of convex objective functions.  We extended ResidualPlanner to support the numerically secure discrete Gaussian noise distribution. We showed how to optimize for the weighted sum of variances objective  in closed form and applied this result to study unintended consequences of such loss functions.  Finally, we proposed the extension \newrp, which can optimize for a more general class of queries beyond simple marginals, including range and prefix-sum queries, significantly broadening its practical utility. 

 While ResidualPlanner achieves proven optimality for marginals when unbiased query answers and Gaussian-distributed noise are required, additional research is needed to determine whether the combination of scalability,  flexible choice of loss functions, and especially optimality can be extended more generally. Future work includes making \newrp optimal for range queries, handling hierarchical structures (e.g., how many people drive sedans vs. vans; out of the sedans, how many are red vs. green, etc.) \cite{tdahdsr,kuo2018differentially,liu2022private}, how to handle attributes with infinite domains \cite{korolova2009releasing,googlesql}, or working with alternative noise distributions like the Laplace distribution. 



\bibliographystyle{plain}
\bibliography{refs}

\clearpage
\appendix



\section{A Run-through of Residual Planner}
In this section, we provide a complete runthrough of ResidualPlanner using a small toy dataset.

\subsection{A Small Dataset and its Vectorized Representation}

In our example, we have a dataset with 3 attributes, so $\numattr=3$. 
 $\attr_1$ takes values `a' or `b'; $\attr_2$ takes values `y' or `n'; $\attr_3$ takes values 1 or 2 or 3.

In this dataset, there are 5 people, and the tabular representation is shown in Table \ref{tab:toy}.
\begin{table}[h]
\begin{center}
\begin{tabular}{|c|c|c|}\hline
$\attr_1$ & $\attr_2$ & $\attr_3$\\\hline
a & n & 2\\
b & n & 3\\
b & y & 3\\
a & n & 2\\
b & y & 3\\\hline
\end{tabular}
\end{center}
\caption{A Toy Dataset $\data$}\label{tab:toy}
\end{table}
For each attribute, we can one-hot encode its attribute values as row vectors. So, for $\attr_1$, the attribute value 'a' is encoded as $[1, 0]$ and 'b' is encoded as $[0, 1]$. For $\attr_2$, the attribute value 'y' is encoded as $[1, 0]$ and 'n' is encoded as $[0, 1]$. For attribute $\attr_3$, the attribute value '1' is encoded as $[1, 0, 0]$, the value '2' is encoded as  $[0, 1, 0]$ and '3' is encoded as $[0,0,1]$. 

The kronecker product representation of a record is the kronecker product of the one-hot encoding of each attributes. So, for example, the record 'an2' is encoded as the kronecker product $[1, 0] \kron [0, 1]\kron[0,1,0]$. When this kronecker product is expanded, it has 12 components. One of the contains a 1 and the rest contain a 0. Thus the expanded kronecker product can be thought of as a one-hot encoding of the entire record. 

Indeed, in the expanded kronecker product, each dimension of the resulting vector is associated with a record. In table \ref{tab:datakron}, we show the kronecker product representation of each record from Table \ref{tab:toy}. The left column of Table \ref{tab:datakron} shows the record and its kronecker representation. The next 12 columns show the  resulting expansion. Each record becomes as 12-dimensional vector and the column labels in Table \ref{tab:datakron} show which record is associated with which index in the 12-dimensional vector.

The sum of the kron representations of all the records is the data vector $\datavec$.  It is again a 12-dimensional vector. At each index $i$, $\datavec[i]$ is the number of people whose record is associated index $i$. For example, the 5th component is associated with the record 'an2' and there are 2 people with that record. For mathematical convenience, $\datavec$ is treated as a column vector, but for display purposes, in Table \ref{tab:datakron} it is written as a row vector.
\begin{table*}[h]
\resizebox{\textwidth}{!}{%
\begin{tabular}{c|r|r|r|r|r|r|r|r|r|r|r|r|}\cline{2-13}
& $ay1$ & $ay2$ & $ay3$ & $an1$ & $an2$  & $an3$
& $by1$ & $by2$ & $by3$ & $bn1$ & $bn2$  & $bn3$\\\cline{2-13}
an2:\mystrut $[1, 0] \kron [0, 1]\kron[0,1,0]$ & 0 &0 &0 &0 &\textcolor{red}{\textbf{1}} &0 &0 &0 &0 &0 &0 &0\\
bn3: $[0, 1] \kron [0, 1]\kron[0,0,1]$ & 0 &0 &0 &0 &0 &0 &0 &0 &0 &0 &0 &\textcolor{red}{\textbf{1}} \\
by3: $[0, 1] \kron [1, 0]\kron[0,0,1]$ & 0 &0 &0 &0 &0 &0 &0 &0 &\textcolor{red}{\textbf{1}} &0 &0 &0\\
an2: $[1, 0] \kron [0, 1]\kron[0,1,0]$ & 0 &0 &0 &0 &\textcolor{red}{\textbf{1}} &0 &0 &0 &0 &0 &0 &0\\
by3: $[0, 1] \kron [1, 0]\kron[0,0,1]$ & 0 &0 &0 &0 &0 &0 &0 &0 &\textcolor{red}{\textbf{1}} &0 &0 &0\\
\cline{2-13}
\multicolumn{1}{c}{}\\\cline{2-13}
Vector of counts $\datavec$\mystrut: &0 &0 &0 &0 &\textcolor{red}{\textbf{2}} &0 &0 &0 &\textcolor{red}{\textbf{2}} &0 &0 &\textcolor{red}{\textbf{1}}\\\cline{2-13}
\end{tabular}
}
\caption{Kron product representations or each record and the whole dataset $\datavec$. Nonzero components are shown in bold red.}\label{tab:datakron}
\end{table*}

\subsection{The Marginal Workload and its Representation as a Query Matrix.}

For this example, we set the marginal workload to consist of 3 marginals $\margworkload=\{\{\attr_1\}, \{\attr_1,\attr_2\}, \{\attr_2,\attr_3\}\}$.

\begin{table}[h]
\centering
\begin{tabular}{c|c|}
\multicolumn{2}{c}{$\margset=\{\attr_1\}$}\\\multicolumn{2}{c}{}\\\cline{2-2}
\textbf{a} & 2\\
\textbf{b} & 3\\\cline{2-2}
\multicolumn{2}{c}{}
\end{tabular}\hspace{1cm}
\begin{tabular}{c|cc|}
\multicolumn{3}{c}{$\margset=\{\attr_1,\attr_2\}$}\\
\multicolumn{1}{c}{}&\multicolumn{1}{c}{\textbf{y}}&\multicolumn{1}{c}{\textbf{n}}\\\cline{2-3}
\textbf{a} & 0 & 2\\
\textbf{b} & 2 & 1\\\cline{2-3}
\multicolumn{3}{c}{}
\end{tabular}\hspace{1cm}
\begin{tabular}{c|ccc|}
\multicolumn{4}{c}{$\margset=\{\attr_2,\attr_3\}$}\\
\multicolumn{1}{c}{}&\multicolumn{1}{c}{\textbf{1}}&\multicolumn{1}{c}{\textbf{2}}&\multicolumn{1}{c}{\textbf{3}}\\\cline{2-4}
\textbf{y} & 0 & 0 & 2\\
\textbf{n} & 0 & 2 & 1\\\cline{2-4}
\multicolumn{3}{c}{}
\end{tabular}
\caption{True answers to the marginal queries in the marginal workload $\margworkload=\{\{\attr_1\}, \{\attr_1,\attr_2\}, \{\attr_2,\attr_3\}\}$.}\label{tab:marginal}
\end{table}
The marginal on attribute set $\margset=\{\attr_1\}$ has only two cells, which correspond to the number of people with $\attr_1=a$ (i.e., 3) and the number with $\attr_1=b$ (i.e., 3). This is called a one-way marginal. The other marginals are two-way marginals because they involve two attributes. For example, the marginal on $\margset=\{\attr_2,\attr_3\}$ has 6 cells. It represents the number of people for each combination of values for $\attr_2$ and $\attr_3$. For example, there are 2 people with $\attr_2=y$ and $\attr_3=3$.

For each set $\margset$, the marginal on those attributes can be represented as a matrix $\marginal_{\margset}$ such that calculating the marginal is equivalent to the matrix-vector multiplication $\marginal_{\margset}\datavec$. 
The construction of the matrix $\marginal_{\margset}$ is straightforward. It is a kronecker product of 3 matrices. Each matrix corresponds to an attribute. If the attribute is in $\margset$ then the corresponding term is the identity matrix, otherwise is is the row vector full of ones.
For example, $\marginal_{\{\attr_1\}}$ is a kron product of 3 matrices: the first matrix corresponds to $\attr_1$ and is the $2\times 2$ identity matrix. The second matrix is actually the vector full of ones because $\attr_2$ is not part of the marginal. This vector has 2 components because $\attr_2$ has 2 possible values. Similarly, the third matrix is the vector full of ones with 3 components.

For the marginals in $\margworkload$, these are the the corresponding matrices:
\begin{align*}
\marginal_{\{\attr_1\}} &=
\left[
\begin{array}{cc}
1 & 0 \\
0 & 1 \\
\end{array}
\right] \kron
\left[
\begin{array}{cc}
1 & 1 \\
\end{array}
\right]\kron\left[
\begin{array}{ccc}
1 & 1 & 1 \\
\end{array}
\right]
\\
&=\left[
\begin{array}{cccccccccccc}
1 & 1 & 1 & 1 & 1 & 1 & 0 & 0 & 0 & 0 & 0 & 0 \\
0 & 0 & 0 & 0 & 0 & 0 & 1 & 1 & 1 & 1 & 1 & 1 \\
\end{array}
\right]
\\
\marginal_{\{\attr_1,\attr_2\}} &=
\left[
\begin{array}{cc}
1 & 0 \\
0 & 1 \\
\end{array}
\right] \kron
\left[
\begin{array}{cc}
\begin{array}{cc}
1 & 0 \\
0 & 1 \\
\end{array}
\end{array}
\right]\kron\left[
\begin{array}{ccc}
1 & 1 & 1 \\
\end{array}
\right]
\\
&=\left[
\begin{array}{cccccccccccc}
1 & 1 & 1 & 0 & 0 & 0 & 0 & 0 & 0 & 0 & 0 & 0 \\
0 & 0 & 0 & 1 & 1 & 1 & 0 & 0 & 0 & 0 & 0 & 0 \\
0 & 0 & 0 & 0 & 0 & 0 & 1 & 1 & 1 & 0 & 0 & 0 \\
0 & 0 & 0 & 0 & 0 & 0 & 0 & 0 & 0 & 1 & 1 & 1 \\
\end{array}
\right]\\
\marginal_{\{\attr_2,\attr_3\}} &=
\left[
\begin{array}{cc}
1 & 1 \\
\end{array}
\right]\kron
\left[
\begin{array}{cc}
1 & 0 \\
0 & 1 \\
\end{array}
\right] \kron
\left[
\begin{array}{ccc}
1 & 0 & 0 \\
0 & 1 & 0 \\
0 & 0 & 1 \\
\end{array}
\right]\\
&=\left[
\begin{array}{cccccccccccc}
1 & 0 & 0 & 0 & 0 & 0 & 1 & 0 & 0 & 0 & 0 & 0 \\
0 & 1 & 0 & 0 & 0 & 0 & 0 & 1 & 0 & 0 & 0 & 0 \\
0 & 0 & 1 & 0 & 0 & 0 & 0 & 0 & 1 & 0 & 0 & 0 \\
0 & 0 & 0 & 1 & 0 & 0 & 0 & 0 & 0 & 1 & 0 & 0 \\
0 & 0 & 0 & 0 & 1 & 0 & 0 & 0 & 0 & 0 & 1 & 0 \\
0 & 0 & 0 & 0 & 0 & 1 & 0 & 0 & 0 & 0 & 0 & 1 \\
\end{array}
\right]
\\
\end{align*}

If we multiply $\marginal_{\{attr_2,\attr_3\}}$ by the data vector $\datavec$ from Table \ref{tab:datakron}, we get:
\begin{align*}
\marginal_{\{attr_2,\attr_3\}}\datavec &=
\left[
\begin{array}{c}
0 \\
0 \\
2 \\
0 \\
2 \\
1 \\
\end{array}
\right]
\end{align*}
Comparing it to the marginals shown in Table \ref{tab:marginal} we see that it is the flattened version of the marginal. That is, we take the first column of the $\{\attr_2,\attr_3\}$ marginal of Table \ref{tab:marginal}, then we put the next column below it, and the third column is placed at the bottom.

\subsection{The Base Mechanisms}
Recall that our workload of desired privacy-preserving marginals  is $\margworkload=\{\{\attr_1\}, \{\attr_1,\attr_2\}, \{\attr_2,\attr_3\}\}$. Its closure, denoted by $\closure(\margworkload)$ is all of its subsets. So,
\begin{align*}
\lefteqn{\closure(\margworkload)} \\
&= \{~\emptyset, ~\{\attr_1\},~\{\attr_2\},~\{\attr_3\}, \{\attr_1,\attr_2\}, \{\attr_2,\attr_3\}~\}
\end{align*}

For each $\margset\in\closure(\margworkload)$ we need to form a base mechanism $\mech_{\margset}$. Each $\mech_{\margset}$ has a free parameter $\sigma^2_{\margset}$ that we are free to choose. Each mechanism $\mech_{\margset}$ has the form $\mech_{\margset}(\datavec;\sigma^2_{\margset})=\resid_{\margset}\datavec + N(\zero,\covar_{\margset})$. That is, on input $\datavec$, the mechanism multiplies it by a special ``residual'' matrix $\resid_{\margset}$ and then adds  correlated Gaussian noise, with zero mean and with covariance matrix $\sigma^2_{\margset}\covar_{\margset}$. The residual and covariance matrices for each base mechanism are shown below.
\begin{align*}
\lefteqn{\mech_{\emptyset}:} \\
&\quad\resid_{\emptyset}= \left[
\begin{array}{cc}
1 & 1 \\
\end{array}
\right]
\kron
\left[
\begin{array}{cc}
1 & 1 \\
\end{array}
\right]
\kron\left[
\begin{array}{ccc}
1 & 1 & 1 \\
\end{array}
\right]\\
&\quad\covar_{\emptyset}=[1]\\\\
\lefteqn{\mech_{\{\attr_1\}}:}\\
&\quad\resid_{\{\attr_1\}}= 
\left[
\begin{array}{cc}
1 & -1 \\
\end{array}
\right]
\kron
\left[
\begin{array}{cc}
1 & 1 \\
\end{array}
\right]
\kron\left[
\begin{array}{ccc}
1 & 1 & 1 \\
\end{array}
\right]\\
&\quad\covar_{\{\attr_1\}}=\left[
\begin{array}{cc}
1 & -1 \\
\end{array}
\right]\left(\left[
\begin{array}{cc}
1 & -1 \\
\end{array}
\right]\right)^T=[2]
\\\\
\lefteqn{\mech_{\{\attr_2\}}:}\\
&\quad\resid_{\{\attr_2\}}=
\left[
\begin{array}{cc}
1 & 1 \\
\end{array}
\right]
\kron
\left[
\begin{array}{cc}
1 & -1 \\
\end{array}
\right]
\kron\left[
\begin{array}{ccc}
1 & 1 & 1 \\
\end{array}
\right]\\
&\quad\covar_{\{\attr_2\}}=\left[
\begin{array}{cc}
1 & -1 \\
\end{array}
\right]\left(\left[
\begin{array}{cc}
1 & -1 \\
\end{array}
\right]\right)^T=[2]\\\\
\lefteqn{\mech_{\{\attr_3\}}:}\\
&\quad\resid_{\{\attr_3\}}=
\left[
\begin{array}{cc}
1 & 1 \\
\end{array}
\right]
\kron
\left[
\begin{array}{cc}
1 & 1 \\
\end{array}
\right]
\kron\left[
\begin{array}{ccc}
1 & -1 & 0 \\
1 & 0 & -1 \\
\end{array}
\right]
\\
&\quad\covar_{\{\attr_3\}}= \left[
\begin{array}{ccc}
1 & -1 & 0 \\
1 & 0 & -1 \\
\end{array}
\right]
\left(\left[
\begin{array}{ccc}
1 & -1 & 0 \\
1 & 0 & -1 \\
\end{array}
\right]
\right)^T=\left[
\begin{array}{cc}
2 & 1 \\
1 & 2 \\
\end{array}
\right]\\\\
\lefteqn{\mech_{\{\attr_1,\attr_2\}}: }\\
&\quad\resid_{\{\attr_1,\attr_2\}}=
\left[
\begin{array}{cc}
1 & -1 \\
\end{array}
\right]
\kron
\left[
\begin{array}{cc}
1 & -1 \\
\end{array}
\right]
\kron\left[
\begin{array}{ccc}
1 & 1 & 1 \\
\end{array}
\right]
\\
&\quad\covar_{\{\attr_1,\attr_2\}}=
\Big(\left[
\begin{array}{cc}
1 & -1 \\
\end{array}
\right]
\kron
\left[
\begin{array}{cc}
1 & -1 \\
\end{array}
\right]\Big)\Big(\left[
\begin{array}{cc}
1 & -1 \\
\end{array}
\right]
\kron
\left[
\begin{array}{cc}
1 & -1 \\
\end{array}
\right]\Big)^T\\&\phantom{\quad\covar_{\{\attr_1,\attr_2\}}}=[4]\\\\
\lefteqn{\mech_{\{\attr_2,\attr_3\}}:}\\ 
&\quad\resid_{\{\attr_2,\attr_3\}}=\left[
\begin{array}{cc}
1 & 1 \\
\end{array}
\right]
\kron
\left[
\begin{array}{cc}
1 & -1 \\
\end{array}
\right]
\kron\left[
\begin{array}{ccc}
1 & -1 & 0 \\
1 & 0 & -1 \\
\end{array}
\right]\\
&\quad\covar_{\{\attr_2,\attr_3\}}=
\left(\left[
\begin{array}{cc}
1 & -1 \\
\end{array}
\right]
\kron\left[
\begin{array}{ccc}
1 & -1 & 0 \\
1 & 0 & -1 \\
\end{array}
\right]\right)\\
&\phantom{\quad\covar_{\{\attr_2,\attr_3\}}}
\quad*\left(\left[
\begin{array}{cc}
1 & -1 \\
\end{array}
\right]
\kron\left[
\begin{array}{ccc}
1 & -1 & 0 \\
1 & 0 & -1 \\
\end{array}
\right]\right)^T\\
&\phantom{\quad\covar_{\{\attr_2,\attr_3\}}}=
\left[
\begin{array}{cc}
4 & 2 \\
2 & 4 \\
\end{array}
\right]
\end{align*}
Note that for any $\margset$, the residual matrix $\resid_{\margset}$ has a similar structure to $\marginal_{\margset}$ except that where $\marginal_{\margset}$ has an identity matrix in its kron product, $\resid_{\margset}$ has a subtraction matrix (e.g. $\left[\begin{smallmatrix}1 & -1\end{smallmatrix}\right]$ or $\left[\begin{smallmatrix}1 & -1 & \phantom{-}0\\1 &\phantom{-}0 & -1\end{smallmatrix}\right]$). Meanwhile the covariance matrix $\covar_{\margset}$ looks like $\resid_{\margset}\resid_{\margset}^T$ except that the vectors full of 1s have been first removed.

How do we interpret the residual matrices? Well, $\resid_{\emptyset}$ is the sum query. In fact the matrix vector multiplication  $\resid_{\emptyset}\datavec$ gives us the total number of people in the data.

Next, $\resid_{\{\attr_1\}}$ tells us the information contained in the marginal on $\{\attr_1\}$ that is not contained in the sum query. If we know the total number of people in the data, then the only new information the marginal gives us is the difference between the number of people with $\attr_1=a$ and the number of people with $\attr_1=b$. In other words, $\resid_{\{\attr_1\}}\datavec$ is this difference. Given this difference, and the total, once can recover the marginal on attribute $\attr_1$.

Similarly, $\resid_{\{\attr_2\}}$ contains the information in the marginal on $\{\attr_2\}$ that is not provided by the sum query. Finally $\resid_{\{\attr_3\}}$ contains the information in the marginal on $\{\attr_3\}$ not provided in the sum query, which is the number of people with $\attr_3=1$ minus the number with $\attr_3=2$, and also the number of people with $\attr_3=1$ minus the number with $\attr_3=3$. The product $\resid_{\{\attr_3\}}\datavec$ returns those two differences as a vector with two components.

Now, $\resid_{\{\attr_1,\attr_2\}}$ and $\resid_{\{\attr_2,\attr_3\}}$ are more complicated, but have the same idea. For example, $\resid_{\{\attr_1,\attr_2\}}$ represents new information that the marginal on $\{\attr_1, \attr_2\}$ provides that is not captures by the sub-marginals (the marginal on $\{\attr_1\}$ and the marginal on $\{\attr_2\}$.

In general, the matrix $\resid_{\margset}$ represents the new information on that the marginal on $\margset$ provides, which is not captured by the marginals on $\margset^\prime$, for $\margset^\prime\subset\margset$ (strict subsets).

Now, Theorem \ref{thm:linspace} tells us that if we take all of the rows of all of the residual matrices, they will be linearly independent. Furthermore, given an attribute set $\margset$, the total number of rows of $\resid_{\margset^\prime}$ for all $\margset^\prime\subseteq\margset$ is the number of rows in $\marginal_{\margset}$. Furthermore, the space spanned by those rows is the same as the space spanned by the rows of  $\marginal_{\margset}$. 

This also means that if we know $\resid_{\margset^\prime}\datavec$ for all $\margset^\prime\subseteq\margset$ then we can figure out $\marginal_{\margset}\datavec$ (and vice versa).

 Now, we want to get privacy-preserving (noisy) answers to the marginal queries contained in the workload  $\margworkload=\{\{\attr_1\}, \{\attr_1,\attr_2\}, \{\attr_2,\attr_3\}\}$ that are as accurate as possible subject to privacy constraints. We quantify accuracy using a regular (Definition \ref{def:regular}) loss function (e.g., sum of the variances of the answers to the marginals) and we quantify privacy by setting privacy parameters for either $(\epsilon,\delta)$-differential privacy, $\rho$-zCDP, or $\mu$-Gaussian differential privacy.

According to Theorem \ref{thm:optimal}, in order to maximize accuracy subject to privacy constraints, we first need to obtain  $\closure(\margworkload)$, the closure of the workload, which is the  set  $\{~\emptyset, ~\{\attr_1\},~\{\attr_2\},~\{\attr_3\}, \{\attr_1,\attr_2\}, \{\attr_2,\attr_3\}~\}$. Then we must carefully choose positive numbers $\sigma^2_{\margset}$ for each $\margset\in\closure(\margworkload)$ -- so that is 6 numbers total. These numbers are chosen \underline{without} looking at the data (we explain how in Section \ref{sec:exampleloss}). Once we have these numbers, we run the mechanisms $\mech_{\margset}(\datavec;\sigma^2_{\margset})$ and return their outputs. In other words, we must release the outputs of:
\begin{itemize}
\item $\mech_{\emptyset}(\datavec;\sigma^2_{\emptyset})$ -- produces 1 number (a vector with just one component)
\item $\mech_{\{\attr_1\}}(\datavec;\sigma^2_{\{\attr_1\}})$ -- produces 1 number (a vector with just one component)
\item $\mech_{\{\attr_2\}}(\datavec;\sigma^2_{\{\attr_2\}})$ -- produces 1 number (a vector with just one component)
\item $\mech_{\{\attr_3\}}(\datavec;\sigma^2_{\{\attr_3\}})$ -- produces 2 numbers (a vector with 2 components)
\item $\mech_{\{\attr_1,\attr_2\}}(\datavec;\sigma^2_{\{\attr_1,\attr_2\}})$ -- produces 1 number (a vector with 1 component)
\item $\mech_{\{\attr_2,\attr_3\}}(\datavec;\sigma^2_{\{\attr_2,\attr_3\}})$ -- produces 2 numbers (a vector with 2 components)
\end{itemize}
Which gives us 8 total (noisy) numbers. In fact, any matrix mechanism for this workload must return at least 8 noisy numbers, by Theorem \ref{thm:optimal}.

From these outputs, one can reconstruct noisy answers to the marginals in $\margworkload$ (actually one can reconstruct noisy answers to any marginal in $\closure(\margworkload)$). We show how to do this in Section \ref{sec:examplerecon}. Then we show how to compute the privacy cost and variances of the algorithm in Section \ref{sec:examplepcost}.

\subsection{Reconstruction}\label{sec:examplerecon}

Let $\outp_{\margset}$ denote the output of $\mech_{\margset}$. Thus, after running  
\begin{itemize}
\item $\mech_{\emptyset}(\datavec;\sigma^2_{\emptyset})$
\item $\mech_{\{\attr_1\}}(\datavec;\sigma^2_{\{\attr_1\}})$
\item $\mech_{\{\attr_2\}}(\datavec;\sigma^2_{\{\attr_2\}})$
\item $\mech_{\{\attr_3\}}(\datavec;\sigma^2_{\{\attr_3\}})$
\item $\mech_{\{\attr_1,\attr_2\}}(\datavec;\sigma^2_{\{\attr_1,\attr_2\}})$
\item and $\mech_{\{\attr_2,\attr_3\}}(\datavec;\sigma^2_{\{\attr_2,\attr_3\}})$
\end{itemize}
we have the noisy answers 

$\outp_{\emptyset},\quad$
 $\outp_{\{\attr_1\}},\quad$
 $\outp_{\{\attr_2\}},\quad$
 $\outp_{\{\attr_3\}},\quad$
 $\outp_{\{\attr_1,\attr_2\}},\quad$
 $\outp_{\{\attr_2,\attr_3\}}$

From these noisy answers we can produce noisy answers for any marginal in $\margworkload$ or even $\closure(\margworkload)$.
To reconstruct a marginal on $\margset$, we need $\outp_{\margset^\prime}$ for all $\margset^\prime\subseteq \margset$  -- this is not a lot as these vectors represent as many noisy numbers as there are cells in the desired histogram. So, for example, if we want to get noisy answers for the marginal on $\{\attr_2,\attr_3\}$ (which has 6 cells), we need to use $\outp_\emptyset$, $\outp_{\{\attr_2\}}$, $\outp_{\{\attr_3\}}$, and $\outp_{\{\attr_2,\attr_3\}}$ (together these $\outp$ vectors represent a total of 6 noisy numbers).

In order to reconstruct the marginal on $\margset$, Algorithm \ref{alg:reconstruction} multiplies each $\outp_{\margset^\prime}$ by a matrix that depends on both $\margset$ and $\margset^\prime$. The algorithm calls this matrix $\mat{U}$, but to make the notation precise for this runthrough, we will call it $\mat{U}_{\margset\leftarrow\margset^\prime}$ (the $\mat{U}$ matrix that multiplies $\outp_{\margset^\prime}$ when reconstructing $\margset$). It turns out that:
$$\marginal_{\margset}\datavec=\sum_{\margset^\prime\subseteq \margset}\mat{U}_{\margset\leftarrow\margset^\prime}\resid_{\margset^\prime}\datavec$$
which means that the marginal on $\margset$ could be recreated if we know the quantities $\resid_{\margset^\prime}\datavec$ (recall $\resid_{\margset^\prime}$ are the matrices used to define our base mechanisms). Now, since $\outp_{\margset^\prime}$ is a noisy version of $\resid_{\margset^\prime}\datavec$, we can get noisy marginal answers by substituting in these noisy values into the above equation. 

For example, to reconstruct a noisy answer to the marginal on $\{\attr_2,\attr_3\}$, we do the following:
\begin{align*}
\lefteqn{\text{Noisy Marginal on $\{\attr_2,\attr_3\}$}}\\&= (\mat{U}_{\{\attr_2,\attr_3\}\leftarrow\emptyset})\outp_{\emptyset}\\ 
&\phantom{=}+
(\mat{U}_{\{\attr_2,\attr_3\}\leftarrow\{\attr_2\}})\outp_{\{\attr_2\}}\\
&\phantom{=}+
(\mat{U}_{\{\attr_2,\attr_3\}\leftarrow\{\attr_3\}})\outp_{\{\attr_3\}}\\
&\phantom{=}+
(\mat{U}_{\{\attr_2,\attr_3\}\leftarrow\{\attr_2,\attr_3\}})\outp_{\{\attr_2,\attr_3\}}\\
&\text{where}\\
\lefteqn{\mat{U}_{\{\attr_2,\attr_3\}\leftarrow\emptyset} }\\&=\left(\frac{1}{2}\one_2\right)\kron\left(\frac{1}{3}\one_3\right)=\left[
\begin{array}{c}
1/2 \\
1/2 \\
\end{array}
\right]\kron\left[
\begin{array}{c}
1/3 \\
1/3 \\
1/3 \\
\end{array}
\right]\\
\lefteqn{\mat{U}_{\{\attr_2,\attr_3\}\leftarrow\{\attr_2\}} }\\&=\left(\submat_{2}^\dagger\right)\kron\left(\frac{1}{3}\one_3\right)=\left[
\begin{array}{c}
1/2 \\
-1/2 \\
\end{array}
\right]\kron\left[
\begin{array}{c}
1/3 \\
1/3 \\
1/3 \\
\end{array}
\right]\\
\lefteqn{\mat{U}_{\{\attr_2,\attr_3\}\leftarrow\{\attr_3\}} }\\&=\left(\frac{1}{2}\one_2\right)\kron\left(\submat_3^\dagger\right)\\
&=\left[
\begin{array}{c}
1/2 \\
1/2 \\
\end{array}
\right]\kron\left[
\begin{array}{cc}
1/3 & 1/3 \\
-2/3 & 1/3 \\
1/3 & -2/3 \\
\end{array}
\right]
\\
\lefteqn{\mat{U}_{\{\attr_2,\attr_3\}\leftarrow\{\attr_2,\attr_3\}})\outp_{\{\attr_2,\attr_3\}} }\\&=\left(\submat_2^\dagger\right)\kron\left(\submat_3^\dagger\right)\\
&=
\left[
\begin{array}{c}
1/2 \\
-1/2 \\
\end{array}
\right]\kron\left[
\begin{array}{cc}
1/3 & 1/3 \\
-2/3 & 1/3 \\
1/3 & -2/3 \\
\end{array}
\right]
\end{align*}
Note $\submat_2^\dagger$ and $\submat_3^\dagger$ are defined in Lemma \ref{lem:psuedo}.

\subsection{Privacy Cost and Marginal Variances}\label{sec:examplepcost}
Recall that for a marginal workload $\margworkload$, we need to run a mechanism $\mech_{\margset}$ for each $\margset\in\closure(\margworkload)$. Theorem \ref{thm:pcost} shows how to compute the privacy cost $\pcost$ of each. In our running example, this means:
\begin{itemize}
\item $\pcost (\mech_{\emptyset}(\datavec;\sigma^2_{\emptyset})) = \frac{1}{\sigma^2_{\emptyset}}$
\item $\pcost (\mech_{\{\attr_1\}}(\datavec;\sigma^2_{\{\attr_1\}})) = \frac{1}{\sigma^2_{\{\attr_1\}}} * \frac{1}{2} $
\item $\pcost (\mech_{\{\attr_2\}}(\datavec;\sigma^2_{\{\attr_2\}})) = \frac{1}{\sigma^2_{\{\attr_2\}}} * \frac{1}{2} $
\item $\pcost (\mech_{\{\attr_3\}}(\datavec;\sigma^2_{\{\attr_3\}})) = \frac{1}{\sigma^2_{\{\attr_3\}}} * \frac{2}{3} $
\item $\pcost(\mech_{\{\attr_1,\attr_2\}}(\datavec;\sigma^2_{\{\attr_1,\attr_2\}})) = \frac{1}{\sigma^2_{\{\attr_1, \attr_2\}}} *\frac{1}{2}* \frac{1}{2}$
\item and $\pcost(\mech_{\{\attr_2,\attr_3\}}(\datavec;\sigma^2_{\{\attr_2,\attr_3\}})) = \frac{1}{\sigma^2_{\{\attr_2, \attr_3\}}} *\frac{1}{2}* \frac{2}{3}$
\end{itemize}

The total privacy cost is,
\begin{align*}
    \frac{1}{\sigma^2_{\emptyset}} + \frac{1}{2}\frac{1}{\sigma^2_{\{\attr_1\}}}   + \frac{1}{2} \frac{1}{\sigma^2_{\{\attr_2\}}}  +  \frac{2}{3} \frac{1}{\sigma^2_{\{\attr_3\}}} \\+  \frac{1}{4} \frac{1}{\sigma^2_{\{\attr_1, \attr_2\}}} + \frac{1}{3}\frac{1}{\sigma^2_{\{\attr_2, \attr_3\}}}
\end{align*}
Thus this is a symbolic expression in terms of the (currently unknown) noise scale parameters $\sigma^2_{\margset}$. According to Definition \ref{def:lgm}, we can convert the privacy cost to the $\rho$ in $\rho$-zCDP by dividing by 2 and we can convert it to the $\mu$ from $\mu$-Gaussian DP by taking the square root.

For our running example with the workload $\margworkload=\{\{\attr_1\}, \{\attr_1,\attr_2\}, \{\attr_2,\attr_3\}\}$, we can express the variance of these marginals (after reconstruction from the noisy $\outp_{\margset}$ answers) also in terms of the noise scale parameters. We do this with the help of Theorem \ref{thm:var}.

\begin{itemize}
\item \underline{Marginal on $\{\attr_1\}$}. This marginal is reconstructed from the noisy answers $\outp_{\emptyset}$ and $\outp_{\{\attr_1\}}$ and so the variance of its cells depends only on $\sigma^2_\emptyset$ and $\sigma^2_{\{\attr_1\}}$. Applying Theorem \ref{thm:var}, get that the variance in each cell of this marginal is the same and equals.
\begin{align*}
\left( \sigma^2_{\emptyset} *\frac{1}{2^2}\right) + \left(\sigma^2_{\{ \attr_1 \}} * \frac{1}{2} \right)
\end{align*}
\item \underline{Marginal on $\{\attr_1,\attr_2\}$}. This marginal is reconstructed from $\outp_{\emptyset}$, $\outp_{\{\attr_1\}}$, $\outp_{\{\attr_2\}}$, and $\outp_{\{\attr_1,\attr_2\}}$ and hence the variance of the cells in the marginal depend on the corresponding 4 noise scale parameters. The cell variance is
\begin{align*}
     \left(\sigma^2_{\emptyset} *\frac{1}{2^2} * \frac{1}{2^2}\right) + \left(\sigma^2_{\{ \attr_1 \}} * \frac{1}{2} * \frac{1}{2^2}\right) \\
     +   \left(\sigma^2_{\{ \attr_2\}} * \frac{1}{2} * \frac{1}{2^2}\right) +  \left(\sigma^2_{\{ \attr_1, \attr_2 \}} * \frac{1}{2} * \frac{1}{2}\right)
\end{align*}
\item \underline{Marginal on $\{\attr_2,\attr_3\}$}. Similarly, this marginal also depends on 4 noise scale parameters as follows:
\begin{align*}
 \left(\sigma^2_{\emptyset} *\frac{1}{2^2} * \frac{1}{3^2}\right) + \left(\sigma^2_{\{ \attr_2 \}} * \frac{1}{2} * \frac{1}{3^2}\right) \\
 +   \left(\sigma^2_{\{ \attr_3\}} * \frac{2}{3} * \frac{1}{2^2}\right) +  \left(\sigma^2_{\{ \attr_2, \attr_3 \}} * \frac{1}{2} * \frac{2}{3}\right)
\end{align*}
\end{itemize}

\subsection{The Sum-of-Variances Loss Function}\label{sec:exampleloss}
Now  we can express the overall privacy cost symbolically in terms of the noise scale parameters. We can also express the variance of each marginal cell symbolically. We can use these symbolic expressions to set up any regular loss function and then run it through a convex optimizer to solve it.

In this section, we give an example for the weighted sum of variances, which is one of the most popular loss functions for the matrix mechanism in research settings (mostly because this loss function is easiest to work with). 

Each marginal has a weight, which we set to be 1 to avoid introducing more symbols, and the objective function is computed by adding up the cell variances in a marginal, multiplying by the weight, and adding up over the workload marginals. The marginal on $\{\attr_1\}$ has two cells (so we multiply the cell variance for this marginal, computed in the previous section, by 2). The marginal on $\{\attr_1,\attr_2\}$ has 4 cells, and the marginal on $\{\attr_2,\attr_3\}$ has 6 cells. Thus, after the dust clears, the sum of the cell variances across the workload marginals is:

\begin{align*}
       \frac{11}{12} \sigma^2_{\emptyset} +  \frac{3}{2} \sigma^2_{\{ \attr_1 \}} + \frac{5}{6} \sigma^2_{\{ \attr_2 \}}\\
       + \sigma^2_{\{ \attr_3 \}}  + \sigma^2_{\{ \attr_1, \attr_2 \}} 
     +  2\sigma^2_{\{ \attr_2, \attr_3  \}}  
\end{align*}

Thus, we can set up the optimization problem: minimize the sum of variances subject to the privacy cost (computed in Section \ref{sec:examplepcost}) being less than some constant $c$:

\begin{align*}
\arg\min_{\substack{\sigma^2_\emptyset,~\sigma^2_\{\attr_1\}\\ \sigma^2_{\{\attr_2\}},~\sigma^2_{\{\attr_3\}}\\\sigma^2_{\{\attr_1,\attr_2\}},~\sigma^2_{\{\attr_2,\attr_3\}}}}
\left(\substack{
\frac{11}{12} \sigma^2_{\emptyset} +  \frac{3}{2} \sigma^2_{\{ \attr_1 \}} + \frac{5}{6} \sigma^2_{\{ \attr_2 \}}\\ + \sigma^2_{\{ \attr_3 \}}  + \sigma^2_{\{ \attr_1, \attr_2 \}}  +  2\sigma^2_{\{ \attr_2, \attr_3  \}}} \right)\\
\text{such that } \left(\substack{\frac{1}{\sigma^2_{\emptyset}} + \frac{1}{2}\frac{1}{\sigma^2_{\{\attr_1\}}}   + \frac{1}{2} \frac{1}{\sigma^2_{\{\attr_2\}}} \\ +  \frac{2}{3} \frac{1}{\sigma^2_{\{\attr_3\}}} +  \frac{1}{4} \frac{1}{\sigma^2_{\{\attr_1, \attr_2\}}} + \frac{1}{3}\frac{1}{\sigma^2_{\{\attr_2, \attr_3\}}}}\right)\leq c
\end{align*}

If we let the coefficient of $\sigma_{\margset}$ be denoted by $v_{\margset}$ and the coefficient of $1/\sigma^2_{\margset}$ be denoted by $p_{\margset}$, then this optimization problem can be written as:
\begin{align*}
\substack{\argmin\\{\sigma^2_{\margset}:~\margset\in\closure(\margworkload)}} & \sum_{\margset\in\closure(\margworkload) } v_{\margset} \sigma^2_{\margset} \\
     s.t. & \sum_{\margset \in\closure(\margworkload)}\frac{p_{\margset}}{\sigma^2_{\margset}}\leq c
\end{align*}

Lemma \ref{lemma:wsum} in Section \ref{subsec:closedvar} shows that the optimal solution is obtained by computing:
\begin{align*}
T &=\left(\sum_{\margset}\sqrt{v_\margset p_\margset}\right)^2/c\\
&=\frac{\left(\sqrt{\frac{11}{12}*1} + \sqrt{\frac{3}{2}*\frac{1}{2}}+\sqrt{\frac{5}{6}*\frac{1}{2}} + \sqrt{2/3} + \sqrt{1/4} + \sqrt{2/3}\right)^2}{c}\\
&\approx 21.18/c\\
\sigma^2_{\margset}&=\sqrt{Tp_{\margset} / (c v_{\margset} )} \approx \sqrt{21.18 p_{\margset}/v_{\margset}}/c\\
\sigma^2_\emptyset &\approx \sqrt{21.18*12/11}/c\approx 4.8/c
\end{align*}
etc.

\section{Optimality Proof of ResidualPlanner}

In this section, we prove the optimality of ResidualPlanner. It takes advantage of the symmetry inherent in marginals and regular loss functions.

The proof sketch is the following. Given one optimal mechanism $\mech$, we can create a variation $\widetilde{\mech}$ of  that does the following. (1) $\widetilde{\mech}$ modifies each input record by  applying some invertible function $f_i$ to each attribute $\attr_i$ (for example, if $\attr_i$ is a  tertiary attribute, we can modify the value of $\attr_i$ for each record  using a function $f_i$ where $f_i(1)= 3$, $f_i(2)=1$, $f_i(3)=2$). This step can be viewed as simply renaming the attribute values within an attribute. (2) Then $\widetilde{\mech}$ runs $\mech$ on the resulting dataset. Note that marginals can be reconstructed from the output of  $\widetilde{\mech}$ by first running the reconstruction one would do for $\mech$ and then inverting the $f_i$ functions on the resulting marginals (i.e., rearranging the cells in each marginal to undo the within-attribute renaming caused by the $f_i$). This variation $\widetilde{\mech}$ has the same privacy properties as $\mech$ and the same loss (due to the regularity condition on the loss). Hence $\widetilde{\mech}$ is also optimal. 
Then we create yet another optimal privacy mechanism $\mech^*$ that splits the privacy budget across all variations of $\mech$ and returns their outputs. It turns out that the privacy cost matrix of $\mech^*$ has eigenvectors that are equal to the rows of the residual matrices $\resid_{\margset}$ used by ResidualPlanner. Rewriting the privacy cost matrix of $\mech^*$ using this eigendecomposition, we create another mechanism (the mechanism that runs the base mechanisms of ResidualPlanner) that has the same privacy cost matrix and the same value
for the loss and hence is optimal.


The rest of this section explains these steps in details with formal proofs and running commentary that helps to better understand the notation and constructs in the proof.

\subsection{Notation Review}
We first start with a review of key notation. Recall that a dataset  $\data=\{\rec_1,\dots,\rec_\datasize\}$ is a collection of records. Each record $\rec_i$ contains attributes $\attr_1,\dots, \attr_{\numattr}$ and each attribute $\attr_j$ can take values $\attrvalue^{(j)}_1,\dots,\attrvalue^{(j)}_{|\attr_j|}$.

An attribute value $\attrvalue^{(j)}_i$ for attribute $\attr_j$ can be represented as a vector using one-hot encoding. Specifically, let $\e^{(j)}_i$ be a row vector of size $|\attr_j|$ with a one in component $i$ and $0$ everywhere else. In this way, $\e^{(j)}_i$ is a representation of $\attrvalue^{(j)}_i$.

A record $\rec$ with attributes $\attr_1=\attrvalue^{(1)}_{i_1}$, $\attr_2=\attrvalue^{(2)}_{i_2}$, $\dots, \attr_{\numattr}=\attrvalue^{(\numattr)}_{i_{\numattr}}$ can thus be represented as the kron product $\e^{(1)}_{i_1}\kron \e^{(2)}_{i_2}\kron \cdots \kron \e^{(\numattr)}_{i_{\numattr}}$. This vector has a 1 in exactly one position and 0s everywhere else. The position of the 1 is the \emph{index} of record $\rec$.

Thus, a data vector $\datavec$ is a vector of integers. The value at index $i$ is the number of times the record associated with index $i$ appears in $\data$.

\subsection{Permutations}
For each attribute $\attr_i$, let $\permspace^{(i)}$ be the set of permutations on the numbers $1,\dots, {|\attr_i|}$, so that each $\perm\in \permspace^{(i)}$ can be interpreted as a permutation (or renaming) of the attributes values of $\attr_i$. We can also view $\perm$ as a function on vectors of size $|\attr_i|$ that permutes their coordinates. That is, the $i^\text{th}$ coordinate of a vector $\vec{y}$ is the $\perm(i)^\text{th}$ coordinate of $\perm(\vec{y})$.

One can select a permutation for each attribute $\perm^{(1)}\in\permspace^{(1)},\dots, \perm^{(\numattr)}\in\permspace^{(\numattr)}$ and use it to define a permutation over records. This permutation maps a record represented by the kron product $\e^{(1)}_{i_1}\kron \e^{(2)}_{i_2}\kron \cdots \kron \e^{(\numattr)}_{i_{\numattr}}$ into \\ $\perm^{(1)}(\e^{(1)}_{i_1})\kron \perm^{(2)}(\e^{(2)}_{i_2})\kron \cdots \kron \perm^{(\numattr)}(\e^{(\numattr)}_{i_{\numattr}})$.   We can think of this permutation $\perm=(\perm^{(1)},\dots, \perm^{(\numattr)})$ as a function that independently renames each attribute value in a record. Thus this permutation can be extended to datavectors $\datavec$. The value of $\datavec$ at the index associated with  record $\rec$ is the value of $\perm(\datavec)$ at the index associated with record $\perm(\rec)$. Another way to look at it is that $\perm(\datavec)$ is the histogram associated with the dataset $\{\perm(\rec_1), \perm(\rec_2),\dots, \perm(\rec_\datasize)\}$. This permutation can be represented as a permutation matrix $\mat{W}_{\perm}$ such that $\mat{W}_{\perm}\datavec = \perm(\datavec)$.

We let $\permspace = \permspace^{(1)}\times \cdots \times \permspace^{(\numattr)}$ be the set of all such permutations. We call this the space of $\emph{renaming}$ permutations since each $\perm\in\permspace$ renames the values of each attribute separately.

Our first result is that permutation does not affect the privacy parameters of a mechanism.

\begin{theoremEnd}[category=sym,proof here]{lemma}\label{lem:permpriv}
Let $\mech(\datavec)\equiv \bmat\datavec + N(\zero,\covar)$ be a mechanism that satisfies $\rho$-zCDP, $(\epsilon,\delta)$-approximate DP, and $\mu$-Gaussian DP. Let $\perm$ be a permutation of the indices of $\datavec$ and $\mat{W}_\perm$ the corresponding permutation matrix. Then the mechanism $\mech_{\perm}(\datavec)\equiv \bmat\mat{W}_\perm \datavec + N(\zero,\covar)$ satisfies $\rho$-zCDP, $(\epsilon,\delta)$-approximate DP, and $\mu$-Gaussian DP (i.e., with the same privacy parameters).
\end{theoremEnd}
\begin{proofEnd}
The privacy cost $\pcost(\mech)$ of $\mech$ is the largest diagonal of $\bmat^T\covar^{-1}\bmat$. The privacy cost $\pcost(\mech_{\perm})$ of $\mech_\perm$ is the largest diagonal of $\mat{W}_{\perm}^T\bmat^T\covar^{-1}\bmat\mat{W}_{\perm}$. The effect of $\mat{W}_{\perm}$ on both sides is to permute the rows and columns of $\bmat^T\covar^{-1}\bmat$ in the same way. Thus the diagonals of $\bmat^T\covar^{-1}\bmat$ and $\mat{W}_{\perm}^T\bmat^T\covar^{-1}\bmat\mat{W}_{\perm}$ are the same up to permutation and hence $\mech$ and $\mech_\pi$ have the same privacy cost and therefore the same privacy parameters.
\end{proofEnd}

The next result is that a renaming permutation preserves the accuracy of a marginal derived from the answer to a mechanism.
\begin{theoremEnd}[category=sym,proof here]{lemma}\label{lem:margacc}
Let $\margworkload=\{\margset_1,\dots,\margset_k\}$ be a workload on marginals.
Let $\mech(\datavec)\equiv \bmat\datavec + N(\zero,\covar)$ be a mechanism whose output can be used to provide unbiased estimates of those marginals. Let $\perm\in \permspace$ be a renaming permutation and $\mat{W}_\perm$ the corresponding permutation matrix. Define $\mech_{\perm}(\datavec)\equiv \bmat\mat{W}_\perm \datavec + N(\zero,\covar)$. Then unbiased answers to $\margworkload$ can be obtained from the output of $\mech_{\perm}$ and for any regular loss function $\loss$ (Definition \ref{def:regular}) $\loss(\varfun(\margset_1;\mech),\dots,\varfun(\margset_k;\mech))=\\ \loss(\varfun(\margset_1;\mech_{\perm}),\dots,\varfun(\margset_k;\mech_{\perm}))$
\end{theoremEnd}
\begin{proofEnd}
For each set of attributes $\margset_i\in\margworkload$, let $\marginal_{\margset_i}$ be the query matrix of the marginal (i.e., the true marginal is computed as $\marginal_{\margset_i}\datavec$). Then the best linear unbiased estimate of the marginal on $\margset_i$ from the output $\outp$ of $\mech$ is $\marginal_{\margset_i}(\bmat^T\covar^{-1}\bmat)^{\dagger}\bmat^T\covar^{-1}\outp$ and $\varfun(\margset_i;\mech)$ is the diagonal of the covariance matrix of this estimate, which is \\ $\marginal_{\margset_i}(\bmat^T\covar^{-1}\bmat)^{\dagger}\marginal_{\margset_i}^T$. Meanwhile, the best linear unbiased estimate of the marginal on $\margset_i$ from the output $\outp^\prime$ of $\mech_{\perm}$ 
is \\ $\marginal_{\margset_i}(\mat{W}_\perm^T\bmat^T\covar^{-1}\bmat\mat{W}_\perm)^{\dagger}\mat{W}^T_\perm\bmat^T\covar^{-1}\outp^\prime$ and $\varfun(\margset_i;\mech)$ is the diagonal of $\marginal_{\margset_i}(\mat{W}_\perm^T\bmat^T\covar^{-1}\bmat\mat{W}_\perm)^{\dagger}\marginal_{\margset_i}^T = \\ \marginal_{\margset_i}\mat{W}_\perm^T(\bmat^T\covar^{-1}\bmat)^{\dagger}\mat{W}_\perm\marginal_{\margset_i}^T$.

We note that $\marginal_{\margset_i}\mat{W}_\perm^T$ is a permutation of the rows of $\marginal_{\margset_i}$ (computing a marginal on a dataset in which attribute values within the same attribute are renamed is the same as computing the marginal on the original dataset and then renaming the marginal cells, which is permutation of the output of the marginal computation).

Therefore the diagonals of $\marginal_{\margset_i}(\bmat^T\covar^{-1}\bmat)^{\dagger}\marginal_{\margset_i}^T$ and \\ $\marginal_{\margset_i}(\mat{W}_\perm^T\bmat^T\covar^{-1}\bmat\mat{W}_\perm)^{\dagger}\marginal_{\margset_i}^T$ are the same up to permutation. Hence the vector $\varfun(\margset_i; \mech)$ is the same as the vector $\varfun(\margset_i; \mech_\perm)$ up to permutation of the components, and hence does not affect a regular loss function $\loss$.
%
%
%
%
\end{proofEnd}

Finally, we show that there exists an optimal mechanism whose privacy cost matrix exhibits symmetries defined by the set of permutations $\permspace$.

\begin{theoremEnd}[category=sym,proof here]{lemma}\label{lem:pcostsym}
Let $\margworkload=\{\margset_1\dots,\margset_{k}\}$ be a workload of marginal queries. Let $\loss$ be a regular loss function.  Let $U$ be the set of all Gaussian linear mechanisms that can provide unbiased answers to the marginals in the $\margworkload$. Let $\gamma$ be a real number.
Then whenever either of the following optimization problems are feasible,
\begin{align*}
\min_{\mech\in U} \pcost(\mech) ~~\text{ s.t. }\loss(\varfun(\margset_1;\mech), \dots, \varfun(\margset_k;\mech))\leq \gamma\\
\min_{\mech\in U} \loss(\varfun(\margset_1;\mech), \dots, \varfun(\margset_k;\mech)) ~~\text{ s.t. } \pcost(\mech)\leq \gamma
\end{align*}
the feasible optimization problem  is minimized by some mechanism of the form $\overline{\mech}(\datavec)\equiv\overline{\bmat}\datavec + N(\vec{0},\overline{\covar})$ whose  privacy cost matrix $\pcostmat\equiv \overline{\bmat}^T\overline{\covar}^{-1}\overline{\bmat}$ has the following symmetries: for all renaming permutations $\perm\in\permspace$ (with $\mat{W}_\perm$ being the associated permutation matrix), we have $\pcostmat = \mat{W}^T_\perm \pcostmat \mat{W}_\perm$ (in other words, permuting the rows has no effect as long as the columns are permuted in the same way).
\end{theoremEnd}
\begin{proofEnd}
Let $\mech_{opt}(\datavec)\equiv \bmat_{opt}\datavec + N(\zero,\covar_{opt})$ be an optimal mechanism to one of these problems. It may not have the required symmetries, but from it we will construct an optimal mechanism that does.

For a permutation $\perm$ (and corresponding permutation matrix $\mat{W}_\perm$) and a positive number $\lambda$, consider the mechanism $\mech_{\perm,\lambda}(\datavec)\equiv \bmat_{opt}\mat{W}_\perm \datavec + N(\zero,\lambda\covar_{opt})$. By Lemma \ref{lem:margacc}, this
mechanism also answers the marginals in $\margworkload$. 

Now consider the mechanism $\overline{\mech}$ which, on input $\datavec$ outputs the result of $\mech_{\perm, |\permspace|}$ for \underline{all} $\perm\in\permspace$.

The query matrix of $\overline{\mech}$ is 
$\overline{\bmat}=\left[\begin{smallmatrix}
\bmat_{opt}\mat{W}_{\perm_1}\\
\vdots\\
\bmat_{opt}\mat{W}_{\perm_{|\permspace|}}
\end{smallmatrix}\right]$
and the covariance matrix $\overline{\covar}$ is a block diagonal matrix with the scaled matrix $|\permspace|\covar_{opt}$ in each block.
Clearly, by Lemma \ref{lem:margacc}, it also provides unbiased answers to the marginals in $\margworkload$. 

First, we claim that the $\pcost(\overline{\mech})\leq\pcost(\mech_{opt})$ so that the privacy parameters are at least as good. Recall $\pcost(\overline{\mech})$ is the largest diagonal entry of: 

\begin{align}
\overline{\bmat}^T\overline{\covar}^{-1}\overline{\bmat} &=
\frac{1}{|\permspace|}\sum_{\perm\in\permspace} \mat{W}_\perm^T\bmat^{T}_{opt}\covar_{opt}^{-1}\bmat_{opt}\mat{W}_\perm,\label{eqn:pcostsymmetry}
\end{align}
Since the privacy cost $\pcost(\mech_{\perm,1})$ is the largest diagonal of $\mat{W}_\perm^T\bmat^{T}_{opt}\covar_{opt}^{-1}\bmat_{opt}\mat{W}_\perm$ and equals $\pcost(\mech_{opt})$, Equation \ref{eqn:pcostsymmetry} (and convexity of the max function) shows that the $\pcost(\overline{\mech})\leq \pcost(\mech_{opt})$.

Next we consider the loss function. Let $\margset_i\in \margworkload$ be a set of attributes and let $\marginal_{\margset_i}$ be the corresponding query matrix for the marginal on $\margset_i$. Then the reconstructed variances of the answers to this marginal, based on the output of $\overline{\mech}$ is:
\begin{align*}
\varfun(\margset_i;\overline{\mech}) =diag\left(\marginal_{\margset_i}(\overline{\bmat}^T\overline{\covar}^{-1}\overline{\bmat})^{\dagger}\marginal_{\margset_i}^T \right)\\
= diag\left(\frac{1}{|\permspace|}\sum_{\perm\in\permspace}\marginal_{\margset_i} \left(\mat{W}_\perm^T\bmat_{opt}^T\covar^{-1}\bmat_{opt}\mat{W}_\perm\right)^\dagger\marginal_{\margset_i}^T\right)\\
=\frac{1}{|\permspace|}\sum_{\perm\in\permspace} \varfun(\margset_i;\mech_{\perm,1})
\end{align*}
For any $\pi\in \Pi$, Lemma \ref{lem:margacc} tells us that \\ $\loss(\varfun(\margset_1;\mech_{opt}),\dots,\varfun(\margset_k;\mech_{opt}))=\\ \loss(\varfun(\margset_1;\mech_{\perm,1}),\dots,\varfun(\margset_k;\mech_{\perm,1}))$ and so regularity of $\loss$ (which includes convexity), means that\\ 
$\loss(\varfun(\margset_1;\overline{\mech}),\dots,\varfun(\margset_k;\overline{\mech}))\leq \\ \loss(\varfun(\margset_1;\mech_{opt}),\dots,\varfun(\margset_k;\mech_{opt}))$.

Thus $\overline{\mech}$ is no worse in privacy or utility than $\mech_{opt}$ and hence is optimal. 

Thus we consider the symmetries of the privacy cost matrix of $\overline{\mech}$, which is given in Equation \ref{eqn:pcostsymmetry}. Clearly it has the desired symmetry property that $\pcostmat =\mat{W}^T_\perm \pcostmat \mat{W}_\perm$ for any $\perm\in\permspace$ as the permutation space $\permspace$ is an algebraic group.

\end{proofEnd}

\subsection{From permutations to interpretations}
Let $\mech_{opt}(\datavec)\equiv\bmat_{opt}\datavec + N(\zero, \covar_{opt})$ be an optimal mechanism that has the symmetries guaranteed by Lemma \ref{lem:pcostsym}.
Our goal is to use the symmetries in the privacy cost matrix $\pcostmat_{opt}\equiv \bmat_{opt}^T\covar_{opt}^{-1}\bmat_{opt}$ to examine the structure of  $\pcostmat_{opt}$.

If $\gamma_{i,j}$ is the $(i,j)^\text{th}$ entry of $\pcostmat_{opt}$ and if there is a renaming permutation that maps $\rec_i$ (the record associated with index $i$) to some $\rec_{i'}$ (at index $i^\prime$) and maps $\rec_j$ to some $\rec_{j^\prime}$ then $\gamma_{i,j}=\gamma_{i^\prime,j^\prime}$. Note that if $\rec_i$ and $\rec_j$ have the same values for attributes $\attr_1$ and $\attr_2$ then  $\rec_{i^\prime}$ and $\rec_{j^\prime}$ must match on the same attributes because renaming permutations just change the names of values within each attribute. Thus we introduce notation for the set of attributes on which two records match:
\begin{definition}[Common Attributes]\label{def:commonattributes}
Define $\shared$ to be the function that takes two records and outputs the set of attributes on which they match. We emphasize that $\shared(\rec_i,\rec_j)$ is a set of attributes, not attribute values.
\end{definition}
This discussion leads to the following result which characterizes the privacy cost matrix of an optimal mechanism.

\begin{theoremEnd}[category=sym,proof here]{lemma}\label{lem:pcostprop}
Under the same conditions as Lemma \ref{lem:pcostsym}, there exists an optimal mechanism with a privacy cost matrix $\pcostmat_{opt}$ for which the following holds. In addition to the symmetry guaranteed by Lemma \ref{lem:pcostsym}, for every subset of attributes $S\subseteq\{\attr_1,\dots,\attr_{\numattr}\}$, there exists a number $c_{S}$ such that $\gamma_{i,j}$, the $(i,j)^\text{th}$ entry of $\pcostmat_{opt}$, is equal to $c_{\shared(\rec_i,\rec_j)}$. In other words, the $(i,j)^\text{th}$ entry is completely determined by the set $\shared(\rec_i,\rec_j)$ (recall  $\rec_i$ the record value associated with index $i$ and $\rec_j$ is the record value associated with index $j$).
\end{theoremEnd}
\begin{proofEnd}
By Lemma \ref{lem:pcostsym}, there exists an optimal mechanism with privacy cost matrix $\pcostmat_{opt}$ that is invariant under renaming permutations of its rows as long as the columns are permuted in the same way. Thus if $\rec_i$ is the record value corresponding to position $i$ and $\rec_j$ is the record value corresponding to position $j$, there exists a renaming permutation that maps $\rec_i$ to some $\rec_{i^\prime}$ and $\rec_j$ to some $\rec_j^\prime$ if and only if the attributes on which $\rec_i$ and $\rec_j$ match are the same as the attributes on which $\rec_{i^\prime}$ and $\rec_{j^\prime}$ match each other (in symbols: $\shared(\rec_i,\rec_j)=\shared(\rec_{i^\prime},\rec_{j^\prime})$). When there exists such a renaming permutation then $\gamma_{i,j}=\gamma_{i^\prime,j^\prime}$. Thus the value of $\gamma_{i,j}$ is completely determined by $\shared(\rec_i,\rec_j)$ and the result follows.
\end{proofEnd}

From Theorem \ref{thm:linspace}, we know that the rows of the matrices of $\resid_{\margset}$, for all $\margset\subseteq\{\attr_1,\dots,\attr_{\numattr}\}$ are a linearly independent basis for $\mathbb{R}^\dimsize$, where $\dimsize=\prod_{i=1}^\numattr |\attr_i|$. Thus we call the rows a \emph{residual basis}.

\begin{definition}\label{def:rbv}
A row vector $\vec{v}$ is a \emph{residual basis vector} if it is a row in $\resid_{\margset}$ for some $\margset\subseteq\{\attr_1,\dots,\attr_{\numattr}\}$. 
\end{definition}

We now provide an interpretation of the residual bases. First, for an attribute $\attr_\ell$, define the vector $\e_{i,j}^{(\ell)}$ to be a vector of length $|\attr_\ell|$ such that the element at position $i$ is 1, the element at position $j$ is -1 and everywhere else is 0. In other words, $\e^{(\ell)}_{i,j}=\e^{(\ell)}_{i} - \e^{(\ell)}_{j}$ (recall $\e^{(\ell)}_i$ is 1 in position $i$ and 0 everywhere else and is a one-hot encoding of the attribute $\attrvalue^{(\ell)}_i$).
Now, each element of the residual basis has the form $\vec{v}^{(1)}\kron\cdots\kron\vec{v}^{(\numattr)}$ where, for each $\ell$, $\vec{v}^{(\ell)}$ is either the vector $\one_{|\attr_\ell|}^T$ or a vector $\e^{(\ell)}_{1,i_\ell}$. When the vector for attribute $\attr_\ell$ is the vector $\one^T_{|\attr_\ell|}$, we say that \underline{all} attribute values of $\attr_{\ell}$ are \emph{selected}. When the vector for $\attr_{\ell}$ is $\e^{(\ell)}_{1,i_\ell}$, then we say attribute value $\attrvalue^{(\ell)}_{1}$ is \emph{positively selected} and $\attrvalue^{(\ell)}_{i_\ell}$ is \emph{negatively selected} (the other attribute values of $\attr_\ell$ are not selected at all). The attributes for which the kron term is not $\one^T_{|\attr_\ell|}$ are called the \emph{discriminative} attributes.

As an example of this notation and terminology, consider  Table \ref{tab:kron}. Suppose we have three attributes: $\attr_1$ takes values `a' or `b'; $\attr_2$ takes values `y' or `n'; $\attr_3$ takes values 1 or 2 or 3.

\begin{table*}[h]
\resizebox{\textwidth}{!}{%
\begin{tabular}{c|r|r|r|r|r|r|r|r|r|r|r|r|}\cline{2-13}
& $ay1$ & $ay2$ & $ay3$ & $an1$ & $an2$  & $an3$
& $by1$ & $by2$ & $by3$ & $bn1$ & $bn2$  & $bn3$\\\cline{2-13}
bn1: $[0, 1]\kron[0,1]\kron[1,0,0]$ &  
   0 & 0 & 0 & 0 & 0 & 0 & 0 & 0 & 0 & 1 & 0 & 0\\
$[1,1]\kron[1, -1]\kron[1, -1, 0]$&   
   1 & -1 & 0  &-1 & 1 & 0 & 1 & -1 & 0 & -1 & 1 & 0\\
$[1,-1]\kron[1,1]\kron[1, 0, -1]$&  
    1 & 0 & -1 & 1 & 0 & -1 & -1  &0 & 1 & -1 & 0 & 1\\\cline{2-13}
\end{tabular}
}
\caption{Kron product representations.}\label{tab:kron}
\end{table*}
In this case, the data vector $\datavec$ would have 12 components. The first component corresponds to the number of appearances of record ``a,y,1'' in the dataset, the second component corresponds to record ``a,y,2'' and so on. The records corresponding to each index of $\datavec$ are listed in order as the column headings in Table \ref{tab:kron}. The first row shows the representation of record ``b,n,1'' which is composed of the second value (b) for $\attr_1$, the second value (n) for $\attr_2$ and the first value (1) for $\attr_3$. Hence its kron representation is $[0, 1]\kron[0,1]\kron[1,0,0]$ and when the kron product is evaluated, the resulting vector has a 1 in the index corresponding to ``bn1'' (10th column) and 0 everywhere else.

The second and third rows show the expansions of two  residual basis vectors $[1,1]\kron[1, -1]\kron[1, -1, 0]$ (its discriminative attributes are $\attr_2$ and $\attr_3$) and $[1,-1]\kron[1,1]\kron[1, 0, -1]$ (its discriminative attributes are $\attr_1$ and $\attr_3$).  Consider again the kron product $[1,1]\kron[1, -1]\kron[1, -1, 0]$. Note that the first part of the kron product, $[1,1]$ refers to the first attribute and selects both of its values (sets them to 1). The second part of the kron product $[1, -1]$ refers to the $\attr_2$  and positively selects the first attribute value 'y'  (sets it to 1) and negatively selected the second attribute value 'n' (sets it to -1). The third part is $[1, -1, 0]$ and it positively selects the first attribute value, negatively selects the second, but  the third attribute value is not selected at all (i.e., the 3rd position is 0). 
These attribute selections can help us determine what the kron product looks like when it is expanded as follows. For the residual basis vector $\vec{v}^{(1)}\kron\cdots\kron\vec{v}^{(\numattr)}$ the value at the index associated with a record $\rec$ is
\begin{itemize}
\item 0 if $\rec$ has an attribute whose value is not selected by the residual basis vector's kron product. In this case we say the residual basis vector assigns a $0$ to record $\rec$.
For example, in the residual basis vector corresponding to kron product $[1,1]\kron[1, -1]\kron[1, -1, 0]$, the third value of the third attribute is not selected. For any record that assigns the attribute value $3$ to $\attr_3$, this residual basis vector assigns a 0 to such a record.  
\item 1 if for every attribute, the value assigned to it by $\rec$  is selected (posititvely or negatively), and the number of negatively selected attribute values is even. In this case we say the residual basis vector assigns a $1$ to record $\rec$.
\item -1 if the attribute value for each attribute is selected, and the number of negatively selected attribute values is odd. In this case we say the residual basis vector assigns a $-1$ to record $\rec$.
\end{itemize}

 For example, for the residual basis vector\\$[1,1]\kron[1, -1]\kron[1, -1, 0]$, the attribute value 3 for $\attr_3$ is not selected. Hence the value at indices corresponding to records an3,bn3,ay3,by3 are all 0 (see Table \ref{tab:kron}). Next, consider the record an2. The value ``a'' is positively selected, ``n'' is negatively selected, and ``2'' is negatively selected. Hence all attributes are selected and an even number of attributes are negatively selected. Therefore the value at the index associated with an2 is 1. Now for the record by2. The ``b'' is positively selected, ``y'' is positively selected, and ``2'' is negatively selected. Hence there are an odd number of negative selections and so the value at the index associated with by2 is -1.

 With this discussion and associated notation, we can now show that each residual basis vector is an eigenvector of the optimal privacy cost matrix, and the eigenvalue only depends on which attributes are discriminative.

\begin{theoremEnd}[category=sym,proof here]{theorem}\label{thm:pcosteigen}
Under the same conditions as Lemma \ref{lem:pcostsym}, there exists an optimal mechanism such that the eigenvectors of its privacy cost matrix $\pcostmat$ are the residual basis vectors (Definition \ref{def:rbv}). Furthermore, if two residual basis vectors $\vec{v}^{(1)}\kron\cdots\kron\vec{v}^{(\numattr)}$ and $\vec{w}^{(1)}\kron\cdots\kron\vec{w}^{(\numattr)}$ have the same discriminative attributes (i.e., for all $i$, $\vec{w}^{(i)}\neq\one^T_{|\attr_i|}$ if and only $\vec{v}^{(i)}\neq\one^T_{|\attr_i|}$) then the two residual basis vectors have the same eigenvalues (in other words, all rows of the same residual matrix have the same eigenvalues).
\end{theoremEnd}
\begin{proofEnd}
Recall from Definition \ref{def:commonattributes} that $\shared(\rec_i,\rec_j)$ is the set of attributes on which $\rec_i$ and $\rec_j$ are equal.

Let $\pcostmat$ be the privacy cost matrix guaranteed by Lemma \ref{lem:pcostprop} with the properties guaranteed by Lemma \ref{lem:pcostprop}, namely that for every subset of attributes $S\subseteq\{\attr_1,\dots,\attr_{\numattr}\}$, there exists a number $c_{S}$ such that $\gamma_{i,j}$, the $(i,j)^\text{th}$ entry of $\pcostmat$, is equal to $c_{\shared(\rec_i,\rec_j)}$ -- the constant associated with the set $\shared(\rec_i,\rec_j)$, where $\rec_i$ the record value associated with index $i$ and $\rec_j$ is the record value associated with index $j$.

Let $\rec_{\ell}$ be a record associated with index $\ell$. We consider the dot product between a residual basis vector $\vec{v}=\vec{v}^{(1)}\kron\cdots\kron\vec{v}^{(\numattr)}$ and the $\ell^\text{th}$ row of $\pcostmat$. 
Since the entries of the $\ell^{th}$ row are
$c_{\shared(\rec_\ell,\rec_1)},\dots, c_{\shared(\rec_\ell,\rec_\dimsize)}$ and the entries of $\vec{v}$ are 0,1,-1, this dot product can be expressed as:
\begin{align}
\sum_{\substack{\rec \text{ assigned}\\\text{value 1 by } \vec{v}}} c_{\shared(\rec_\ell, \rec)} 
-
\sum_{\substack{\rec \text{ assigned}\\\text{value -1 by } \vec{v}}} c_{\shared(\rec_\ell, \rec)}\label{eqn:pcostbasisdot} 
\end{align}
We analyze this in three cases.

\vspace{0.5cm}
\noindent\textbf{Case 1: $\vec{v}$ assigns a $0$ to $\rec_\ell$.} In this case, there is an attribute for which $\rec_\ell$ has a value that is not selected. Without loss of generality, we may assume this is the first attribute $\attr_1$ so that $\vec{v}^{(1)}=\e_{1,i}$ (the vector with a 1 at the first index and -1 at the $i^\text{th}$ index for some $i$ and 0 everywhere else) and the value of $\attr_1$ for $\rec_\ell$ is therefore not $\attrvalue^{(1)}_{1}$ or $\attrvalue^{(1)}_{i}$ (because $\rec_\ell$ got assigned 0 by $\vec{v}$ due to attribute $\attr_1$). Now, if a record $\rec$ appears in the left summation of Equation \ref{eqn:pcostbasisdot} then its value for $\attr_1$ is either $\attrvalue^{(1)}_{1}$ or $\attrvalue^{(1)}_{i}$ and it does not match $\rec_\ell$ on the first attribute. But this means that we can transform $\rec$ into a record $\rec^\prime$ by replacing $\attrvalue^{(1)}_{1}$ and $\attrvalue^{(1)}_{i}$ with each other. This $\rec^\prime$ would be on the right hand side of the summation (because we are flipping the sign of the selection by $\vec{v}$ of attribute $\attr_1$ in $\rec^\prime$). Furthermore $\rec^\prime$ also does not match $\rec_\ell$ on $\attr_1$ and therefore $\rec$ matches $\rec_\ell$ on exactly the same attributes as $\rec^\prime$ matches $\rec_\ell$. Thus $\shared(\rec_\ell,\rec)=\shared(\rec_\ell,\rec^\prime)$. Thus the summation term from record $\rec$ is cancelled out by $\rec^\prime$ in Equation \ref{eqn:pcostbasisdot}. Using the same argument, we see that every term in the left summation is canceled out by a unique term in the right summation, and vice versa. Hence, \uline{if $\vec{v}$ assigns a 0 to record $\rec_\ell$ (i.e., has a 0 in index $\ell$ when its kron product representation is expanded) then the dot product between $\vec{v}$ and the $\ell^{\text{th}}$ row of $\pcostmat$ is 0.}

\vspace{0.5cm}
\noindent\textbf{Case 2: $\vec{v}$ assigns a $1$ to $\rec_\ell$.} In this case, every attribute of $\rec_\ell$ has a value that is (either positively or negatively) selected by $\vec{v}$ and  an even number are negatively selected. Our goal is to show that if some other record $\rec_t$ is also assigned a $1$ by $\vec{v}$, then the dot product between $\vec{v}$ and $\ell^{\text{th}}$ row of $\pcostmat$ is the same as the dot product between $\vec{v}$ and the $t^\text{th}$ row of $\pcostmat$. 
That is, we want to show:

\begin{align}
\sum_{\substack{\rec \text{ assigned}\\\text{value 1 by } \vec{v}}} c_{\shared(\rec_\ell, \rec)} 
-
\sum_{\substack{\rec \text{ assigned}\\\text{value -1 by } \vec{v}}} c_{\shared(\rec_\ell, \rec)}
&= \nonumber \\
\sum_{\substack{\rec \text{ assigned}\\\text{value 1 by } \vec{v}}} c_{\shared(\rec_t, \rec)} 
-
\sum_{\substack{\rec \text{ assigned}\\\text{value -1 by } \vec{v}}} c_{\shared(\rec_t, \rec)}\label{eqn:pcostbasisdotequal} 
\end{align}

Let $S$ be the set of attributes on which $\rec_\ell$ and $\rec_t$ disagree. Now define a mapping $\phi$ between records such that $\phi$ only modifies attributes in $S$. For each attribute $\attr$ in $S$, it maps the value that record $\rec_\ell$ has into the value that $\rec_t$ has an vice versa. (For example, suppose $S=\{\attr_1,\attr_2\}$ and $\rec_\ell$ has values $\attrvalue^{(1)}_2$ and $\attrvalue^{(2)}_3$ for those attributes, respectively, and suppose that $\rec_t$ has values  $\attrvalue^{(1)}_4$ and $\attrvalue^{(2)}_5$ for those attributes. Then $\phi$ changes $\attrvalue^{(1)}_2$ in $\attr_1$ to $\attrvalue^{(1)}_4$ and changes $\attrvalue^{(1)}_4$ into $\attrvalue^{(1)}_2$; for $\attr_2$ it changes $\attrvalue^{(2)}_3$ into $\attrvalue^{(2)}_5$ and changes $\attrvalue^{(2)}_5$ into $\attrvalue^{(2)}_3$. Thus $\phi(\rec_\ell)=\rec_t$ and $\phi(\rec_t)=\rec_\ell$ and $\phi$ is its own inverse. Furthermore, \uline{for any record $\rec$, $\shared(\rec_\ell,\rec)=\shared(\phi(\rec_\ell),\phi(\rec))=\shared(\rec_t,\phi(\rec))$} since renaming attribute values the same way in two records does not affect the set of attributes on which they match (and the last equality is because $\phi(\rec_\ell)=\rec_t$).

We next note that since $\rec_t$ and $\rec_\ell$ are both assigned $1$ by $\vec{v}$, then they must differ on an even number of discriminative attributes of $\vec{v}$ (if they differ on a discriminative attribute, one must have a value that is positively selected and the other must have a value that is negatively selected -- there cannot be a 0 because $\rec_\ell$ and $\rec_t$ are not assigned a 0 by $\vec{v}$). Therefore, due to its definition, $\phi$ modifies an even number of discriminative attributes and therefore \uline{for any record $\rec$, both $\rec$ and $\phi(\rec)$ get assigned the same value by $\vec{v}$.} 

Putting these facts together, we get:
\begin{align*}
\lefteqn{\sum_{\substack{\rec \text{ assigned}\\\text{value 1 by } \vec{v}}} c_{\shared(\rec_\ell, \rec)} 
-
\sum_{\substack{\rec \text{ assigned}\\\text{value -1 by } \vec{v}}} c_{\shared(\rec_\ell, \rec)} }\\
&=\sum_{\substack{\phi(\rec) \text{ assigned}\\\text{value 1 by } \vec{v}}} c_{\shared(\rec_\ell, \rec)} 
-
\sum_{\substack{\phi(\rec) \text{ assigned}\\\text{value -1 by } \vec{v}}} c_{\shared(\rec_\ell, \rec)} \\&\quad \text{ since $\phi$ doesn't change the summation set}\\
&=\sum_{\substack{\phi(\rec) \text{ assigned}\\\text{value 1 by } \vec{v}}} c_{\shared(\phi(\rec_\ell), \phi(\rec))} 
-
\sum_{\substack{\phi(\rec) \text{ assigned}\\\text{value -1 by } \vec{v}}} c_{\shared(\phi(\rec_\ell), \phi(\rec))}\\&\quad\text{ since $\phi$ preserves the outcome of $\shared$}\\
&=\sum_{\substack{\phi(\rec) \text{ assigned}\\\text{value 1 by } \vec{v}}} c_{\shared(\rec_t, \phi(\rec))} 
-
\sum_{\substack{\phi(\rec) \text{ assigned}\\\text{value -1 by } \vec{v}}} c_{\shared(\rec_t, \phi(\rec))}\\&\quad\text{ since $\phi(\rec_\ell)=\rec_t$}\\
&=\sum_{\substack{\rec' \text{ assigned}\\\text{value 1 by } \vec{v}}} c_{\shared(\rec_t, \rec')} 
-
\sum_{\substack{\rec' \text{ assigned}\\\text{value -1 by } \vec{v}}} c_{\shared(\rec_t, \rec')}\\&\quad\text{ renaming the summation variable from $\phi(\rec)$ to $\rec^\prime$}\\
\end{align*}
and that proves Equation \ref{eqn:pcostbasisdotequal}

\vspace{0.5cm}
\noindent\textbf{Case 3: $\vec{v}$ assigns a $-1$ to $\rec_\ell$.} In this case, every attribute of $\rec_\ell$ has a value that is (either positively or negatively) selected by $\vec{v}$ and  an odd number are negatively selected.
Our goal is to show that if some other record $\rec_t$ is  assigned a $1$ by $\vec{v}$, then the dot product between $\vec{v}$ and $\ell^{\text{th}}$ row of $\pcostmat$ is the \underline{negative of} the dot product between $\vec{v}$ and the $t^\text{th}$ row of $\pcostmat$. 
That is, we want to show:

\begin{align}
\sum_{\substack{\rec \text{ assigned}\\\text{value 1 by } \vec{v}}} c_{\shared(\rec_\ell, \rec)} 
-
\sum_{\substack{\rec \text{ assigned}\\\text{value -1 by } \vec{v}}} c_{\shared(\rec_\ell, \rec)}
&=\nonumber \\
-\sum_{\substack{\rec \text{ assigned}\\\text{value 1 by } \vec{v}}} c_{\shared(\rec_t, \rec)} 
+
\sum_{\substack{\rec \text{ assigned}\\\text{value -1 by } \vec{v}}} c_{\shared(\rec_t, \rec)}\label{eqn:pcostbasisdotneg} 
\end{align}
As in the previous case, we define $\phi$ in the same way and reasoning as before we see that \uline{for any record $\rec$, \\
$\shared(\rec_\ell,\rec)=\shared(\phi(\rec_\ell),\phi(\rec))=\shared(\rec_t,\phi(\rec))$} and since now $\phi$ must change an odd number of discriminative attributes (since $\rec_\ell$ and $\rec_t$ are assigned -1 and 1 by $\vec{v}$) then  \uline{for any record $\rec$, the value assigned  to  $\rec$ by $\vec{v}$ is the negative of the value assigned to $\phi(\rec)$ by $\vec{v}$.} Thus we have:

\begin{align*}
\lefteqn{\sum_{\substack{\rec \text{ assigned}\\\text{value 1 by } \vec{v}}} c_{\shared(\rec_\ell, \rec)} 
-
\sum_{\substack{\rec \text{ assigned}\\\text{value -1 by } \vec{v}}} c_{\shared(\rec_\ell, \rec)} }\\
&=\sum_{\substack{\phi(\rec) \text{ assigned}\\\text{value -1 by } \vec{v}}} c_{\shared(\rec_\ell, \rec)} 
-
\sum_{\substack{\phi(\rec) \text{ assigned}\\\text{value +1 by } \vec{v}}} c_{\shared(\rec_\ell, \rec)}\\&\quad\text{ since $\phi$ flips the summation sets}\\
&=\sum_{\substack{\phi(\rec) \text{ assigned}\\\text{value -1 by } \vec{v}}} c_{\shared(\phi(\rec_\ell), \phi(\rec))} 
-
\sum_{\substack{\phi(\rec) \text{ assigned}\\\text{value +1 by } \vec{v}}} c_{\shared(\phi(\rec_\ell), \phi(\rec))}\\&\quad\text{ since $\phi$ preserves the outcome of $\shared$}\\
&=\sum_{\substack{\phi(\rec) \text{ assigned}\\\text{value -1 by } \vec{v}}} c_{\shared(\rec_t, \phi(\rec))} 
-
\sum_{\substack{\phi(\rec) \text{ assigned}\\\text{value +1 by } \vec{v}}} c_{\shared(\rec_t, \phi(\rec))}\\&\quad\text{ since $\phi(\rec_\ell)=\rec_t$}\\
&=\sum_{\substack{\rec' \text{ assigned}\\\text{value -1 by } \vec{v}}} c_{\shared(\rec_t, \rec')} 
-
\sum_{\substack{\rec \text{ assigned}\\\text{value +1 by } \vec{v}}} c_{\shared(\rec_t, \rec')}\\&\quad\text{ renaming the summation variable from $\phi(\rec')$ to $\rec^\prime$}\\
\end{align*}
and that proves Equation \ref{eqn:pcostbasisdotneg}.

Thus what these 3 cases show us are that there exists some constant $\beta$ such that:
\begin{itemize}
\item If the i$^\text{th}$ position of the expansion of $\vec{v}$ is 0 (i.e., $\rec_i$ is assigned 0 by $\vec{v}$), then the i$^\text{th}$ position of $\pcostmat\vec{v}$ is also 0 (the dot product between the $i^\text{th}$ row and $\vec{v}$ is 0).
\item If the i$^\text{th}$ position of the expansion of $\vec{v}$ is 1 (i.e., $\rec_i$ is assigned 1 by $\vec{v}$), then the i$^\text{th}$ position of $\pcostmat\vec{v}$ is $\beta$ (the dot product between the $i^\text{th}$ row and $\vec{v}$ is $\beta$).
\item If the i$^\text{th}$ position of the expansion of $\vec{v}$ is -1 (i.e., $\rec_i$ is assigned -1 by $\vec{v}$), then the i$^\text{th}$ position of $\pcostmat\vec{v}$ is $-\beta$ (the dot product between the $i^\text{th}$ row and $\vec{v}$ is $-\beta$).
\end{itemize}
Thus $\vec{v}$ is an eigenvector of $\pcostmat$ with eigenvalue $\beta$. That proves the first part of the theorem.

The next part of the theorem is to show that if two residual basis vectors have the same discriminative attributes, then they have the same eigenvalue. So let $\vec{v}=\vec{v}^{(1)}\kron\cdots\kron\vec{v}^{(\numattr)}$ and $\vec{w}=\vec{w}^{(1)}\kron\cdots\kron\vec{w}^{(\numattr)}$ be two residual basis vectors that have the same discriminative attributes. Define a renaming permutation $\perm$ as follows:
\begin{itemize}
\item For an attribute $\attr_\ell$ that is not discriminative for $\vec{v}$ (and hence also not for $\vec{w}$), $\perm$ does not rename its values (i.e., it acts as the identity for those attribute values).
\item For a discriminative attribute $\attr_\ell$, let $\e_{1, i_\ell}$ be the kron component for $\vec{v}$ (i.e., $\vec{v}^{(\ell)}=\e_{1, i_\ell}$) and let $\e_{1, j_\ell}$ be the kron component for $\vec{w}$. Note the indices $i_\ell$ and $j_\ell$ are not equal to 1. In this case, we make $\perm$ do the following renamings:
\begin{itemize}
\item $\attrvalue_{i_\ell}\rightarrow \attrvalue_{j_\ell}$
\item $\attrvalue_{j_\ell}\rightarrow\attrvalue_{i_\ell}$
\item The remaining attribute values are unchanged.
\end{itemize}
\end{itemize}
By considering which records are assigned 1,-1 and 0 by $\vec{v}$ and $\vec{w}$, it is clear that $\perm$ converts $\vec{v}$ into $\vec{w}$ (and vice versa). Let $\mat{W}$ be the matrix representation of the renaming permutation $\perm$, so that $\mat{W}\vec{v}=\vec{w}$ and $\mat{W}^T\vec{w}=\vec{v}$ (a permutation matrix is orthogonal, so its inverse is its transpose).
Thus, letting $\beta$ denote the eigenvalue of $\vec{v}$ with respect to $\pcostmat$, we have:
\begin{align*}
\beta\vec{v} &= \pcostmat \vec{v} \\
&= \pcostmat\mat{W}^T\vec{w}\\
&= \mat{W}^T\pcostmat\mat{W}\mat{W}^T\vec{w} \\&\quad \text{due to the symmetry from Lemma \ref{lem:pcostsym}}\\
&= \mat{W}^T\pcostmat\vec{w},
\intertext{since $\mat{W}^T$ is the inverse of $\mat{W}$ and so}
\beta\vec{w}=\beta \mat{W}\vec{v} &= \mat{W}\mat{W}^T\pcostmat\vec{w}=\pcostmat\vec{w}
\end{align*}
and thus $\vec{w}$ has the same eigenvalue as $\vec{v}$.
\end{proofEnd}

Thus each residual basis matrix $\resid_{\margset}$ has a useful property: its rows are linearly independent and are part of the same  eigenspace (linear space of vectors with the same eigenvalue) of the privacy cost matrix $\pcostmat$ of an optimal mechanism. This allows us to prove the main result:

\printProofs[optmain]

\section{The other proofs for ResidualPlanner base mechanisms}
\printProofs[base]

\section{ResidualPlanner reconstruction proofs}
\printProofs[reconstruct]


\section{ResidualPlanner Computational Complexity Proofs}
\printProofs[complexity]

\section{Discrete Noise Proofs}
\printProofs[discrete]

\section{Proofs for \newrp}
\printProofs[extension]

\section{Proofs related to Trends in Cell Fairness Section}
\printProofs[fair]

\section{Additional Experiments}
In this section, we present additional experiments. Following \cite{mckenna2018optimizing}, the experiments use the following type of workloads:
\begin{itemize}
    \item All $k$-way marginals.
    \item All $\leq 3$-way marginals. This includes all $0$-way marginal (the total sum), all $1$-way marginals, all $2$-way marginals, and all $3$-way marginals.
    \item Small marginals. This includes any $k$-way marginal that has at most 5000 cells.
\end{itemize}
We also use these  metrics:
\begin{itemize}
    \item RMSE: The total variance is the sum of the variances of the reconstructed cells in each marginal in the workload. Root Mean Squared Error is obtained by taking the total variance, dividing by the total number of cells in the workload marginals, then taking the square root. The SVD Bound (SVDB for short) \cite{li2013optimal} provides a theoretical lower bound on RMSE for any matrix mechanism. For marginals, the SVDB is tight, but its computation is not scalable. 
    \item MaxVar: compute the variance of each reconstructed cell for each marginal in the workload, then take the maximum of these. 
    \item Running time (in seconds) of the different stages of the algorithms (select and reconstruct).
\end{itemize}

Unless otherwise stated, ResidualPlanner uses the open-source ECOS optimizer \cite{domahidi2013ecos} for solving the optimization problem it generates for the select step.

For all experiments, we require all mechanisms to have privacy cost $\pcost(\mech) = 1$. By definition \ref{def:lgm}, $\mech$ satisfies $\rho$-zCDP with $\rho=1/2$ \cite{xiao2020optimizing} and satisfies $\mu$-Gaussian DP with $\mu=1$ \cite{fdp,xiao2020optimizing}.  

Each experiment is repeated 5 times, we report the mean value of these 5 results and a confidence interval consisting of $\pm 2$ standard deviations. This is most useful for running time, as the variance loss metrics have negligible variance across all algorithms.

\subsection{Scalability}
In this section, we study the scalability of ResidualPlanner. This is done using the Synth$-n^d$ dataset, where $d$ is the number of attributes and $n$ is the domain size of each attribute.  We use all $\leq 3$-way marginals as a fixed workload and vary $n$ or $d$ to get the computation time for HDMM and ResidualPlanner.

\subsubsection{Varying Attribute Domain Size $n$ in the Selection Step.}
This experiment considers what happens when the attribute domain size $n$ get larger. 
We fix the number of attributes $d=5$ and vary the domain size $n$ for each attribute, where $n$ ranges from 2 to 1024. We evaluate the running time and accuracy of the selection step

Table \ref{tab:time-n-rmse} shows the running time for the selection step  of HDMM and ResidualPlanner. The RMSE  on the workload that the selection step guarantees is also measured.  Both HDMM and ResidualPlanner have no trouble here. HDMM is nearly optimal in RMSE and ResidualPlanner is optimal, as shown by agreement with the SVD Bound. ResidualPlanner is faster, but both methods are fast in this experiment setting.

\begin{table*}[h]
\centering
\caption{Selection step on Synth$-n^d$ dataset where $d=5$ and $n$ varies. The workload is all $\leq $ 3-way marginals. Metrics are running time and RMSE.}
\label{tab:time-n-rmse}
\begin{tabular}{|c|c| c||c| c| c|}
\hline
$n$ & $Time_{HDMM}$ & $Time_{ResPlan}$ & $ RMSE_{HDMM}$ & $RMSE_{ResPlan}$ & SVDB\\
\hline
2 & $0.069 \pm 0.018$ & $0.001 \pm 0.000$ & 1.903 & 1.890  &1.890\\
4 & $0.064 \pm 0.006$ & $0.001 \pm 0.000$&2.685 & 2.681 &2.681\\
8 & $0.070 \pm 0.021$ & $0.001 \pm 0.000$ &3.156 & 3.156 & 3.156\\
16 & $0.076 \pm 0.020$ &$0.001 \pm 0.000$ &3.367 & 3.366 & 3.366\\
32 & $0.105 \pm 0.020$ &$0.001 \pm 0.000$ &3.422 & 3.423 &3.423\\
64 & $0.114 \pm 0.033$ &$0.001 \pm 0.000$ &3.408 & 3.407&3.407\\
128 & $0.137 \pm 0.048$ &$0.001 \pm 0.000$ &3.371 & 3.367 &3.367\\
256 & $0.187 \pm 0.050$ &$0.001 \pm 0.000$ &3.331 & 3.322 & 3.322\\
512 & $0.183 \pm 0.020$ &$0.001 \pm 0.000$ &3.294 & 3.283 & 3.283\\
1024 & $0.353 \pm 0.058$ &$0.001 \pm 0.000$ &3.328 & 3.251 &3.251\\
\hline
\end{tabular}
\end{table*}

Table \ref{tab:time-n-maxvar} shows the running time and Max Variance comparison for the selection step. HDMM can only optimize for RMSE, not max variance, so this table shows that RMSE is not a good substitute when one needs to optimize for Max Variance. \begin{table*}[h]
\centering
\caption{Selection step on Synth$-n^d$ dataset where $d=5$ and $n$ varies. The workload is all $\leq $ 3-way marginals. Metrics are running time and Max Variance. }
\label{tab:time-n-maxvar}
\begin{tabular}{|c|c| c||c| c|}
\hline
$n$ & $Time_{HDMM}$ & $Time_{ResPlan}$ & $ MaxVar_{HDMM}$ & $MaxVar_{ResPlan}$ \\
\hline
2 & $0.069 \pm 0.018$ & $0.008 \pm 0.001$   & 8.091 &  4.148 \\
4 &  $0.064 \pm 0.006$ & $0.008 \pm 0.001$& 44.693 &  9.760 \\
8 & $0.070 \pm 0.021$ & $0.008 \pm 0.001$ &  180.343& 15.643  \\
16 & $0.076 \pm 0.020$ & $0.008 \pm 0.001$ &  588.115&  20.067 \\
32 & $0.105 \pm 0.020$ &$0.008 \pm 0.001$ &  1649.341&  22.811 \\
64 &  $0.114 \pm 0.033$ & $0.008 \pm 0.001$ & 5560.807&  24.345 \\
128 & $0.137 \pm 0.048$ &$0.008 \pm 0.001$ &  12229.480&  25.157 \\
256 & $0.187 \pm 0.050$ & $0.008 \pm 0.001$&  8168.716&  25.574 \\
512 & $0.183 \pm 0.020$ & $0.008 \pm 0.001$ &  32159.958&  25.786 \\
1024 & $0.353 \pm 0.058$ & $0.008 \pm 0.001$ &  277825.955&   25.893\\
\hline
\end{tabular}
\end{table*}

\subsubsection{Impact of varying the number of attributes in the Selection Step.}
Next, we fix the domain size of each attribute to be $n=10$ and vary the number of attributes $d$, where $d$ ranges from 2 to 200. This experiment can test some of the limits of ResidualPlanner. While HDMM cannot perform selection when the number of attributes is 20 or larger, ResidualPlanner has no trouble optimizing RMSE even for 200 attributes. However, optimizing for Max Variance is much more difficult. ResidualPlanner can do this for $d=100$ but the underlying optimization took more than 1 hour for $d=200$ and we killed the process.

Table \ref{tab:time-d-rmse} shows the running time and RMSE comparison for the selection step. The running time of HDMM increases sharply and it quickly runs out of memory. At the same point, the SVD Bound can no longer be computed. Meanwhile, ResidualPlanner continues to run efficiently.

\begin{table*}[h!]
\centering
\caption{
Selection step on Synth$-n^d$ dataset where  $n=10$ and $d$ varies. The workload is all $\leq $ 3-way marginals. Metrics are running time and RMSE. }
\label{tab:time-d-rmse}
\begin{tabular}{|c|c| c||c| c| c|}
\hline
$d$ & $Time_{HDMM}$ & $Time_{ResPlan}$ & $ RMSE_{HDMM}$ & $RMSE_{ResPlan}$ & SVDB\\
\hline
2 & $0.013 \pm 0.003 $ & $0.001 \pm 0.0008 $ & 1.379 & 1.379 & 1.379\\
4 & $0.028 \pm 0.007 $ & $0.002 \pm 0.001$ & 2.346 & 2.345 &2.345\\
6 & $0.065 \pm 0.012 $ & $0.002 \pm 0.0008$ & 4.278 &4.275 &4.275\\
8 & $ 0.167\pm 0.019$&  $0.004 \pm0.001$ & 6.726 &6.638 
 &6.638\\
10 & $0.639 \pm 0.059 $ & $0.009 \pm 0.001$  & 9.629 &9.348 &9.348\\
12 & $4.702 \pm 0.315 $ & $0.015 \pm 0.001$ & 12.904 &12.359 &12.359\\
14 & $46.054 \pm 12.735 $ &  $0.025 \pm 0.002$ & 16.506 &15.642 &15.642 \\
15 & $201.485 \pm 13.697 $ & $0.030 \pm 0.017$ & 18.421 &17.378 &17.378 \\
\hline
20 & Out of memory & $0.079 \pm 0.017$ & Out of memory & 26.916 &Out of memory\\
30 & Out of memory & $0.247 \pm 0.019$ & Out of memory & 49.713 &Out of memory\\
50 & Out of memory & $1.207 \pm 0.047$ & Out of memory & 107.258 &Out of memory\\
100 & Out of memory & $9.913 \pm 0.246$ & Out of memory & 303.216 &Out of memory\\
200 & Out of memory & $80.120 \pm 1.502$ & Out of memory & 855.330 &Out of memory\\
\hline
\end{tabular}
\end{table*}

Table \ref{tab:time-d-mavar} shows the running time and Max Variance comparison on the Selection step. Optimizing for Max Variance is much harder for ResidualPlanner compared to RMSE and we killed the process for $d=200$. Meanwhile, HDMM is not able to run at $d=20$ (we emphasize again, it optimizes for RMSE even if one cares about Max Variance). There is an interesting phenomenon with HDMM that takes place for $d$ between $8$ and $15$. In this case, HDMM always produces a max variance of 1000. This maximum is always achieved for the sum query (a zero-dimensional marginal) for the following reason. For $d$ beween $8$ and $15$, HDMM decides to add noise to all 3-way marginals and nothing else (even though the workload is all $\leq 3$ marginals). The privacy loss
budget is split equally among them. Thus, each of the ${d \choose 3}$ marginals it measures gets $N(0, {d\choose 3})$ noise. The sum query gets reconstructed as follows. For any single noisy 3-way marginal, one can estimate the sum by adding up the cells in the marginal. Since each cell has variance ${d\choose 3}$ and there are $n^3=1,000$ cells, the sum estimate from a single 3-way marginal has a variance of $1000{d\choose 3}$. But one can obtain an independent estimate to the sum query from each of the ${d\choose 3}$ noisy 3-way marginals. By averaging these noisy estimates, one can obtain an estimate of the sum query with variance $1,000$.


\begin{table*}[h!]
\centering
\caption{Selection step on Synth$-n^d$ dataset where  $n=10$ and $d$ varies. The workload is all $\leq $ 3-way marginals. Metrics are running time and Max Variance. }
\label{tab:time-d-mavar}
\begin{tabular}{|c|c| c||c| c|}
\hline
$d$ & $Time_{HDMM}$ & $Time_{ResPlan}$ & $ MaxVar_{HDMM}$ & $MaxVar_{\algoname}$ \\
\hline
2 & $0.013 \pm 0.003 $ &  $0.007 \pm 0.001$ &  13.745&  3.306\\
4 &  $0.028 \pm 0.007 $  & $0.010 \pm 0.005$ &  132.620& 10.480 \\
6 &  $0.065 \pm 0.012 $ &$0.009 \pm 0.001$ &  461.132&  26.904\\
8 &  $ 0.167\pm 0.019$ & $0.015 \pm 0.003$  &  1000.000&  56.961\\
10 &  $0.639 \pm 0.059 $ & $0.018 \pm 0.001$ &  1000.000&  105.031\\
12 &   $4.702 \pm 0.315 $ & $0.028 \pm 0.001$ &  1000.000& 175.496 \\
14 &   $46.054 \pm 12.735 $ & $0.041 \pm 0.001$ &  1000.000& 272.738 \\
15 &  $201.485 \pm 13.697 $  & $0.050 \pm 0.001$ &  1000.000&  332.769 \\
\hline
20 & Out of memory  & $0.123 \pm 0.023$ & Out of memory  & 768.941 \\
30 & Out of memory & $0.461 \pm 0.024$ & Out of memory  &  2540.440\\
50 & Out of memory  & $4.011 \pm 0.112$ & Out of memory  & 11597.037 \\
100 & Out of memory  & $121.224 \pm 3.008$ & Out of memory  & 91960.917 \\
\hline
\end{tabular}
\end{table*}

\subsubsection{Scalability of the Reconstruction Step.}
We conduct similar experiments, but now we measure the time in the reconstruction step. To complement the reconstruction scalability experiments from the main paper on the Synth$-n^d$ synthetic dataset, we first 
 fix the number of attributes $d=5$ and vary the domain size $n$ for each attribute, where $n$ ranges from 2 to 512. The reconstruction time for ResidualPlanner does not depend on the metric that the select step was optimized for. Again we compare with HDMM \cite{mckenna2021hdmm} and a version of HDMM with improved reconstruction scalability called HDMM+PGM \cite{mckenna2021hdmm,mckenna2019graphical} (the PGM settings used 50 iterations of  its Local-Inference estimator, as the default 1000 was too slow).
Table \ref{tab:time-reconstruct} shows the results. 
Again, at some point HDMM runs out of memory while ResidualPlanner runs efficiently. HDMM runs of out memory because of choices it had made in the selection step. When $n=128$ it decided to measure a 5-way marginal, which is so large (requiring $128^5$ space) that it caused HDMM and HDMM+PGM to have memory issues. 
\begin{table*}[h!]
\centering
\caption{Running time (in seconds) of the reconstruction step on Synth$-n^d$ dataset where $d=5$ and $n$ varies. The workload is all $\leq$ 3-way marginals. }
\label{tab:time-reconstruct}
\begin{tabular}{|c|c|c| c|}
\hline
$n$ &HDMM & HDMM + PGM & \algoname \\
\hline
2 & $ 0.005 \pm 0.002 $ & $ 2.466 \pm 0.278 $ & $ 0.008\pm 0.002$ \\
4  & $ 0.005\pm 0.000 $ & $ 1.894 \pm 0.146 $ & $ 0.011\pm 0.008$ \\ 
8  & $ 0.008 \pm 0.000 $ & $ 1.871 \pm 0.122 $ & $0.011 \pm 0.008$ \\
16  & $ 0.064 \pm 0.036 $ & $ 1.936 \pm 0.131 $ & $ 0.016\pm 0.001$ \\
32  & $ 1.924 \pm 0.060$ & $ 3.211 \pm 0.220 $ & $ 0.045\pm 0.007$ \\
64  & $ 56.736	\pm 1.460 $ & $ 12.574 \pm 0.512 $ & $0.217 \pm 0.021$ \\
\hline  
128  & Out of memory & Out of memory & $ 1.244\pm0.059 $ \\
256  & Out of memory & Out of memory & $ 12.090\pm 0.504$ \\
512  & Out of memory & Out of memory & $ 166.045\pm 13.803$ \\
\hline
\end{tabular}
\end{table*}

We next fix $n=3$ and vary $d$.  Table \ref{tab:time-reconstruct-phase-n3} shows ResidualPlanner is clearly faster. Furthermore, HDMM and HDMM+PGM are hampered by the failure of the selection step (when selection fails, there is nothing to reconstruct). It is interesting to compare HDMM+PGM behavior when $n=3$ in Table \ref{tab:time-reconstruct-phase-n3} with $n=10$ in Table \ref{tab:time-reconstruct-phase} from the main paper. Clearly  HDMM+PGM is faster for $n=10$ than $n=3$. This counterintuitive result can be explained by the complex workings of HDMM as follows. When $n=3$, the selection step in HDMM returns some 4-way marginals. But when $n=10$, HDMM only returns $\leq 3$-way marginals. The 4-way marginals make the reconstruction step harder for both HDMM and HDMM + PGM.

\begin{table*}[h!]
\centering
\caption{\textbf{Time for Reconstruction Step in seconds} on Synth$-n^d$ dataset. $n=3$ and the number of attributes $d$ varies. The workload consists of all marginals on $\leq 3$ attributes each. Times are reported with $\pm 2$ standard deviations. Reconstruction can only be performed if the select step completed.}
\label{tab:time-reconstruct-phase-n3}
\begin{tabular}{|c|c|c|c|}\hline
$d$ & \multicolumn{1}{|p{0.25\textwidth}|}{\centering HDMM} & \multicolumn{1}{|p{0.25\textwidth}|}{\centering HDMM + PGM} & \multicolumn{1}{|p{0.19\textwidth}|}{\centering ResidualPlanner}\\\hline
2  & $ 0.001 \pm 0.0001 $ & $ 0.256\pm 0.030 $  & $0.005 \pm 0.002$ \\
6 &  $0.009 \pm 0.001 $ & $ 3.293\pm 0.253 $  & $0.020 \pm 0.004$ \\
10 &  $ 0.334\pm 0.010 $ & $51.568 \pm 3.391 $   & $0.086 \pm 0.004$ \\
12 & $3.882 \pm 0.101$  & $ 180.708 \pm  $ 5.437  & $0.153 \pm 0.002$ \\
14 &  $55.856 \pm 0.361$ & $ 314.252 \pm 3.991$  & $0.280 \pm 0.072$ \\
15 & $231.283 \pm 0.554$  & $ 713.526 \pm 4.957 $  & $0.307 \pm 0.005$ \\
20 & Unavailable (select step failed) & Unavailable (select step failed) & $0.758 \pm 0.023$ \\
30 & Unavailable (select step failed)& Unavailable (select step failed)& $2.700 \pm 0.200$ \\
50 & Unavailable (select step failed)& Unavailable (select step failed)& $12.480 \pm 0.208$ \\
100 & Unavailable (select step failed)& Unavailable (select step failed)& $99.787 \pm 2.113$ \\
\hline
\end{tabular}
\end{table*}

\subsection{Comparison on Real Datasets.}
In this section, we  compare RMSE and Max Variance on the real datasets: CPS, Adult, and Loans. The different workloads are 1-way, 2-way, 3-way, 4-way, 5-way marginals, all $\leq 3$-way marginals, and Small Marginals.

\subsubsection{RMSE Comparisons}
We  provide an expanded comparison of RMSE on the 3 real datasets from the main paper. Here we add more workloads.
Table \ref{tab:cps-rmse}, \ref{tab:adult-rmse} and \ref{tab:loans-rmse} show the comparison of RMSE on the CPS, Adult, and Loans datasets respectively.

We notice that ResidualPlanner matches the theoretical SVD Bound while HDMM is slightly worse, but still accurate. We conclude that when optimizing RMSE, the main advantage of ResidualPlanner is superior scalability.

\begin{table*}[h]
\centering
\caption{Comparison of RMSE on CPS(5D) dataset.}
\label{tab:cps-rmse}
\begin{tabular}{|c|c|c|c|}
\hline
Workload & HDMM & \algoname & SVDB
 \\
\hline
1-way Marginals & 1.756& 1.744 & 1.744  \\
2-way Marginals & 2.103& 2.035 &  2.035\\
3-way Marginals & 2.089& 2.048 &  2.048\\
4-way Marginals & 1.648& 1.627 &  1.627 \\
5-way Marginals & 1.000& 1.000 &  1.000\\
$\leq$ 3-way Marginals & 2.301& 2.276 &  2.276\\
Small Marginals & 2.525 & 2.525 & 2.525 \\
\hline
\end{tabular}
\end{table*}

\begin{table*}[h]
\centering
\caption{Comparison of RMSE on Adult(14D) dataset. }
\label{tab:adult-rmse}
\begin{tabular}{|c|c|c|c|}
\hline
Workload & HDMM & \algoname & SVDB
 \\
\hline
1-way Marginals & 3.081& 3.047 & 3.047 \\
2-way Marginals & 6.504& 6.359 &  6.359\\
3-way Marginals & 11.529& 10.515 &  10.515\\
4-way Marginals & 16.618& 14.656 &  14.656\\
5-way Marginals & 20.240& 17.844 &  17.844\\
$\leq$ 3-way Marginals & 11.555& 10.665 &  10.665\\
Small Marginals & 10.006 & 9.945 &  9.945\\
\hline
\end{tabular}
\end{table*}

\begin{table*}[h]
\centering
\caption{Comparison of RMSE on Loans(12D) dataset. }
\label{tab:loans-rmse}
\begin{tabular}{|c|c|c|c|}
\hline
Workload & HDMM & \algoname & SVDB
 \\
\hline
1-way Marginals & 2.903& 2.875 &  2.875\\
2-way Marginals & 5.747& 5.634 &  5.634\\
3-way Marginals & 9.478& 8.702 &  8.702\\
4-way Marginals & 12.537& 11.267 &  11.267\\
5-way Marginals & 14.872& 12.678 &  12.678\\
$\leq$ 3-way Marginals & 9.406& 8.876 &  8.876\\
Small Marginals & 8.262 & 8.206 &  8.206\\
\hline
\end{tabular}
\end{table*}

\subsubsection{Max Variance}

The next comparison is on optimization for Max Variance. We repeat that HDMM only optimizes for RMSE and this shows that optimizing for RMSE is highly suboptimal when one cares about max variance.

In contrast to RMSE, where the optimization problem generated by ResidualPlanner's selection step can be solved in closed form, for Max Variance, the optimization needs a convex solver. Hence we include comparisons between the open source ECOS \cite{domahidi2013ecos} optimizer to the commercial Gurobi optimizer \cite{gurobi}.
Thus, our results have columns labeled ResidualPlanner+ECOS and ResidualPlanner+Gurobi.

Tables \ref{tab:cps-maxvar}, \ref{tab:adult-maxvar} and \ref{tab:loans-maxvar} show the results for the CPS, Adult, and Loans datasets, respectively.
There is one item to note about numerical stability. Although Gurobi is generally faster and more numerically stable, the differences do not matter much. Situations where EOCS was worse are highlighted in red. For example, in  Table \ref{tab:cps-maxvar} for the CPS dataset,
the dataset has only 5 attributes, so a 5-way marginal is basically the entire dataset. The optimal mechanism for 5-way marginals simply adds $N(0,1)$ noise to each cell and optimizing for RMSE is equal to optimizing Max Variance for this special case. As we see, the Max Variance for ResidualPlanner+ECOS is $1.008$ which is $0.8\%$ worse than optimal. The reason for this is the numerical precision with which ECOS can solve the optimization problem that ResidualPlanner gives it. In general, however, it looks like open source optimizers should work fairly reliably for them to be used in real applications of ResidualPlanner.

\begin{table*}[h!]
\centering
\caption{Comparison of Max Variance on CPS(5D) dataset.}
\label{tab:cps-maxvar}
\begin{tabular}{|c|c|c|c|c|}
\hline
Workload & HDMM & \algoname + ECOS & \algoname + Gurobi
 \\
\hline
1-way Marginals & 13.672& 4.346  & 4.346\\
2-way Marginals & 47.741&  7.897  & 7.897\\
3-way Marginals & 71.549&  7.706  & 7.706\\
4-way Marginals & 15.538&  \addcolor{4.142}  & 4.141\\
5-way Marginals & 1.000&  \addcolor{1.008}  & 1.000\\
$\leq$ 3-way Marginals &   415.073& 13.216 & 13.216\\
Small Marginals & 223.579 &   11.774 & 11.774\\
\hline
\end{tabular}
\end{table*}
\begin{table*}[h!]
\centering
\caption{Comparison of Max Variance on Adult(14D) dataset.}
\label{tab:adult-maxvar}
\begin{tabular}{|c|c|c|c|c|}
\hline
Workload & HDMM & \algoname + ECOS & \algoname + Gurobi
 \\
\hline
1-way Marginals & 41.772&  12.047 & 12.047\\
2-way Marginals & 599.843&  67.802  & 67.802\\
3-way Marginals & 5675.238&  236.843  & 236.843\\
4-way Marginals & 26959.322&   575.213 & 575.213\\
5-way Marginals & 79817.002&  1030.948  & 1030.948\\
$\leq$ 3-way Marginals &   6677.253& 253.605 & 253.605\\
Small Marginals & 2586.980 &   126.902 & 126.902\\
\hline
\end{tabular}
\end{table*}
\begin{table*}[h!]
\centering
\caption{Comparison of Max Variance on Loans(12D) dataset.}
\label{tab:loans-maxvar}
\begin{tabular}{|c|c|c|c| c|}
\hline
Workload & HDMM & \algoname + ECOS & \algoname + Gurobi
 \\
\hline
1-way Marginals & 33.256&  10.640 & 10.640\\
2-way Marginals & 437.478&  52.217  & 52.217\\
3-way Marginals & 3095.997&  156.638  & 156.638\\
4-way Marginals & 13776.417&   320.778 &320.778\\
5-way Marginals & 26056.289&  \addcolor{474.244}  & 474.243\\
$\leq$ 3-way Marginals & 4317.709  & 180.817 &180.817\\
Small Marginals & 2330.883 &   89.873 &89.873\\
\hline
\end{tabular}
\end{table*}



%



\end{document}